\documentclass[prb,aps,superscriptaddress,showpacs,floatfix,twocolumn]{revtex4}
\usepackage{graphicx,epsfig}
\usepackage{ulem}
\usepackage{amssymb}
\usepackage{amsfonts}
\usepackage{amsmath}
\usepackage{graphicx}
\usepackage[usenames]{color}

\newcommand{\bk}{{\bf k}}

\newcommand{\bq}{{\bf q}}

\newcommand{\bw}{{\bf w}}

\providecommand{\U}[1]{\protect\rule{.1in}{.1in}}

\begin{document}

\title{Electron-phonon heat exchange in layered nano-systems}
\author{D. V. Anghel}
\affiliation{Horia Hulubei National Institute For Physics And Nuclear Engineering,
RO-077125, M\u agurele, Romania}

\author{S. Cojocaru}
\affiliation{Horia Hulubei National Institute For Physics And Nuclear Engineering,
RO-077125, M\u agurele, Romania}


\begin{abstract}

We analyze the heat power $P$ between electrons and phonons in thin metallic films deposited on free-standing dielectric membranes in a temperature range in which the phonon gas has a quasi two-dimensional distribution.
The quantization of the electrons wavenumbers in the direction perpendicular to the film surfaces lead to the formation of quasi two-dimensional electronic sub-bands.
The electron-phonon coupling is treated in the deformation potential model and, if we denote by $T_e$ the electrons temperature and by $T_{ph}$ the phonons temperature, we find that $P\equiv P^{(0)}(T_{e})-P^{(1)}(T_{e},T_{ph})$; $P^{(0)}$ is the power ``emitted'' by the electron system to the phonons and $P^{(1)}$ is the power ``absorbed'' by the electrons from the phonons.

Due to the quantization of the electronic states, $P$ vs $(d,T_e)$ and $P$ vs $(d,T_{ph})$ show very strong oscillations with $d$, forming sharp crests almost parallel to the temperature axes. In the valleys between the crests, $P \propto T_e^{3.5} - T_{ph}^{3.5}$.
From valley to crest, $P$ increases by more than one order of magnitude and on the crests $P$ does not have a simple power law dependence on temperature.

The strong modulation of $P$ with the thickness of the film may provide a way to control the electron-phonon heat power and the power dissipation in thin metallic films. Eventually the same mechanism may be used to detect small variations of $d$ or surface contamination. On the other hand, the surface imperfections of the metallic films may make it difficult to observe the oscillations of $P$ with $d$ and eventually due to averaging the effects the heat flow would have a more smooth dependence on the thickness in real experiments.
\end{abstract}

\maketitle




\section{Introduction}

In a recent paper Nquyen et al. \cite{PhysRevApplied.2.054001.2014.Nguyen}
reported remarkable cooling properties of normal
metal-insulator-superconductor (NIS) tunnel junctions refrigerators, by
reaching electronic temperatures of 30~mK or below, from a bath temperature
of 150~mK, at a cooling power of 40~pW. Such micro-refrigerators have great
potential for applications, since they can be mounted directly on chips for
cooling qubits or ultra-sensitive detectors, like micro-bolometers or
micro-calorimeters.


The principle of operation of NIS micro-refrigerators has been explained in
several publications (e.g. \cite{ApplPhysLett.65.3123.Nahum,ApplPhysLett.68.1996.1996.Leivo,APL76.2782.2000.Pekola,RevModPhys.78.217.2006.Giazotto,RepProgrPhys.75.046501.2012.Muhonen,PhysRevB.88.075428.2013.Kauppila}) and consists basically in cooling of a normal metal island by evacuating
the ``hot'' electrons (from above the Fermi sea) into a superconductor while
injecting ``cold'' electrons (below the Fermi sea) from another
superconductor using a pair of symmetrically biased NIS tunnel junctions. If
the normal metal island is deposited on a chip, then it can serve as a
refrigerator by cooling the chip through electron-phonon interaction. The
efficiency of the electron cooling process is controlled by the bias
voltages of the NIS junctions, whereas the success of the chip refrigeration
strongly depends on the electron-phonon coupling. Moreover, due to the
strong temperature dependence of the electric current through the junctions
at fixed bias voltage (or the strong variation of voltage with temperature
at fixed current) the NIS junctions can also serve as thermometers. Because
of this, if the normal metal island absorbs radiation,
the device turns into a very sensitive radiation detector \cite{RevModPhys.78.217.2006.Giazotto,ApplPhysLett.82.293.2003.Anghel}.

When it works as a detector, the normal metal island can be kept at the
nominal working temperature either by cooling it directly through NIS
junctions (eventually through the thermometer junctions), or indirectly
through electron-phonon coupling to a cold substrate \cite{PhysicaBCondMatt.280.485.2000.Pekola,ApplPhysLett.78.556.2001.Anghel,JLowTempPhys.123.197.2001.Anghel}.
Therefore in any situation, i.e. when the NIS junctions work as coolers, thermometers, or radiation detectors, the electron-phonon coupling plays a central role in the functionality of the device.

\begin{figure}[t]
\begin{center}
\includegraphics[height=6.2cm,width=7.6cm]{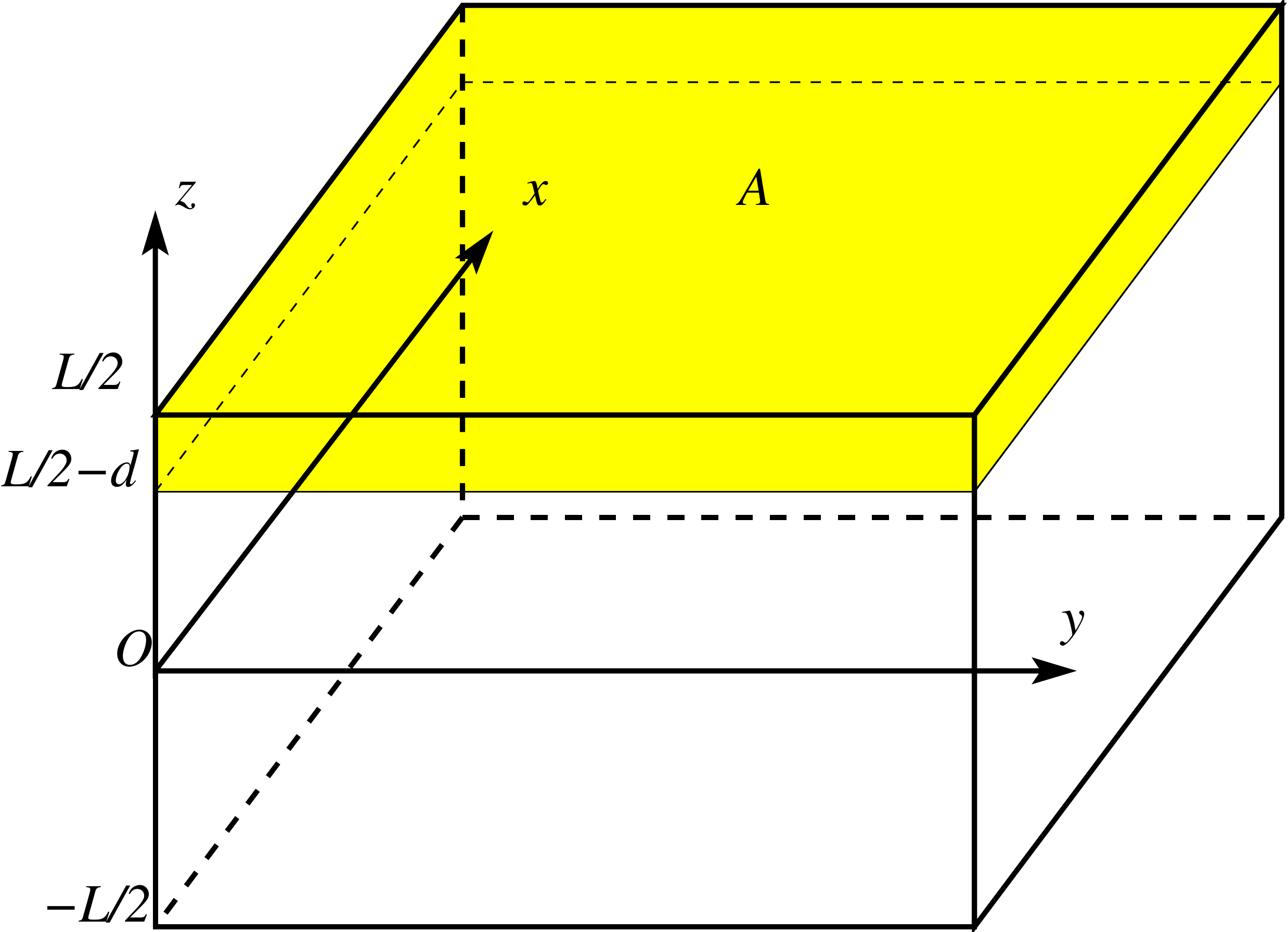}
\end{center}
\caption{The schematic model of the system. On a dielectric membrane with
parallel surfaces perpendicular on the $z$ axis is deposited an uniform
metallic film. The surfaces of the membrane cut the $z$ axis at $-L/2$ and $L/2-d$, wheres the upper surface of the film cuts the $z$ axis at $L/2$. In
the $(x,y)$ plane the area of the device is $A\ (\gg L^2)$. The whole system
is treated like a homogeneous elastic continuum, with the elastic properties
and the density of silicon nitride (SiN$_x$).}
\label{fig1}
\end{figure}

A typical experimental setup \cite{PhysicaBCondMatt.280.485.2000.Pekola,RevModPhys.78.217.2006.Giazotto} is depicted in Fig. \ref{fig1} and consists of a Cu film of thickness of the order of
10~nm deposited on a dielectric silicon-nitride (SiN$_{x}$) membrane of
thickness of the order of 100~nm. The Cu film is the normal metal island and
is connected to superconducting Al leads through NIS tunnel junctions. When
it is used as a radiation detector, to reach the sensitivity required by
astronomical observations, the working temperature of the device should be
in the range of hundreds of mK or below \cite{PhysRevApplied.2.054001.2014.Nguyen,ApplPhysLett.78.556.2001.Anghel,JLowTempPhys.123.197.2001.Anghel}. At such temperatures the phonon gas in the layered structure formed by the
normal metal island and the supporting membrane undergoes a dimensionality
cross-over from a three-dimensional (3D) gas (at higher temperatures) to a
quasi two-dimensional (2D) gas (at lower temperatures) \cite{ApplPhysLett.72.1305.1998.Leivo,PhysRevLett.81.2958.1998.Anghel,PhysRevB.70.125425.2004.Kuhn}.

In the stationary regime one may assume that the electrons in the metallic
layer have a Fermi distribution characterized by an effective temperature $T_{e}$, whereas the phonons have a Bose distribution of effective temperature $T_{ph}$.
In 3D bulk systems the heat flux between the electrons and phonons have been
calculated by Wellstood et al. \cite{PhysRevB.49.5942.1994.Wellstood} and
has been shown to vary as $T_{e}^{5}-T_{ph}^{5}$ at low temperatures. Such a
model is not justified for our devices and finite size effects \cite{Stroscio_Dutta:book} have to be taken into account.

Phenomenologically, a number of experimental studies have been interpreted
by assuming that the heat flux has a temperature dependence proportional to $T_e^{x} - T_{ph}^x$, where $x<5$ \cite{PRL.99.145503.2007.Karvonen,JPhysConfSer.92.012043.2007.Karvonen, PhysRevLett.68.1156.1992.DiTusa}. On
the other hand, a theoretical investigation of the surface effects for a
thin metallic film deposited on a half-space (semi-infinite) dielectric
showed that the value of $x$ is actually larger than 5 \cite{PhysRevB.72.224301.2005.Qu}. Qualitatively, the growth of the exponent $x$ in the lowest part of the measured temperature range has been later
found in some experiments when metallic films were deposited on bulky substrates \cite{JPhysConfSer.92.012043.2007.Karvonen}.

The electron scattering rate caused by interaction with 2D (Lamb) phonon
modes in a semiconductor quantum well (QW) has been studied in Refs. [\onlinecite{PhysRevB.51.9930.1995.Bannov,Stroscio_Dutta:book}] and in a double
heterostructure QW including the piezoelectric coupling in Ref. [\onlinecite{PhysRevB.65.205315.2002.Glavin}].

The electron-phonon heat transfer in monolayer and bilayer graphene was studied in Ref. [\onlinecite{PhysRevB.81.245404.2010.Viljas}] and a temperature dependence of the form $T_e^4 - T_{ph}^4$ was found in the low temperature regime. 
For a quasi one-dimensional geometry (metallic nanowire) the electron-phonon
power flux was studied theoretically in [\onlinecite{PhysRevB.77.033401.2008.Hekking}] and a $T^{3}$ dependence was obtained.
It is argued that a general temperature dependence of the form $T_e^{s + 2} - T_{ph}^{s + 2}$ should be valid, where $s$ is the smaller of dimensions of the electron and phonon system.\cite{PhysRevB.81.245404.2010.Viljas}
However, in experimental studies on quasi 1D Al nanowires with $65\times90$~nm$^2$ cross-section,
a better fit is achieved with the standard exponential $x=5$.\cite{ApplPhysLett.94.073101.2009.Muhonen}

In this context it is interesting to analyze theoretically the temperature
dependence of the heat transfered between electrons and phonons in the
typical experimental setup of Fig.~\ref{fig1}.
We assume that the metallic layer is made of Cu and the supporting membrane
is silicon nitride (SiN$_x$). We first observe \cite{PhysRevB.70.125425.2004.Kuhn} that the
dimensionality crossover of the phonon gas in a 100~nm thick SiN$_x$
membrane occurs around a scaling temperature $T_C \equiv c_l \hbar/(2k_B L)
\approx 237$~mK. Below this temperature one may expect a quasi-2D behavior
of the phonon system (rigorously, for $T \ll T_C$), whereas above it
(rigorously, at $T \gg T_C$) a 3D model would suffice.

We carry out our analysis employing a QW picture of the metallic film \cite{PhysRevB.51.9930.1995.Bannov,Stroscio_Dutta:book,PhysRevB.65.205315.2002.Glavin,JPhysCondMatt.20.035213.2008.Wu} by taking into account the discretization of the components of the electrons wavenumbers perpendicular to the membrane's surfaces, whereas the 3D Fermi gas picture of the electron system is used in an accompanying paper.
We work in a temperature range $T < 200$~mK in which the phonon gas is quasi-2D  and we observe that the electron-phonon heat flux $P$ cannot be simply described by a single power-law dependence, $P \propto T_e^x - T_{ph}^x$. While in some ranges of $d$ we have $P^{(0)} \propto T_e^{3.5} - T_{ph}^{3.5}$, in general the heat flux has a very strong oscillatory behavior as a function of $d$.
At certain regular intervals the power flux increases sharply with the film thickness by at least one order of magnitude (Fig. \ref{log10_P_tot_0o01_0o2K_9o9_11nm_L100nm}) and in the regions of increased heat flux the exponent of the temperature dependence is not well defined (Fig. \ref{log10_P_tot_0o01_0o2K_9o9_11nm_L100nm} b).


\section{Electron-phonon interaction}

\subsection{The electron gas}

The electron system is described as a gas of free fermions confined in the metallic film (Fig. \ref{fig1}). The electron wavevector will be denoted by $\bk \equiv (\bk_\parallel, k_z)$, where $\bk_\parallel$ and $k_z$ are the components of $\mathbf{k}$ perpendicular and parallel to $z$, respectively.
The wavefunctions satisfy the Dirichlet boundary conditions on the film surfaces (at $L/2-d$ and $L/2$) and periodic boundary conditions in the $(xy)$ plane, $\psi_{\mathbf{k}_\parallel,n} (\mathbf{r},t) = \phi_n (z) e^{i (\mathbf{k}_\parallel \mathbf{r}_\parallel - \epsilon_{\mathbf{k}_\parallel,n} t/\hbar)}/\sqrt{A}$, where $n = 1,2,\ldots$, $k_z = \pi n/d$, and
\begin{equation}
  \phi_{n} ( z )  =\sqrt{\frac{2}{d}} \sin \left[ \left(z+d-\frac{L}{2}\right) k_z \right] . \label{QW}
\end{equation}
$A$ is the area of the device surface. The electron energy is
\begin{equation}
\epsilon_\mathbf{k} = \frac{\hbar^2 k^2}{2m_e} = \frac{\hbar^2 k_\parallel^2}{2m_e} + \frac{\hbar^2 \pi^2}{2m_e d^2} n^2 \equiv \epsilon_{k_\parallel,n} ,
\label{def_eps_kn}
\end{equation}
where $m_e$ is the electron mass, $k\equiv |\bk|$, and $k_\parallel \equiv |\bk_\parallel|$.
The electrons are distributed over energy sub-bands defined by $n$. The electron field operators are
\begin{subequations} \label{def_Psi}
\begin{eqnarray}
\Psi(\mathbf{r},t) &=& \sum_{\mathbf{k}_\parallel,n} \psi_{\mathbf{k}%
_\parallel,n} (\mathbf{r},t) c_{\mathbf{k}_\parallel,n} ,  \label{def_Psi1}
\\
\Psi^\dag(\mathbf{r},t) &=& \sum_{\mathbf{k}_\parallel,n} \psi^*_{\mathbf{k}_\parallel,n} (\mathbf{r},t) c^\dag_{\mathbf{k}_\parallel,n} , \label{def_Psi2}
\end{eqnarray}
\end{subequations}
where $c^\dag_{\mathbf{k}_\parallel,n}$ and $c_{\mathbf{k}_\parallel,n}$ are
the electron creation and annihilation operators on the state $(\mathbf{k}_\parallel,n)$.

\subsubsection{Numerical estimations} \label{subsub_num_est}

To better understand the physical problem, let us compute the magnitude of some important quantities related to the electron gas.
The Cu Fermi energy is $E_F \approx 7$~eV; the highest sub-band populated at 0~K is denoted as $\epsilon_{n_F}$, where
\begin{equation}
  n_F \equiv \left\lfloor \frac{\sqrt{2m_eE_F}}{\pi \hbar} d \right\rfloor  \label{def_nF}
\end{equation}
and $\lfloor x\rfloor$ is the biggest integer smaller than $x$. For $d = 10$ and 20~nm, $n_F=43$ and 86, respectively.
The Cu Fermi temperature is $T_F = E_F/k_B \approx 81231$~K, whereas $(\epsilon_2-\epsilon_1)/k_B \approx 131$~K and $(\epsilon_{44}-\epsilon_{43})/k_B \approx 3796$~K for $d = 10$~nm.

\subsection{The phonon gas}

For simplicity, we treat the whole system--supporting membrane and metallic
film (Fig. \ref{fig1})--as a single elastic continuum of thickness $L$, area
$A$, and volume $V_{ph} = LA$. The elastic modes of such a system \cite{Stroscio_Dutta:book,PhysRevB.70.125425.2004.Kuhn,JPhysA.40.10429.2007.Anghel,Auld:book}
have the form $\mathbf{w}_{\mathbf{q}_\parallel \xi}(z) e^{i (\mathbf{q}%
_\parallel \mathbf{r}_\parallel - \omega_{\mathbf{q}_\parallel \xi} t)} /\sqrt{A}$, where $\mathbf{q}_\parallel$ and $\mathbf{r}_\parallel$ are the
components of the wavevector and position vector, respectively, parallel to
the $(x,y)$ plane. The functions $\mathbf{w}_{\mathbf{q}_\parallel \xi}(z)$
are normalized on the interval $z\in [-L/2,L/2]$, namely $\int_{-L/2}^{L/2} \bw_{\bq_\parallel \xi}(z)^{\dag} \bw_{\bq_\parallel \xi'}(z)dz = \delta_{\xi,\xi^{\prime}}$.

The elastic modes are divided into three main categories: horizontal shear ($h$), dilatational or symmetric ($s$) with respect to the mid-plane, and
flexural or antisymmetric ($a$) with respect to the mid-plane. The
quantization of the elastic modes in the $z$ direction leads to the
formation of phonon branches (or sub-bands), such that a phonon mode is
identified by its symmetry $\alpha = h, s, a$, sub-band number $\nu = 1, 2,
\ldots$, and $\mathbf{q}_\parallel$; $\xi$ represents the pair $(\alpha,\nu)$.

The $h$ modes are simple transversal modes. The free boundary conditions imposed at the upper and lower surfaces of the membrane quantize the $z$-component of the wavevector at the values $q_{t h n} = n \pi/L$, $n = 0,1,\ldots$, whereas $q_{l h n} \equiv 0$.
If we denote by $q_\parallel \equiv |\bq_\parallel|$, the phonon frequency is $\omega_{q_\parallel,h,\nu} = c_t \sqrt{q_\parallel^2 + q_{t,h,\nu}^2}$, where by $c_t$ and $c_l$ we denote the transversal and longitudinal sound velocities, respectively. The sound velocities are determined by the Lam\'e coefficients $\lambda$ and $\mu$, and the density $\rho$ of SiN$_x$,
\begin{equation}
  c_{t} = \frac{\mu}{\rho}, \quad c_{l}=\frac{\lambda+2\mu}{\rho}. \label{ct_cl}
\end{equation}

The $s$ and $a$ modes are superpositions of longitudinal and transversal modes, oscillating in a plane perpendicular to the surfaces. The quantization relations for $q_{l \alpha}$ and $q_{t \alpha}$--which are the $z$ components of the wavevectors corresponding to the longitudinal and transversal oscillations, respectively--are \cite{Auld:book}
\begin{equation}
  \frac{- 4q_\parallel^{2}q_{l\alpha}q_{t\alpha}}{(q_\parallel^{2}-q_{t\alpha}^{2})^{2}} = \left[ \frac{\tan(q_{t\alpha}L/2)} {\tan(q_{l\alpha}L/2)} \right]^{\pm 1} , \label{6}
\end{equation}
where $+1$ and $-1$ exponents on the right hand side correspond to the $\alpha = s$ and $\alpha = a$, respectively.
The closure equation is secured by the Snell's law,
\begin{equation}
  \omega_{q_\parallel,\xi}=c_{l}\sqrt{q_{l,\xi}^{2}+q_\parallel^{2}}=c_{t}\sqrt{q_{t,\xi}^{2}+q_\parallel^{2}}, \label{8}
\end{equation}
and the branch number $\nu$ is determined from Eqs. (\ref{6}) and (\ref{8}) in the limit $q_\parallel \to 0$.

In the second quantization, the displacement field operator will be denoted by
\begin{eqnarray}
\mathbf{u} ( \mathbf{r}, t ) &=& \sum_{\xi,\mathbf{q}_\parallel} \sqrt{
\frac{\hbar}{2 \rho \omega_{\mathbf{q}_\parallel \xi} }} \left[ a_{\bq_\parallel \xi} + a_{-\mathbf{q}_\parallel \xi}^{\dag} \right]  \notag \\
&& \times \mathbf{w}_{\mathbf{q}_\parallel \xi}(z) e^{i (\bq_\parallel \mathbf{r}_\parallel - i\omega_{\mathbf{q}_\parallel \xi}t )} ,
\label{displ_operator}
\end{eqnarray}
where $a_{\mathbf{q}_\parallel \xi}^\dag$ and $a_{\mathbf{q}_\parallel \xi}$
are the phonon creation and annihilation operators.

We introduce the notation $J \equiv c_{t}^{2}/c_{l}^{2}$ ($J<1/2$ from Eq. \ref{ct_cl}). We shall write $q_t$ instead of $q_{t,\xi}$ and $q_l$ instead of $q_{l,\xi}$, when this does not lead to confusions.
Explicitly, the normalized displacement fields of the $s$ and $a$ waves are \cite{JPhysA.40.10429.2007.Anghel}
\begin{subequations} \label{def_wsa_xz}
\begin{eqnarray}
  w_{\bq_\parallel,s,\nu,x} &=& N_s iq_{t}\left[  2q_\parallel^{2}\cos\left(  \frac{q_{t}L}{2}\right)  \cos\left(q_{l}z\right)  +(q_{t}^{2}-q_\parallel^{2}) \right. \nonumber \\
  && \left. \times \cos\left(  \frac{q_{l}L}{2}\right) \cos\left(  zq_{t}\right)  \right]  , \label{def_ws_x} \\
  w_{\bq_\parallel,s,\nu,z} &=& N_s q_\parallel\left[  -2q_{t}q_{l}\cos\left(  \frac{q_{t}L}{2}\right) \sin\left(  q_{l}z\right)  +(q_{t}^{2}-q_\parallel^{2}) \right. \nonumber \\
  && \left. \times \cos\left(\frac{q_{l}L}{2}\right)  \sin \left( zq_{t}\right)  \right]  , \label{def_ws_z} \\
  w_{\bq_\parallel,a,\nu,x} &=& N_a iq_{t}\left[  2q_\parallel^{2}\sin\left(  \frac{q_{t}L}{2}\right)  \sin\left(q_{l}z\right) + (q_{t}^{2}-q_\parallel^{2}) \right. \nonumber \\ 
  && \left. \times \sin\left(  \frac{q_{l}L}{2}\right) \sin\left(  zq_{t}\right)  \right]  , \label{def_wa_x} \\
  w_{\bq_\parallel,a,\nu,z} &=& N_a q_\parallel\left[  2q_{t}q_{l}\sin\left(  \frac{q_{t}L}{2}\right)  \cos\left(q_{l}z\right) - (q_{t}^{2}-q_\parallel^{2}) \right. \nonumber \\
  && \left. \times \sin \left(\frac{q_{l}L}{2}\right) \cos\left(  zq_{t}\right)  \right]  , \label{def_wa_z}
\end{eqnarray}
\end{subequations}
where we took $\bq_\parallel$ along the $x$ axis.

We are interested in the low temperatures limit, so $q_\parallel$ is small (rigorously it should be $L q_\parallel \ll 1$) in the lowest phonon sub-band ($\nu = 0$). Under these conditions, $q_{t,s,0} \equiv q_{t,s}$ is real, whereas $q_{l,s,0} \equiv i p_{l,s}$, $q_{t,a,0} = i p_{t,a}$, and $q_{l,a,0} = i p_{l,a}$ are imaginary. The component $q_{t,h,\nu}$ is always real as stated above. With these specifications we can rewrite Eqs. (\ref{def_wsa_xz}),
\begin{subequations} \label{def_wsa_xz_Q2}
\begin{eqnarray}
  w_{\bq_\parallel,s,\nu,x} &=& N_s iq_{t}\left[ 2q_\parallel^{2}\cos\left( \frac{q_{t}L}{2}\right) \cosh\left(p_{l}z\right)  +(q_{t}^{2}-q_\parallel^{2}) \right. \nonumber \\
  && \left. \times \cosh\left( \frac{p_{l}L}{2}\right) \cos (zq_{t})  \right] , \label{def_ws_x_Q2} \\
  w_{\bq_\parallel,s,\nu,z} &=& N_s q_\parallel\left[  2 q_{t}p_{l} \cos\left( \frac{q_{t}L}{2}\right) \sinh (p_{l}z)  +(q_{t}^{2}-q_\parallel^{2}) \right. \nonumber \\
  && \left. \times \cosh\left(\frac{p_{l}L}{2}\right) \sin(zq_{t}) \right]  , \label{def_ws_z_Q2} \\
  w_{\bq_\parallel,a,\nu,x} &=& N_a p_{t} \left[ 2q_\parallel^{2} \sinh\left( \frac{p_{t}L}{2}\right)  \sinh(p_{l}z)  - (p_{t}^{2} + q_\parallel^{2}) \right. \nonumber \\
  && \left. \times \sinh \left( \frac{p_{l}L}{2}\right) \sinh(zp_{t}) \right]  , \label{def_wa_x_Q2} \\
  w_{\bq_\parallel,a,\nu,z} &=& - i N_a q_\parallel \left[ 2 p_{t}p_{l}\sinh\left(\frac{p_{t}L}{2}\right) \cosh(p_{l}z) \right. \nonumber \\
  && \left. - (p_{t}^{2}+q_\parallel^{2}) \sinh\left(\frac{p_{l}L}{2}\right) \cosh(zp_{t}) \right] . \label{def_wa_z_Q2}
\end{eqnarray}
\end{subequations}
In the same limit, the normalization constants are
\begin{subequations}\label{eqn_N_complex}
\[
	N_h^{-2} = V, \ {\rm for}\ m=0 \quad {\rm and}\quad N_h^{-2} = V/2, \ {\rm for}\ m>0 ,
\]
\begin{eqnarray}
  N_s^{-2} &=& A \left\{
  4 q_t^2 q_\parallel^2 \cos^2(q_t L/2) \left[(p_l^2 + q_\parallel^2) \frac{\sinh(p_lL)}{2 p_l} \right. \right. \nonumber \\
  && \left. - (p_l^2-q_\parallel^2) \frac{L}{2} \right] + (q_t^2-q_\parallel^2)^2 \cosh^2(p_lL/2) \nonumber \\
  && \times \left[ (q_t^2+q_\parallel^2) \frac{L}{2} + (q_t^2-q_\parallel^2) \frac{\sin(q_tL)}{2q_t}\right] \nonumber\\
  && + 4q_\parallel^2 q_t(q_t^2-q_\parallel^2) \cosh^2(p_l L/2) \sin(q_td) 
    \left.\vphantom{\frac{1}{2}}\right\} \label{eqn_N_s_complex}
  %
\end{eqnarray}
\begin{eqnarray}
N_a^{-2}
 &=& A \left\{
     4 p_t^2 q_\parallel^2 \sinh^2(p_tL/2) \left[
     (p_l^2 + q_\parallel^2)\frac{\sinh(p_lL)}{2p_l} \right. \right. \nonumber \\
     && \left. + (p_l^2 - q_\parallel^2)\frac{L}{2}\right] +(p_t^2 + q_\parallel^2)^2 \sinh^2(p_lL/2) \nonumber \\
     && \times \left[
     (p_t^2 + q_\parallel^2)\frac{\sinh(p_tL)}{2p_t} - (p_t^2 - q_\parallel^2) \frac{L}{2}\right]\nonumber\\
  && - 4 q_\parallel^2 p_t (p_t^2 + q_\parallel^2) \sinh^2(p_lL/2) \sinh(p_tL)
     \left.\vphantom{\frac{1}{2}}\right\} \label{eqn_N_a_complex}
	%
\end{eqnarray}
\end{subequations}

\subsection{The interaction}

Electron-phonon interaction in the volume of the film is described by the
deformation potential Hamiltonian \cite{Ziman:book},
\begin{equation}
H_{def}=\frac{2}{3} E_{F} \int_{V_{el} = A\times d} d^{3}\mathbf{r}\,
\Psi^{\dagger}(\mathbf{r})\,\Psi(\mathbf{r}) \nabla\cdot\mathbf{u} (\mathbf{r%
}) ,  \label{def_def_pot}
\end{equation}
where $E_{F}$ is the Fermi energy. From Eq. (\ref{def_def_pot}) we notice that all the transversal components of the $h$, $s$, and $a$ modes are not contributing to the electron-phonon interaction. Plugging Eqs. (\ref{displ_operator}) and (\ref{def_Psi}) into (\ref{def_def_pot}) we get
\begin{eqnarray}
H_{def} &=& \sum_{\mathbf{k}_\parallel, \mathbf{q}_\parallel, \xi, n,
n^{\prime }} \left[ g_{\xi,\mathbf{q}_\parallel}^{n^{\prime },n} c_{\mathbf{k%
}_\parallel + \mathbf{q}_\parallel, n^{\prime }}^{\dagger} c_{\mathbf{k}%
_\parallel, n} a_{\mathbf{q}_\parallel \xi} \right.  \notag \\
&& + \left. \left(g_{\xi,\mathbf{q}_\parallel}^{n^{\prime },n}\right)^* c_{%
\mathbf{k}_\parallel - \mathbf{q}_\parallel n^{\prime }}^{\dagger} c_{%
\mathbf{k}_\parallel, n} a_{\mathbf{q}_\parallel \xi}^{\dagger} \right] .
\label{3}
\end{eqnarray}
where $n$ and $n'$ in the notation $g_{\xi,\bq_\parallel}^{n',n}$ refer to the electrons states (see Eqs. \ref{QW} and \ref{def_eps_kn}). The coupling constants are
\begin{eqnarray}
  g_{\xi,\bq_\parallel}^{n',n} &=& \frac{2}{3} E_{F} N_{\alpha} \sqrt{\frac{\hbar}{2 \rho \omega_{\bq_\parallel,\xi}}} \int_{L/2-d}^{L/2} \phi_{n'}^*(z) \phi_{n}(z)
  \nonumber \\
  && \times \left[ i \bq_\parallel \cdot \bw_{\bq_\parallel,\xi}(z) + \frac{\partial w_{\bq_\parallel,\xi,z}(z)}{\partial z} \right]  dz . \label{3a}
\end{eqnarray}
Because the $h$ modes do not contribute to the interaction (the divergence of any transversal displacement field is zero) and we are interested in the low temperature limit, in the following we shall take into account in the expression of $g_{\xi,\bq_\parallel}^{n',n}$ (\ref{3a}) only the modes with $\xi = (s,0)$ and $(a,0)$.

The energy transferred from electrons to phonons in a unit of time (heat flux) is
\begin{eqnarray}
P &=&2\sum_{\mathbf{k}_{\parallel };\mathbf{k}_{\parallel }^{\prime },%
\mathbf{q}_{\parallel },\xi }\hbar \omega _{\mathbf{q}_{\parallel
},\xi }\left[ \Gamma _{\xi ,n,n^{\prime }}^{\mathrm{em}}(\mathbf{k}%
_{\parallel };\mathbf{k}_{\parallel }^{\prime },\mathbf{q}_{\parallel
})\right.  \notag \\
&&\left. -\Gamma _{\xi ,n,n^{\prime }}^{\mathrm{ab}}(\mathbf{k}%
_{\parallel };\mathbf{k}_{\parallel }^{\prime },\mathbf{q}_{\parallel })%
\right] ,  \label{15}
\end{eqnarray}
where the factor 2 comes from the electrons spin degeneracy.

As we wrote in the introduction, due to the weak coupling between electrons and phonons we may assume that the electron system has an equilibrium Fermi distribution corresponding to a temperature $T_e$ and the phonons have a Bose distribution of effective temperature $T_{ph}$.
We shall use the notations $\beta_e \equiv 1/(k_BT_e)$ and $\beta_{ph} \equiv 1/(k_BT_{ph})$, and $\epsilon_{q_\parallel,\xi} \equiv \hbar\omega_{q_\parallel,\alpha,0}$ ($\xi$ may be omitted). With these notations and applying the Fermi's golden rule, we obtain the emission and absorption rates $\Gamma^{\rm em}$ and $\Gamma^{\rm abs}$,
\begin{subequations} \label{16}
\begin{eqnarray}
  && \Gamma_{\xi,n,n'}^{\rm em} (\bk_\parallel; \bk_\parallel - \bq_\parallel, \bq_\parallel) = \frac{2\pi}{\hbar} |g_{\xi,\bq_\parallel}^{n',n}|^{2} \left[n(\beta_{ph} \epsilon_{q_\parallel}) + 1\right] \nonumber \\
  && \times f(\beta_e \epsilon_{k_\parallel,n'}) [1-f(\beta_e \epsilon_{\bk_\parallel-\bq_\parallel, n})] \nonumber \\
  && \times \delta(\epsilon_{\bk_\parallel-\bq_\parallel,n} - \epsilon_{\bk_\parallel,n'}+ \epsilon_{q_\parallel}) , \label{16a} \\
  && \Gamma_{\xi,n,n'}^{\rm ab} (\bk_\parallel - \bq_\parallel; \bk_\parallel,\bq_\parallel) = \frac{2\pi}{\hbar} |g_{\xi,\bq_\parallel}^{n',n}|^{2} n(\beta_{ph} \epsilon_{q_\parallel}) \nonumber \\
  && \times [1 - f(\beta_e \epsilon_{k_\parallel,n'})] f(\beta_e \epsilon_{\bk_\parallel-\bq_\parallel, n}) \nonumber \\
  && \times \delta(\epsilon_{\bk_\parallel-\bq_\parallel,n} - \epsilon_{\bk_\parallel,n'}+ \epsilon_{q_\parallel}). \label{16b}
\end{eqnarray}
\end{subequations}
where $f(x)$ and $n(x)$ are the Fermi and Bose distributions, respectively. Using the identity $f(x) [1 - f(x-y)] = n(y) [f(x-y) - f(x)]$ and taking into account the $\delta$ functions in Eq. (\ref{16}) we can write
\begin{eqnarray}
  && \Gamma_{\xi,n,n'}^{\rm em} (\bk_\parallel; \bk_\parallel - \bq_\parallel, \bq_\parallel) - \Gamma_{\xi,n,n'}^{\rm ab} (\bk_\parallel - \bq_\parallel; \bk_\parallel, \bq_\parallel) \nonumber \\
  && = \frac{2\pi}{\hbar} |g_{\xi,\bq_\parallel}^{n',n}|^{2} [f(\beta_e \epsilon_{\bk_\parallel-\bq_\parallel, n}) - f(\beta_e \epsilon_{k_\parallel,n'}) ] \nonumber \\
  && \times [n(\beta_e \epsilon_{q_\parallel}) - n(\beta_{ph} \epsilon_{q_\parallel})] \label{dif_Gammas}
\end{eqnarray}
and the power flux becomes
\begin{equation}
  P = P^{(0)} \left(  T_{e}\right) - P^{(1)} ( T_e, T_{ph}) . \label{17}
\end{equation}

\subsection{Long wavelength approximation} \label{subsect_LWA}

We find the relevant low temperature asymptotic expressions by expanding all the quantities in a Taylor series in $q_\parallel$. From Eqs. (\ref{6}) and (\ref{8}) we derive expressions for $q_{ts}$, $p_{ls}$, $p_{ta}$, and $p_{la}$, which we then use to calculate all the other quantities.
For the symmetric modes it is sufficient to express $q_{t,s}$ and $p_{l,s}$ to the lowest order in $q_\parallel$, whereas for the antisymmetric modes we have to express $p_{t,a}$ and $p_{l,a}$ up to the third order. In this way we obtain
\begin{subequations} \label{9s}
\begin{eqnarray}
  q_{t,s} &\approx& q_\parallel\sqrt{3-4J} , \label{9s_qt} \\
  p_{l,s} &\approx& q_\parallel (1-2J) \label{9s_pl} \\
  \omega^{s} &\approx& 2q_\parallel c_{t}\sqrt{1-J} = 2q_\parallel c_{l}\sqrt{J(1-J)} , \label{9s_omega}\\
  N_{s}^{-2} &\approx& 16q_\parallel^{6}(3-4J) (1-J)^{2}, \label{9s_Ns}
\end{eqnarray}
\end{subequations}
\begin{subequations} \label{9a} 
\begin{eqnarray}
  p_{l,a} &\approx& q_\parallel \left\{ 1-\frac{L^2}{6} J(1-J) q_\parallel^2 + \frac{2 J (1-J)}{9} \right. \nonumber \\
  && \left. \times \left[ \frac{27 - 20J}{5} - J(1-J) \right] \left( \frac{L}{2} \right)^4 q_\parallel^4 \right\} \label{chil_ord_3} \\
  p_{t,a} &\approx& q_\parallel \left\{ 1-\frac{L^2}{6} (1-J) q_\parallel^2 + \frac{2(1-J)}{9} \right. \nonumber \\
  && \left. \times\left[ \frac{27-20 J)}{5} - (1-J) \right] \left( \frac{L}{2} \right)^4 q_\parallel^4 \right\} \label{chit_ord_3} \\
  %
  %
  \omega_a &\approx& c_t L \sqrt{\frac{1 - J}{3}} q_\parallel^2 \left[ 1 - \frac{L^2}{120} (27 - 20 J) q_\parallel^2 \right] \label{omega_ord_3} \\
  N_{a}^{-2} &\approx& q_\parallel^{12}L^{6} \frac{(1-J)^{2}}{36} . \label{9a_Na}
\end{eqnarray}
\end{subequations}

Plugging Eqs. (\ref{def_wsa_xz_Q2}), (\ref{eqn_N_complex}),
(\ref{9s}), and (\ref{9a}) into (\ref{3a}) we may write in the first stage
\begin{subequations} \label{mod_gsa}
\begin{eqnarray}
  |g_{s,\bq_\parallel}^{n',n}|^{2} &=& |N_{s}|^{2} [q_{t} q_\parallel (q_\parallel^{2} - p_{l}^{2})]^{2} \frac{8\hbar E_{F}^{2}}{9 \rho \omega_{s}} \cos^{2}(Lq_{t}/2) \nonumber \\
  && \times S (n',n,s,p_{l})  ,  \label{mod_gs} \\
  |g_{a,\bq_\parallel}^{n',n}|^{2} &=& |N_{a}|^{2} [p_{t} q_\parallel (q_\parallel^{2} - p_{l}^{2})]^{2} \frac{8\hbar E_{F}^{2}}{9 \rho \omega_{a}} \sinh^{2}(Lp_{t}/2) \nonumber \\
  && \times S (n',n,a,p_{l}) , \label{mod_ga}
\end{eqnarray}
\end{subequations}
where the overlap integral $S(n',n,\alpha,p_{l})$ is in cases $s$ and $a$
\begin{subequations} \label{S_sa}
\begin{equation}
  S (n',n,s,p_{l})  = \left| \int\limits_{L/2-d}^{L/2}\phi_{n'}^*(z) \phi_{n}(z) \cosh(zp_{l})\, dz\right|^{2}, \label{S_s}
\end{equation}
\begin{equation}
  S (n',n,a,p_{l}) = \left| \int\limits_{L/2-d}^{L/2}\phi_{n'}^*(z) \phi_{n}(z) \sinh(zp_{l})\, dz\right|^{2}. \label{S_a}
\end{equation}
\end{subequations}
Using Eqs. (\ref{QW}) and (\ref{9s}) and expressing all the quantities in terms of the frequency $\omega$ we obtain from Eq. (\ref{S_s}) in the first relevant order
\begin{subequations} \label{S_s_lim}
\begin{eqnarray}
  && S ( n_1, n_2, s, \omega ) \approx 1, \qquad  {\rm if} \  n_1 - n_2 = 0 \label{lim_Ss_nn} \\
  && \approx \frac{\omega^4 d^4}{16 \pi^4 c_l^4} \frac{(1 - 2J)^4}{J^2(1 - J)^2} \bigg[ \frac{ 1 }{(n_1-n_2)^2} - \frac{1}{(n_1 + n_2)^2} \bigg]^2, \nonumber \\
  && {\rm if}\ n_1 - n_2 = 2m \label{lim_Ss_n1n2_2k} \\
  && \approx \frac{\omega^4 (L - d/2)^2 d^2}{4 \pi^4 c_l^4} \frac{(1-2J)^4}{J^2(1-J)^2} \bigg[ \frac{ 1 }{(n_1-n_2)^2} -  \frac{ 1 }{(n_1+n_2)^2} \bigg]^2 , \nonumber \\
  && {\rm if}\ n_1 - n_2 = 2m + 1 , \label{lim_Ss_n1n2_2kp1}
\end{eqnarray}
\end{subequations}
whereas from (\ref{S_a}) we get
\begin{subequations} \label{Sa_limT0}
\begin{eqnarray}
  && S ( n_1,n_2, a, \omega ) \approx \frac{\sqrt{3}}{4} \frac{1}{c_l \sqrt{J(1-J)}} \frac{(L-d)^2}{L} \omega , \nonumber \\
  && {\rm if} \  n_1 - n_2 = 0 \label{Sa1_limT0} \\
  && \approx \frac{9\sqrt{3}}{4 \pi^4} \frac{d^4 (L-d)^2}{L^3 c_l^3 [J(1-J)]^{3/2}} \omega^3 \bigg[ \frac{ 1 }{(n_1-n_2)^2} - \frac{ 1 }{(n_1+n_2)^2} \bigg]^2 , \nonumber \\
  && {\rm if}\ n_1 - n_2 = 2m \label{Sa2_limT0} \\
  && \approx \frac{4\sqrt{3}}{\pi^4} \frac{d^2}{\sqrt{J(1-J)} L c_l} \omega \bigg[ \frac{ 1 }{(n_1-n_2)^2} - \frac{ 1 }{(n_1+n_2)^2} \bigg]^2 , \nonumber \\
  && {\rm if}\ n_1 - n_2 = 2m + 1 \label{Sa3_limT0}
\end{eqnarray}
\end{subequations}
Using these approximations we can now calculate the low temperature heat flux.

\section{Electron-phonon heat flux}

%
We separate $P^{(0)}$ and $P^{(1)}$ into the symmetric and the antisymmetric parts, according to which type of phonons contribute to the process: $P^{(0)} \equiv P^{(0)}_s + P^{(0)}_a$ and $P^{(1)} \equiv P^{(1)}_s + P^{(1)}_a$.
We work in the low temperature limit, using the approximations of Section \ref{subsect_LWA}.

\subsection{Calculation of $P^{(0)}$} \label{subsec_HeatFl}

\subsubsection{Contribution of the $s$ modes} \label{subsubsec_HeatFl_S}

Using (\ref{9s}) and (\ref{mod_gsa}) and changing from summations over $\bk_\parallel$ and $\bq_\parallel$ to integrals, for the symmetric part we have
\begin{eqnarray}
  P_s^{(0)} &=& 2 \frac{A^2}{(2\pi)^3}\sum_{n'} \sum_{n} \int_0^\infty dk_\parallel\, k_\parallel \int_0^\infty dq_\parallel q_\parallel \int_0^{2\pi} d\phi \nonumber \\
  && \times |N_{s}|^{2} \left| q_{t}q\left( q^{2} - p_{l}^{2} \right)  \right|^{2} \frac{16 \pi E_{F}^{2}}{9 \rho \omega_{s}} \cos^{2} \left( \frac{Lq_{t}}{2} \right) \nonumber \\
  && \times S ( n^{\prime},n, s, q_{l} ) \hbar \omega_{q_\parallel} n(\beta_e \epsilon_{q_\parallel}) \big\{ f[\beta_e (\epsilon_{k_\parallel} - \epsilon_{q_\parallel})] \nonumber \\
  && - f(\beta_e \epsilon_{k_\parallel}) \big\} \delta(\epsilon_{\bk_\parallel -  \bq_\parallel} - \epsilon_{\bk_\parallel} + \epsilon_{q_\parallel}) \nonumber \\
  &=& \frac{2A}{\pi^2 L} \frac{\hbar E_{F}^{2}}{18 \rho c_l^2} \frac{J}{1-J} \sum_{n'} \sum_{n} \int_0^\infty dk_\parallel \, k_\parallel \int_0^\infty dq_\parallel\, q_\parallel \nonumber \\
  && \int_0^{2\pi} d\phi \, \omega^2 S ( n^{\prime},n, s, q_{l} ) n(\beta_e \epsilon_{ph}) \big\{ f[\beta_e (\epsilon_k - \epsilon_{ph})] \nonumber \\
  && - f(\beta_e \epsilon_k) \big\} \delta(\epsilon_{\mathbf{k} - \mathbf{\mathbf{q}_{||}}}-\epsilon_{\mathbf{k}}+\hbar\omega_{s}) \label{Ps_em1}
\end{eqnarray}
The energy conservation gives the angles $\phi$ (which may take two values):
%
\begin{eqnarray}
  0 &=& \frac{\hbar^2 (k_\parallel^2 + k_{z1}^2)}{2m_e} - \frac{\hbar^2 (k_\parallel^2 + q_\parallel^2 - 2k_\parallel q_\parallel \cos \phi + k_{z2}^2)}{2m_e} \nonumber \\
  && - \hbar \omega_s \label{eqs_cos_phi}
\end{eqnarray}
%
We change the variables in Eq. (\ref{Ps_em1}) into electron and phonon energies and denote them by $\epsilon_{ph}$ and $\epsilon_e$, respectively. Then $\epsilon_e$ takes values from $\epsilon_{n'}$ to $\infty$. Nevertheless, the energy conservation imposes extra constraints on the electron energy, such that the lower limit increases from $\epsilon_{n'}$ to
\begin{equation}
  \epsilon_{n_1}^{(s)} (n_2,\epsilon_{ph}) = \epsilon_{n_1} + \delta \epsilon_{n_1}^{(s)} (n_2, \epsilon_{ph}) \label{def_eps1p}
\end{equation}
where
%
\begin{eqnarray}
	\delta \epsilon_{n_1}^{(s)} (n_2,\epsilon_{ph}) &=& \frac{ 2 m_e J (1-J) c_l^2 }{ \epsilon_{ph}^2 } \left[ \epsilon_{ph} - \epsilon_{n_1} + \epsilon_{n_2} \right. \nonumber \\
	&& \left. + \frac{\epsilon_{ph}^2}{m_e c_l^2 8 J (1-J)} \right]^2 ; \label{delta_eps1_23S}
\end{eqnarray}
in general we shall use the shorthand notations $\epsilon_{n_1}^{(s)}$ and $\delta \epsilon_{n_1}^{(s)}$, when this does not lead to confusions.

Taking the three cases (\ref{S_s_lim}) for $n'$ and $n$ we write $P_s^{(0)}$ as the sum of three terms:
\begin{eqnarray}
  P_s^{(0)} &=& A \frac{m_e E_{F}^{2}}{36\pi^2 L \rho c_l^4 \hbar^5} \frac{1}{(1-J)^2} ( S^{(0)}_{1s} + S^{(0)}_{2s} + S^{(0)}_{3s} ) \nonumber\\
  &\equiv& P^{(0)}_{1s} + P^{(0)}_{2s} + P^{(0)}_{3s}. \label{Ps_em_S1S2S3}
\end{eqnarray}
The terms $S^{(0)}_{1s}$, $S^{(0)}_{2s}$, and $S^{(0)}_{3s}$ are the summations and integrals taken separately from Eq. (\ref{Ps_em1}), so
\begin{eqnarray}
  S^{(0)}_{1s} &=& 2 c_l \sqrt{2 m_e J (1-J)} \sum_{n'} \int_0^\infty d\epsilon_{ph} \int_{\epsilon^{(s)}_{n'}}^\infty d\epsilon_e \nonumber \\
  && \times \frac{\epsilon_{ph}^2 \, n(\beta_e \epsilon_{ph})}{\sqrt{\epsilon_e - \epsilon^{(s)}_{n'}(n',\epsilon_{ph})}} \big\{ f[\beta_e (\epsilon_e - \epsilon_{ph})] - f(\beta_e \epsilon_e) \big\} \nonumber \\
  &=& 2 c_l \sqrt{2 m_e J (1-J)} (k_BT_e)^{7/2} \sum_{n'} \int_0^\infty dx_{ph} \frac{x_{ph}^2}{e^{x_{ph}} - 1} \nonumber \\
  &\times& \int\limits_0^\infty \frac{dx}{\sqrt{x}} \left\{ \frac{1}{e^{x - (y-x^{(s)}_{n'}(n',x_{ph},T_e)) - x_{ph}} + 1} \right. \nonumber \\
  && \left. - \frac{1}{e^{x - (y-x^{(s)}_{n'}(n',x_{ph},T_e))} + 1} \right\} \label{S1s_approx}
\end{eqnarray}
where $x_{ph} = \beta_e \epsilon_{ph}$, $x_e = \beta_e \epsilon_e$, $x_{n'} = \beta_e \epsilon_{n'}$, $x^{(s)}_{n'}(n,x_{ph},T_e) = \beta_e \epsilon^{(s)}_{n'}(n,\epsilon_{ph})$, 
$y = \beta_e \epsilon_F$, $x = x_e - x^{(s)}_{n'}(n,x_{ph})$, and
\begin{eqnarray}
  && y - x^{(s)}_{n'}(n',x_{ph}, T_e) = y - x_{n'} - \frac{ 2 m_e c_l^2 J (1-J) }{ k_BT_e x_{ph}^2 } \nonumber \\
  && \times \left[ x_{ph} + \frac{k_BT_e x_{ph}^2}{8 m_e c_l^2 J (1-J)} \right]^2 . \label{ymx1p_1s}
\end{eqnarray}

\begin{figure}[t]
  \begin{center}
  \includegraphics[width=8cm]{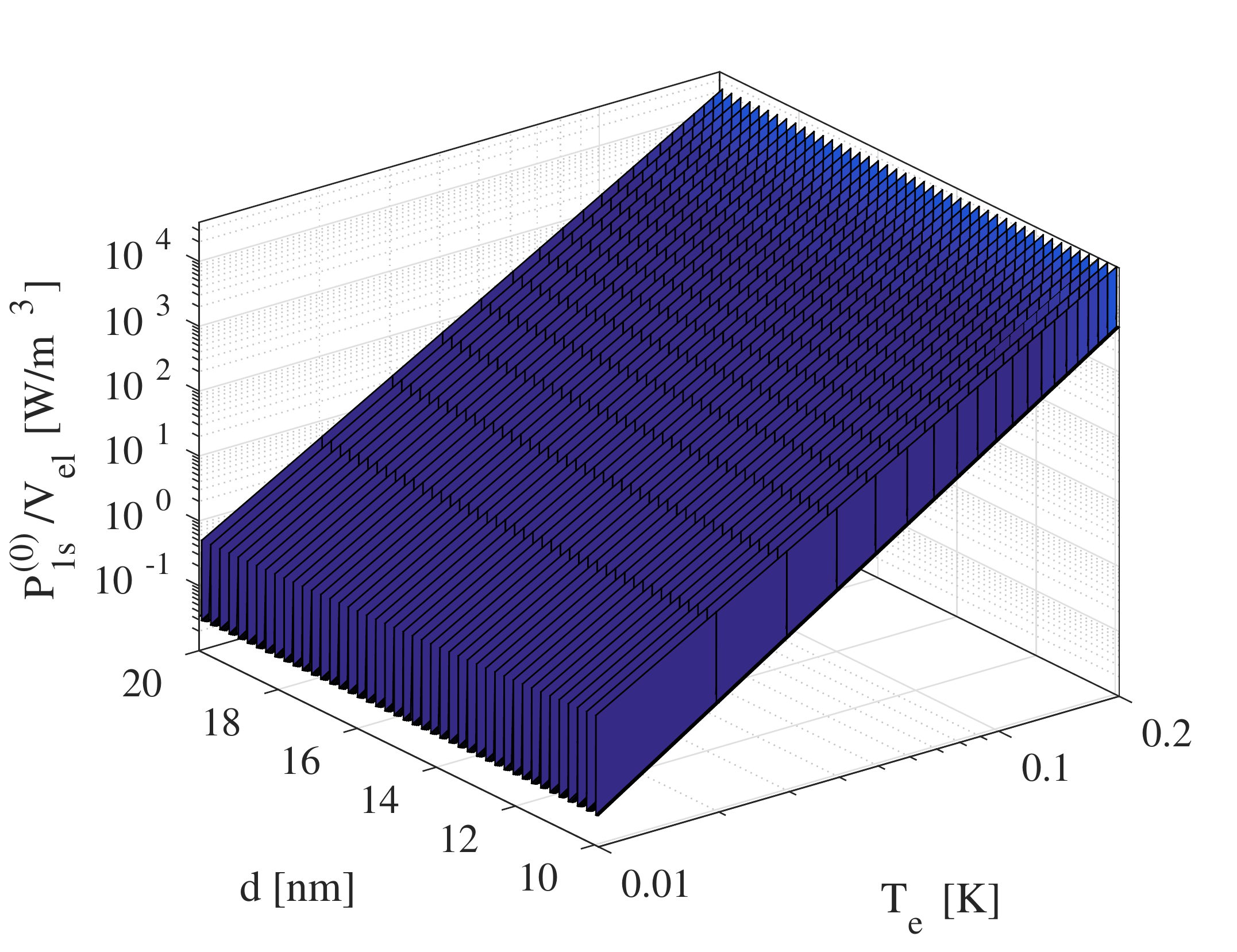} \\
  \includegraphics[width=8cm]{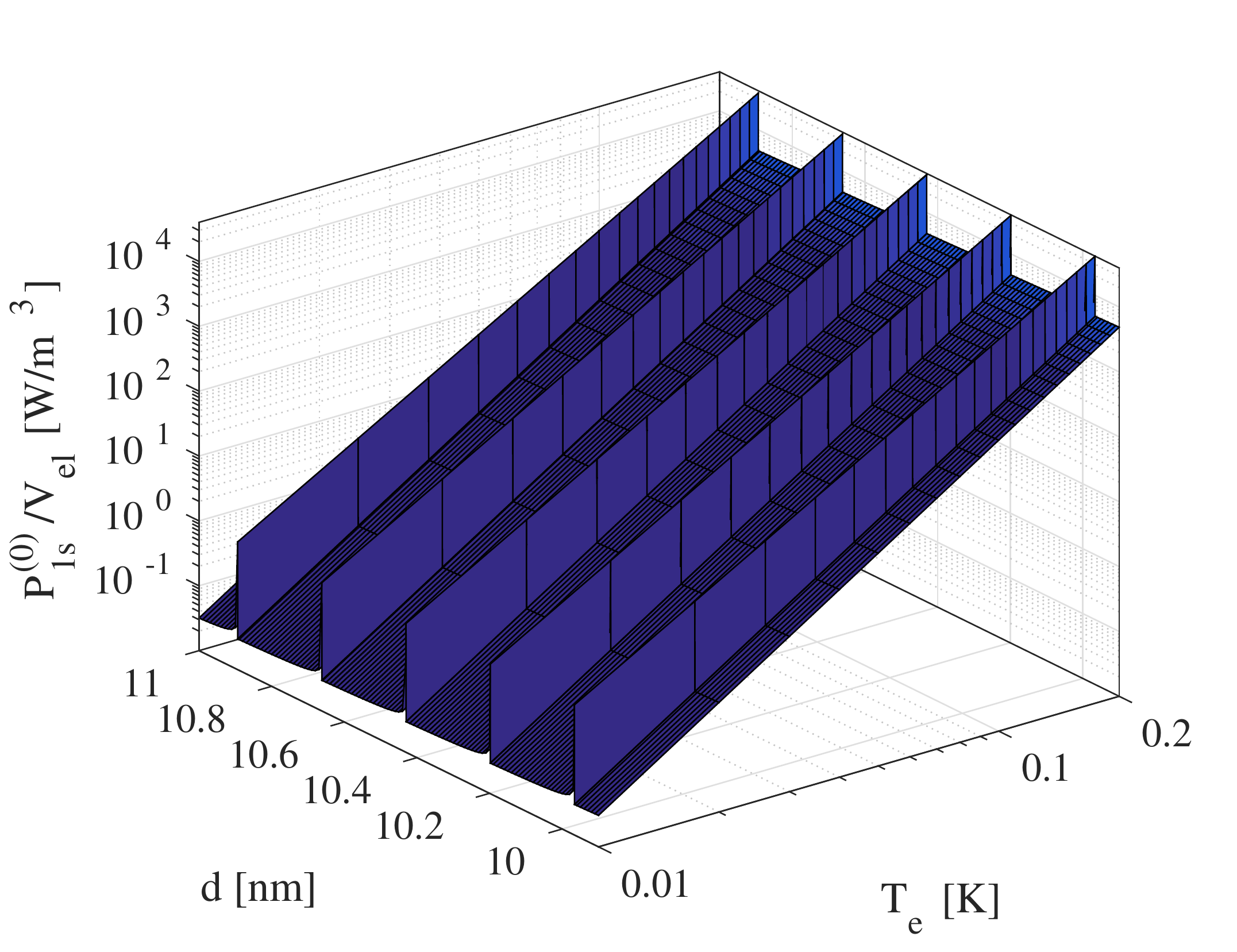}
  \caption{$P_{1s}/V_{el}$ (Eqs. \ref{Ps_em_S1S2S3} and \ref{S1s_approx}) vs $T_e$ and $d$ for $L=100$~nm. The lower plot is a zoom in the upper figure. $P^{(0)}_{1s}$ forms narrow crests approximatively in the $T_e$ direction.
  }
  \label{S_1s_0o01_2K_9o9_11o1nm}
  \end{center}
\end{figure}

From Eqs. (\ref{S1s_approx}) and (\ref{ymx1p_1s}) we calculate $P^{(0)}_{1s}(T_e)$, which is plotted in Fig. \ref{S_1s_0o01_2K_9o9_11o1nm} as a function of $T_e$ and $d$. We observe very sharp crests forming in narrow $d$ intervals, which correspond to $E_F \approx \epsilon_{n_F}$, as explained in Appendix \ref{app_S1s}.
In in the regions between the crests
\begin{eqnarray}
  \frac{P^{(0)}_{1s}}{A} &\approx& (k_BT_e)^{4} \frac{\pi^2}{135\sqrt{2}} \frac{m_e^{3/2} E_{F}^{2}}{\hbar^5 L \rho c_l^3} \sqrt{\frac{J}{(1-J)^3}} \nonumber \\
  && \times \sum_{n = 1}^{n_F} \frac{1}{\sqrt{E_F - \epsilon_n}} , \label{Ps_em_S1s_valey}
\end{eqnarray}
whereas on the crests the heat flux may increase by more than one order of magnitude and is not proportional, in general, to a single power of $T_e$.

The contribution coming from the electrons which are scattered between sub-bands of indexes that differ by an even integer ($n'-n=2k$) is (see Eq. \ref{lim_Ss_n1n2_2k})
\begin{eqnarray}
  S^{(0)}_{2s} &=& \sum_{n'} \sum_{n}^{n' - n = 2k} \int_0^\infty d\epsilon_{ph} \int_{\epsilon_1}^\infty d\epsilon_e \int_0^{2\pi} d\phi \, \epsilon_{ph}^3 \nonumber \\
  && \times S( n', n, s, q_{l} )  n(\beta_e \epsilon_{ph}) \big\{ f[\beta_e (\epsilon_e - \epsilon_{ph})] - f(\beta_e \epsilon_e) \big\} \nonumber \\
  && \times \delta(\epsilon_{\bk_\parallel - \bq_\parallel,n} - \epsilon_{\bk_\parallel,n'} + \epsilon_{ph}) \nonumber \\
  &=& \frac{\sqrt{2 m_e} d^4}{8 \pi^4 \hbar^4 c_l^3} \frac{(1 - 2J)^4}{[J(1 - J)]^{3/2}} (k_BT)^{15/2} \sum_{n'} \sum_{n}^{n' - n = 2k} \nonumber \\
  && \times \bigg[ \frac{ 1 }{(n' - n)^2} - \frac{1}{(n'+ n)^2} \bigg]^2 \int_0^\infty \frac{x_{ph}^6 \, dx_{ph}}{e^{x_{ph}}-1} \nonumber \\
  && \times \int_{0}^\infty \frac{dx}{\sqrt{x}} \left\{ \frac{1}{e^{x - (y-x^{(s)}_{n'}) - x_{ph}} + 1} \right. \nonumber \\
  && \left. - \frac{1}{e^{x - (y-x^{(s)}_{n'})} + 1} \right\} \nonumber \\
\end{eqnarray}
where $x^{(s)}_{n'} \equiv x^{(s)}_{n'}(n,x_{ph},T_e)$ and
\begin{eqnarray}
  && y - x^{(s)}_{n'} = y - x_{n'} - \beta_e \frac{ 2 m_e c_l^2 J (1-J) }{ x_{ph}^2 } \nonumber \\
  && \times \left[ x_{ph} - x_1 + x_2 + \frac{k_BT_e x_{ph}^2}{8 m_e c_l^2 J (1-J)} \right]^2 \label{ymx1p_2s}
\end{eqnarray}

\begin{figure}[t]
  \centering
  \includegraphics[width=8cm]{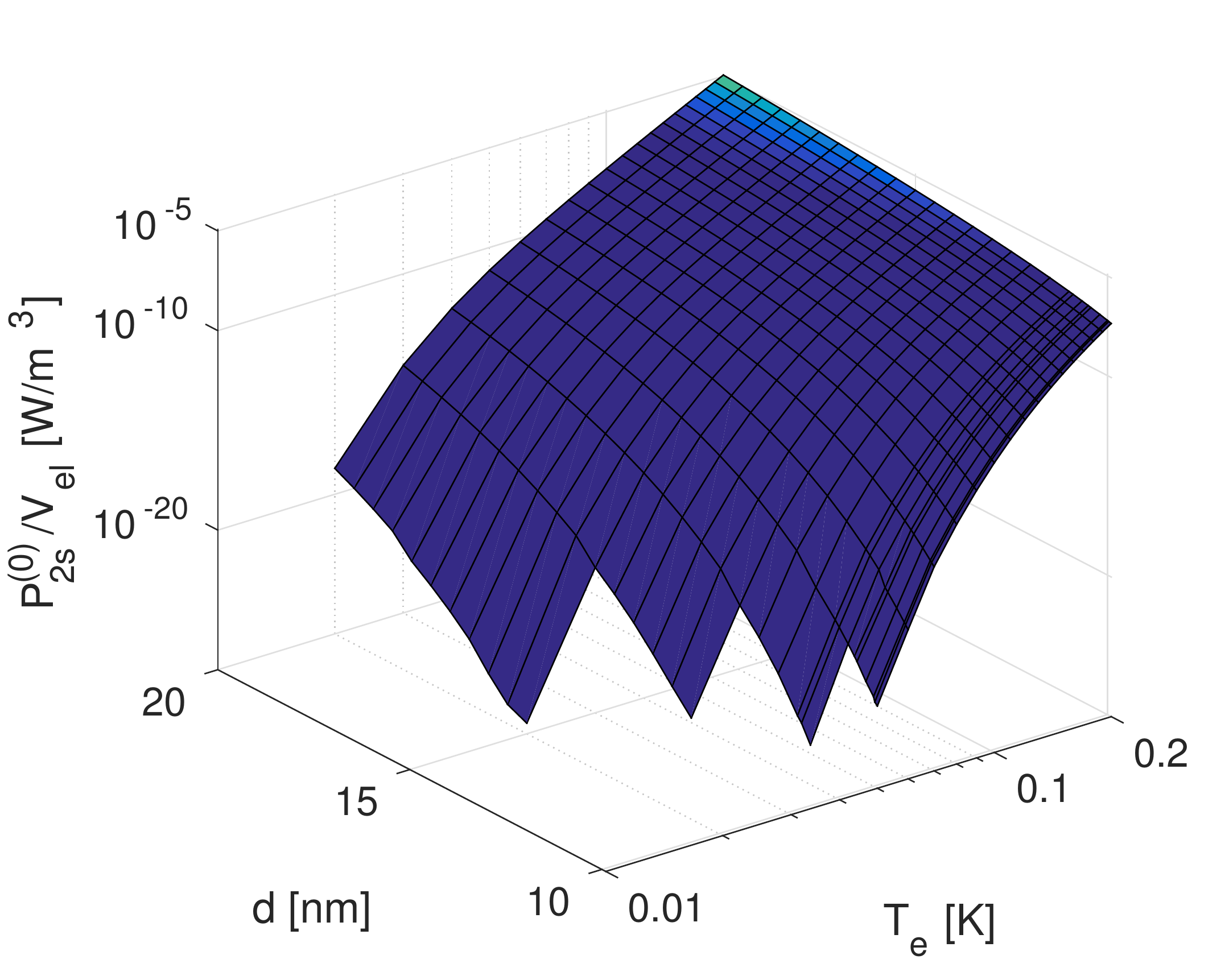} \\
  \includegraphics[width=8cm]{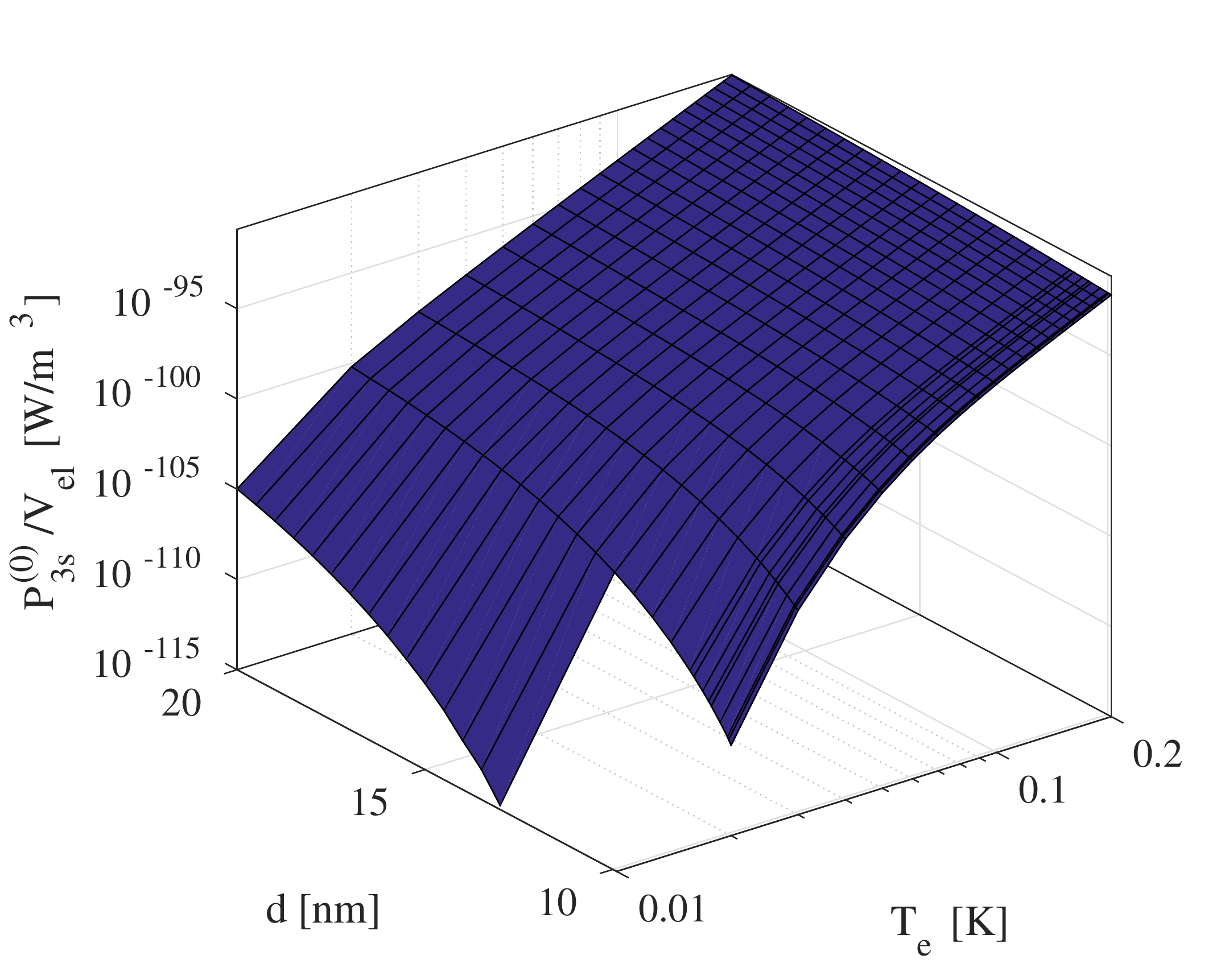} \\
  \caption{$P_{2s}$ (top) and $P_{3s}$ (bottom), for comparison with $S^{(0)}_{1s}$ (Fig. \ref{S_1s_0o01_2K_9o9_11o1nm}) and $S^{(0)}_{1a}$ (Fig. \ref{S_1a_0o01_2K_9o9_10o6nm_100nm}). Both plots are calculated for $L=100$~nm.}
  \label{S_23s}
\end{figure}

For $S^{(0)}_{3s}$ we have
\begin{eqnarray}
  S^{(0)}_{3s} &=& \sum_{n',n}^{n'-n=2k+1} \int_0^\infty d\epsilon_{ph} \int_{\epsilon_1}^\infty d\epsilon_e \int_0^{2\pi} d\phi \, \epsilon_{ph}^3 S ( n^{\prime},n, s, q_{l} ) \nonumber \\
  && \times n(\beta_e \epsilon_{ph}) \big\{ f[\beta_e (\epsilon_e - \epsilon_{ph})] - f(\beta_e \epsilon_e) \big\} \nonumber \\
  && \times \delta(\epsilon_{\bk_\parallel - \bq_\parallel,n} - \epsilon_{\bk_\parallel,n'} +\epsilon_{ph}) \nonumber \\
  &=& \frac{ (L - d/2)^2 d^2 \sqrt{2 m_e}}{2 \pi^4 \hbar^4 c_l^3} \frac{(1-2J)^4}{[J(1-J)]^{3/2}} (k_BT_e)^{15/2} \nonumber \\
  && \times \sum_{n'} \sum_{n_2}^{n'-n_2 = 2k+1}  \bigg[ \frac{ 1 }{(n' - n_2)^2} -  \frac{ 1 }{(n' + n_2)^2} \bigg]^2 \nonumber \\
  && \times \int_0^\infty \frac{dx_{ph}\, x_{ph}^6}{e^{x_{ph}} - 1} \int_{0}^\infty \frac{dx_e}{\sqrt{x}} \left[ \frac{1}{e^{x - (y - x^{(s)}_{n'}) - x_{ph}} + 1} \right. \nonumber \\
  && \left. - \frac{1}{e^{x - (y - x^{(s)}_{n'})} + 1} \right] . \label{S3s_calc1}
\end{eqnarray}
where again $x^{(s)}_{n'} \equiv x^{(s)}_{n'}(n,x_{ph},T_e)$.
$S^{(0)}_{2s}$ and $S^{(0)}_{3s}$ are plotted in Fig. \ref{S_23s} (top and bottom, respectively). Comparing these plots with Fig. \ref{S_1s_0o01_2K_9o9_11o1nm} we see that $S^{(0)}_{2s}$ and $S^{(0)}_{3s}$ practically do not contribute to $P^{(0)}$.

\subsubsection{Contribution of the $a$ modes} \label{subsubsec_HeatFl_a}

For the antisymmetric modes we have
\begin{eqnarray}
  P_a^{(0)} &=& 2 \frac{A^2}{(2\pi)^3}\sum_{n'} \sum_{n} \int_0^\infty dk_\parallel\, k_\parallel \int_0^\infty dq_\parallel q_\parallel \int_0^{2\pi} d\phi \nonumber \\
  && \times |N_{a}|^{2} \left| q_{t}q_\parallel\left( q_\parallel^{2} - p_{l}^{2} \right)  \right|^{2} \frac{16 \pi E_{F}^{2}}{9 \rho \omega_{a}} \sin^{2} \left( \frac{Lq_{t}}{2} \right) \nonumber \\
  && \times S ( n',n_2, a, q_{l} ) \hbar \omega_a n(\beta_e \epsilon_{q_\parallel}) \big\{ f[\beta_e (\epsilon_{k_\parallel,n'} - \epsilon_{ph})] \nonumber \\
  && - f(\beta_e \epsilon_{k_\parallel,n'}) \big\} \delta(\epsilon_{\bk_\parallel - \bq_\parallel,n}-\epsilon_{\bk,n'} + \hbar\omega_{a}) \nonumber \\
  &=& \frac{2 A}{3\pi^2 L^3} \frac{m_e E_F^2}{\rho c_l^2 \hbar^3} \frac{J}{(1-J)} \sum_{n'} \sum_{n} \int_{\epsilon_n'}^\infty d\epsilon_e \int_0^\infty d\epsilon_{ph} \, \epsilon_{ph} \nonumber \\
  && \times \int_0^{2\pi} d\phi \, S ( n',n, a, q_{l} ) n(\beta_e \epsilon_{ph}) \big\{ f[\beta_e (\epsilon_k - \epsilon_{ph})] \nonumber \\
  && - f(\beta_e \epsilon_k) \big\} \delta(\epsilon_{\bk_\parallel - \bq_\parallel}-\epsilon_\bk + \hbar\omega_{a})
  \label{Pa_em1}
\end{eqnarray}
Plugging in the energy conservation from the $\delta$ function, the power flux emitted from electrons to the antisymmetric phonons can be divided in three parts, in the same way as we did before,
\begin{eqnarray}
  P_a^{(0)} &=& \frac{2 A}{3\pi^2 L^3} \frac{m_e E_F^2}{\rho c_l^2 \hbar^3} \frac{J}{(1-J)} (S^{(0)}_{1a} + S^{(0)}_{2a} + S^{(0)}_{3a}) \nonumber \\
  &\equiv& P_{1a}^{(0)} + P_{2a}^{(0)} + P_{3a}^{(0)} . \label{Pa_em_sum}
\end{eqnarray}
For the intra-band transitions we have
\begin{eqnarray}
  && S^{(0)}_{1a} = \frac{1}{2} \sqrt{\frac{m_e}{2\hbar^3}} \left[\frac{3}{ J (1-J)}\right]^{1/4} \frac{(L-d)^2}{\sqrt{c_l L}} (k_BT_e)^{3} \sum_{n'} \nonumber \\
  && \times \int_0^\infty \frac{dx_{ph} \, x_{ph}^{3/2}}{e^{x_{ph}} - 1} \int_{0}^\infty \frac{dx}{\sqrt{x}} \left\{ \frac{1}{e^{x - (y-x^{(a)}_{n'}(n',x_{ph})) - x_{ph}} + 1} \right. \nonumber \\
  && \left. - \frac{1}{e^{x - (y-x^{(a)}_{n'}(n',x_{ph}))} + 1} \right\} , \label{S1a_form1}
\end{eqnarray}
where the variables for the antisymmetric modes are
%
\begin{eqnarray}
  && \epsilon^{(a)}_{n'}(n,\epsilon_{ph}) \equiv \epsilon_{n'} + \frac{m_e L c_l}{2 \hbar} \sqrt{\frac{J(1-J)}{3}} \frac{1}{\epsilon_{ph}} \left\{ \epsilon_n - \epsilon_{n'} \right. \nonumber \\ 
  && \left. + \epsilon_{ph} \left[1 + \frac{\hbar}{2 m_e L c_l} \sqrt{ \frac{3}{J(1-J)} } \right] \right\}^2 \equiv \epsilon_{n'} + \delta \epsilon^{(a)}_{n'}(n,\epsilon_{ph}) , \nonumber
\end{eqnarray}
$x^{(a)}_{n'}(n,x_{ph}) \equiv \beta_e \epsilon^{(a)}_{n'}(n,\epsilon_{ph})$, and
\begin{eqnarray}
  && \delta x^{(a)}_{n'}(n,x_{ph}) = \beta_e \delta\epsilon^{(a)}_{n'}(n,\epsilon_{ph}) = \frac{m_e L c_l}{2 \hbar x_{ph}} \sqrt{\frac{J(1-J)}{3}} \nonumber \\
  && \times \left\{ x_n - x_{n'} + x_{ph} \left[1 + \frac{\hbar}{2 m_e L c_l} \sqrt{ \frac{3}{J(1-J)} } \right] \right\}^2 . \label{epsp_np_a}
\end{eqnarray}
We note that in the case of scattering on the antisymmetric modes $\delta x^{(a)}_{n'}(n,x_{ph})$ depends on $T_e$ only through $x_n$, $x_{n'}$, and $x_{ph}$.

\begin{figure}[t]
  \centering
  \includegraphics[width=8cm,bb=0 0 607 438,keepaspectratio=true]{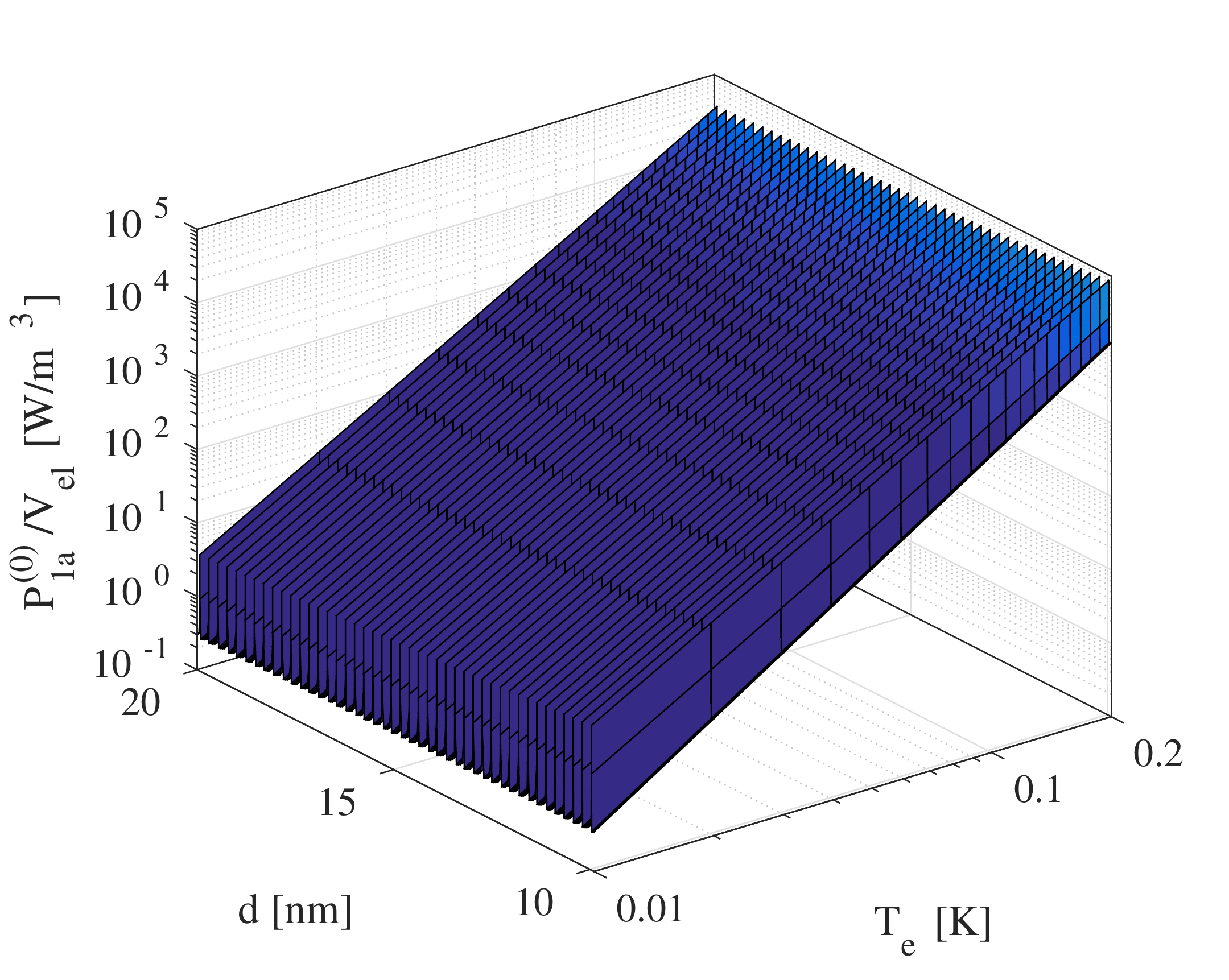} \\
  \includegraphics[width=8cm,bb=0 0 607 438,keepaspectratio=true]{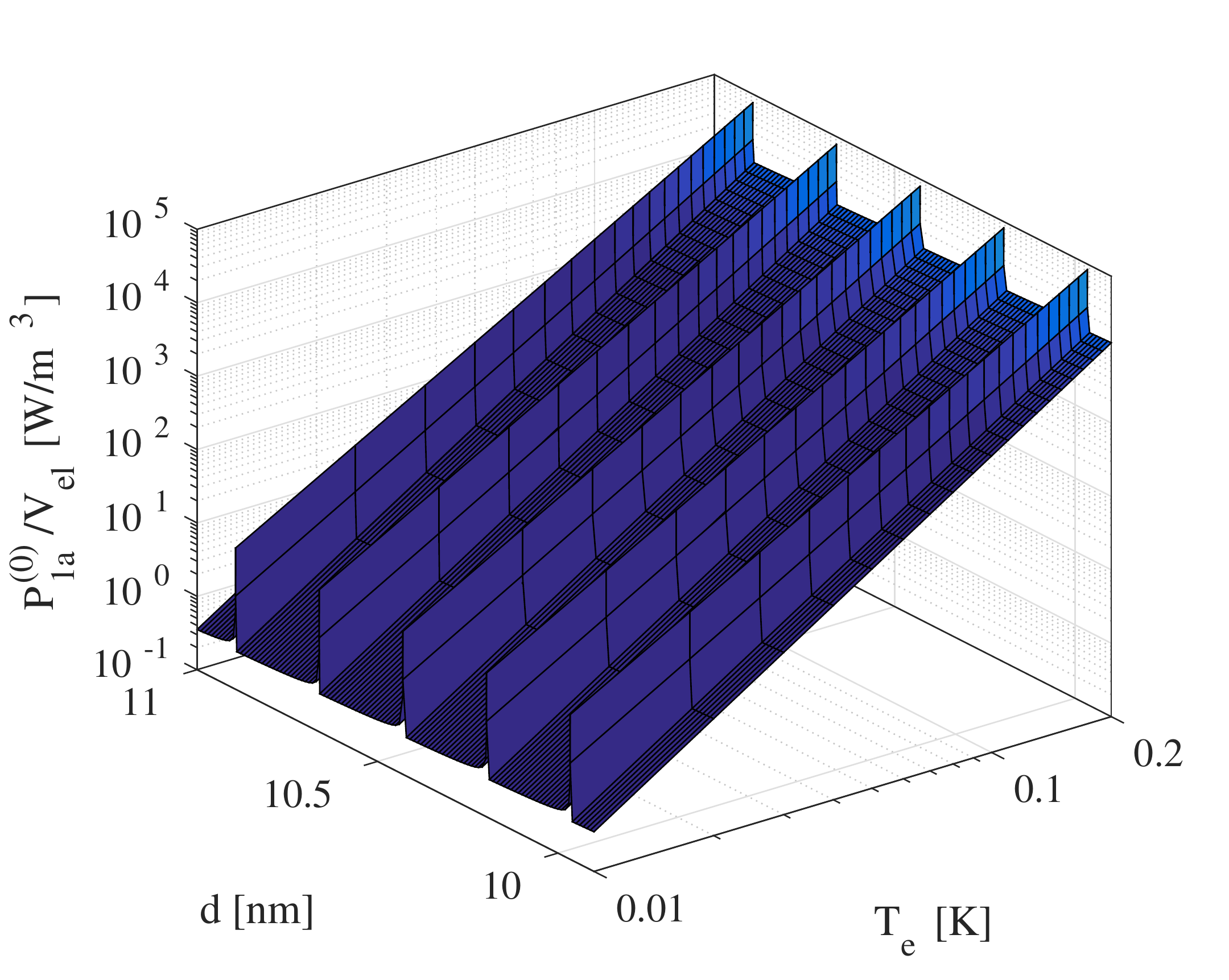}
  \caption{
  $P_{1a}/V_{el}$ (Eqs. \ref{Pa_em_sum} and \ref{S1a_form1}) vs $T_e$ and $d$ for $L=100$~nm for $L=100$~nm. The lower plot is a zoom in the upper figure. $P^{(0)}_{1a}$ forms narrow crests approximatively in the $T_e$ direction.
  }
  \label{S_1a_0o01_2K_9o9_10o6nm_100nm}
\end{figure}

The function $P_{1a}/V_{el}$ is plotted in Fig. \ref{S_1a_0o01_2K_9o9_10o6nm_100nm}. We notice similar crests as in Fig. \ref{S_1s_0o01_2K_9o9_11o1nm}, which correspond to $E_F \approx \epsilon_{n_F}$. The nature of these crests is discussed in Appendix~\ref{app_S1a}.
In the regions between the crests the thermal power is described by
\begin{eqnarray}
  \frac{P_{1a}^{(0)}}{A} &=& \frac{3^{1/4} 5}{8 \sqrt{2} \pi^{3/2}}\zeta\left( \frac{7}{2} \right) \frac{m_e^{3/2} E_F^2 (L-d)^2}{\hbar^{9/2} \rho c_l^{5/2} L^{7/2}} \frac{J^{3/4}}{(1-J)^{5/4}} \nonumber \\
  && \times (k_BT_e)^{7/2} \sum_n^{n\le n_{n_F}} \frac{1}{\sqrt{E_F - \epsilon_n}} , \label{Pa_em_S1s_valey}
\end{eqnarray}
whereas on the crests the heat power may increase by more than one order of magnitude and is not proportional, in general, to a single power of $T_e$.

\begin{figure}[t]
  \centering
   \includegraphics[width=8cm]{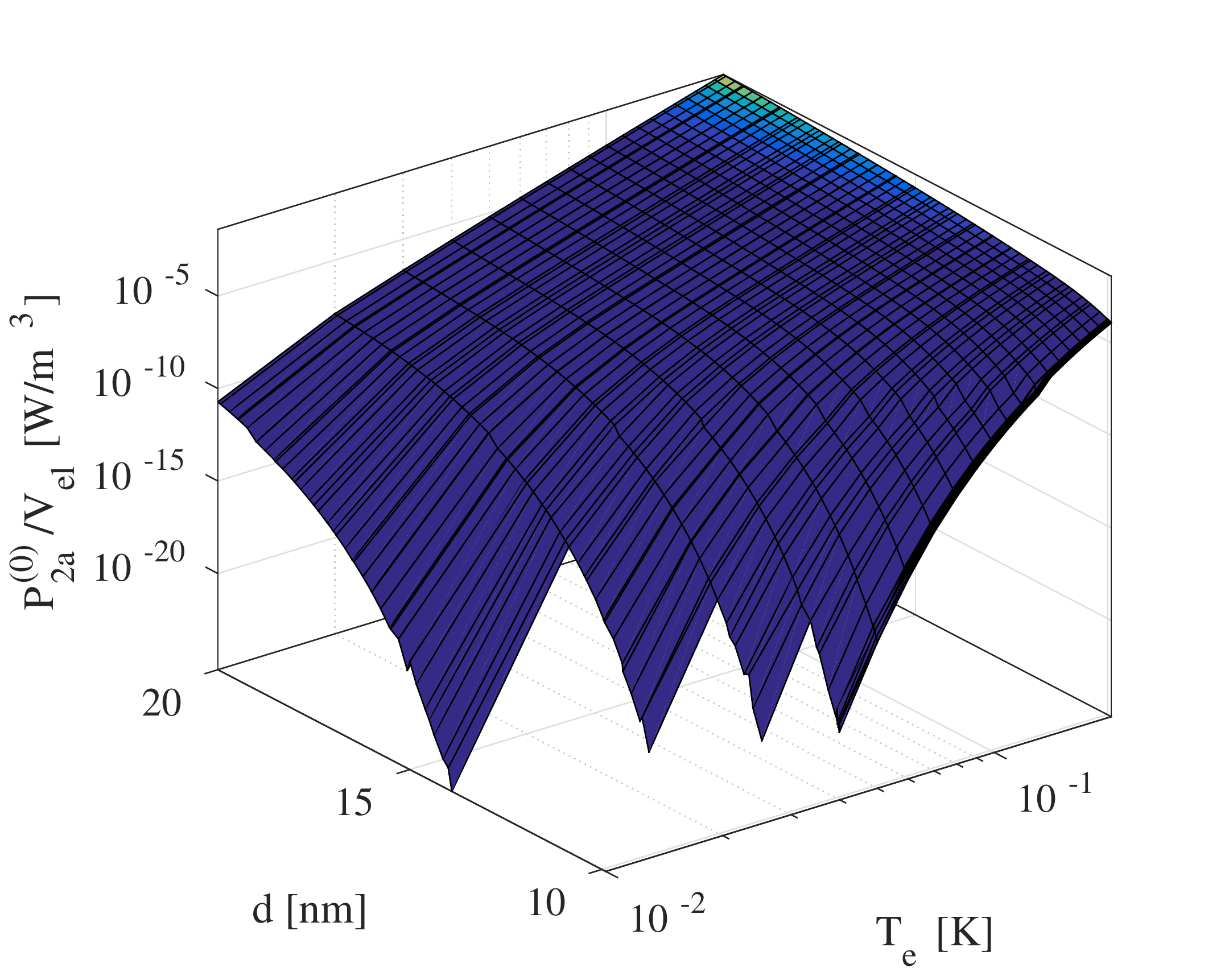} \\
  \includegraphics[width=8cm]{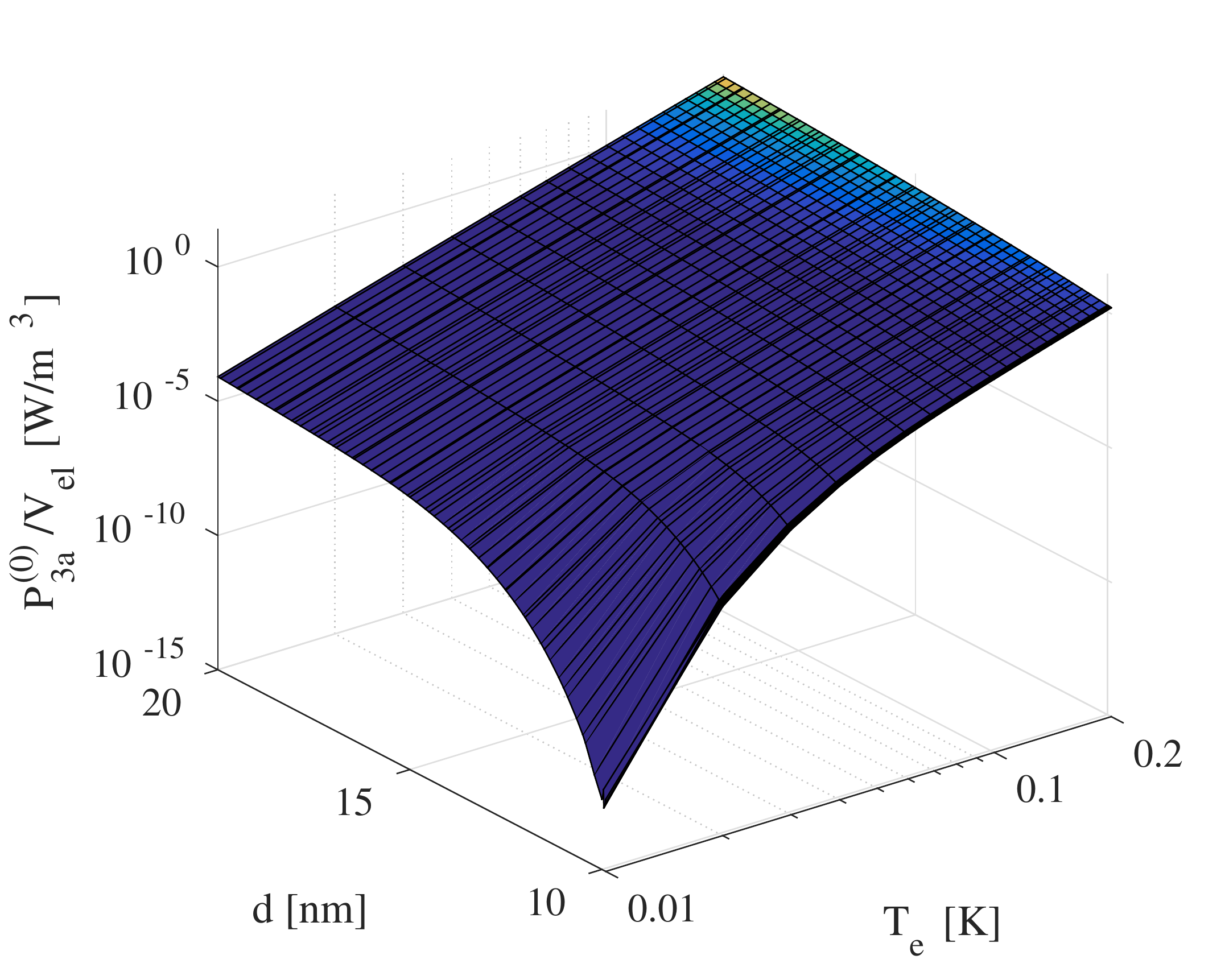}
 \caption{$P_{2s}$ (top) and $P_{3s}$ (bottom) for comparison with $S^{(0)}_{1s}$ (Fig. \ref{S_1s_0o01_2K_9o9_11o1nm}) and $S^{(0)}_{1a}$ (Fig. \ref{S_1a_0o01_2K_9o9_10o6nm_100nm}). Both plots are calculated for $L=100$~nm.}
  \label{S_23a}
\end{figure}

Using Eqs. (\ref{Sa2_limT0}) and (\ref{Sa3_limT0}) we write
\begin{eqnarray}
  && S^{(0)}_{2a} = \frac{9}{2 \pi^4} \sqrt{\frac{m_e}{2\hbar^7}} \frac{d^4 (L-d)^2}{(L c_l)^{5/2} J(1-J)} \left[ \frac{3}{J (1-J)} \right]^{1/4} \nonumber \\
  && \times \sum_{n',n}^{n'-n = 2k} \bigg\{ \frac{ 1 }{(n'-n)^2} - \frac{ 1 }{(n'+n)^2} \bigg\}^2 (k_BT_e)^5 \nonumber \\
  && \times \int_0^\infty \frac{dx_{ph} \, x_{ph}^{7/2}}{e^{x_{ph}} -1} \int_{0}^\infty \frac{dx}{\sqrt{x}} \left\{ \frac{1}{e^{x - (y-x^{(a)}_{n'}(n,x_{ph})) - x_{ph}} + 1} \right. \nonumber \\
  && \left. - \frac{1}{e^{x - (y-x^{(a)}_{n'}(n,x_{ph}))} + 1} \right\} \label{S2a_form1}
\end{eqnarray}
and
\begin{eqnarray}
  && S^{(0)}_{3a} =  \frac{4\sqrt{2}}{\pi^4} \sqrt{\frac{m_e}{\hbar^3}} \left[ \frac{3}{J (1-J)} \right]^{1/4} \frac{d^2}{\sqrt{L c_l}} (k_BT_e)^3 \nonumber \\
  && \times \sum_{n',n}^{n'-n = 2k+1}  \bigg\{ \frac{ 1 }{(n'-n)^2} - \frac{ 1 }{(n'+n)^2} \bigg\}^2 \nonumber \\
  && \times \int_0^\infty \frac{dx_{ph} \, x_{ph}^{3/2}}{e^{x_{ph}} - 1} \int_{0}^\infty \frac{dx}{\sqrt{x}} \left[ \frac{1}{e^{x - (y-x^{(a)}_{n'}(n,x_{ph})) - x_{ph}} + 1} \right. \nonumber \\
  && \left. - \frac{1}{e^{x - (y-x^{(a)}_{n'}(n,x_{ph}))} + 1} \right] . \label{S3a_form1}
\end{eqnarray}
The functions $S^{(0)}_{2a}$ and $S^{(0)}_{3a}$ are plotted in Fig. \ref{S_23a} (top and bottom, respectively) and we see that they do not practically contribute to the heat power.

\begin{figure}[b]
  \centering
  \includegraphics[width=8cm,bb=0 0 607 438,keepaspectratio=true]{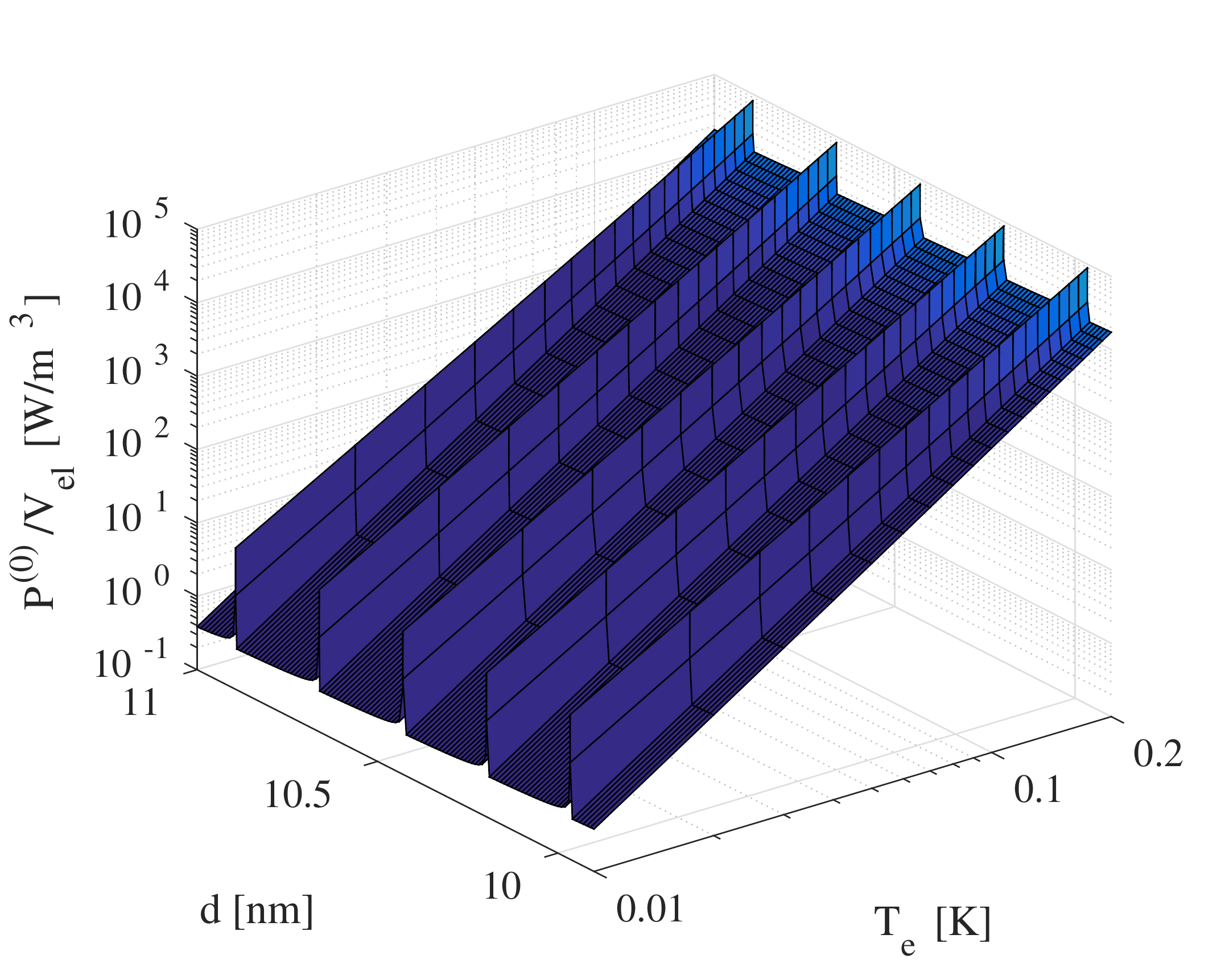}
  \caption{$P^{(0)}/V_{el}$ for $L=100$~nm.}
  \label{log10_P0_vol_0o01_0o2K_9o9_11nm_L100nm}
\end{figure}

\begin{figure}[t]
  \centering
  \includegraphics[width=8cm,bb=0 0 607 438,keepaspectratio=true]{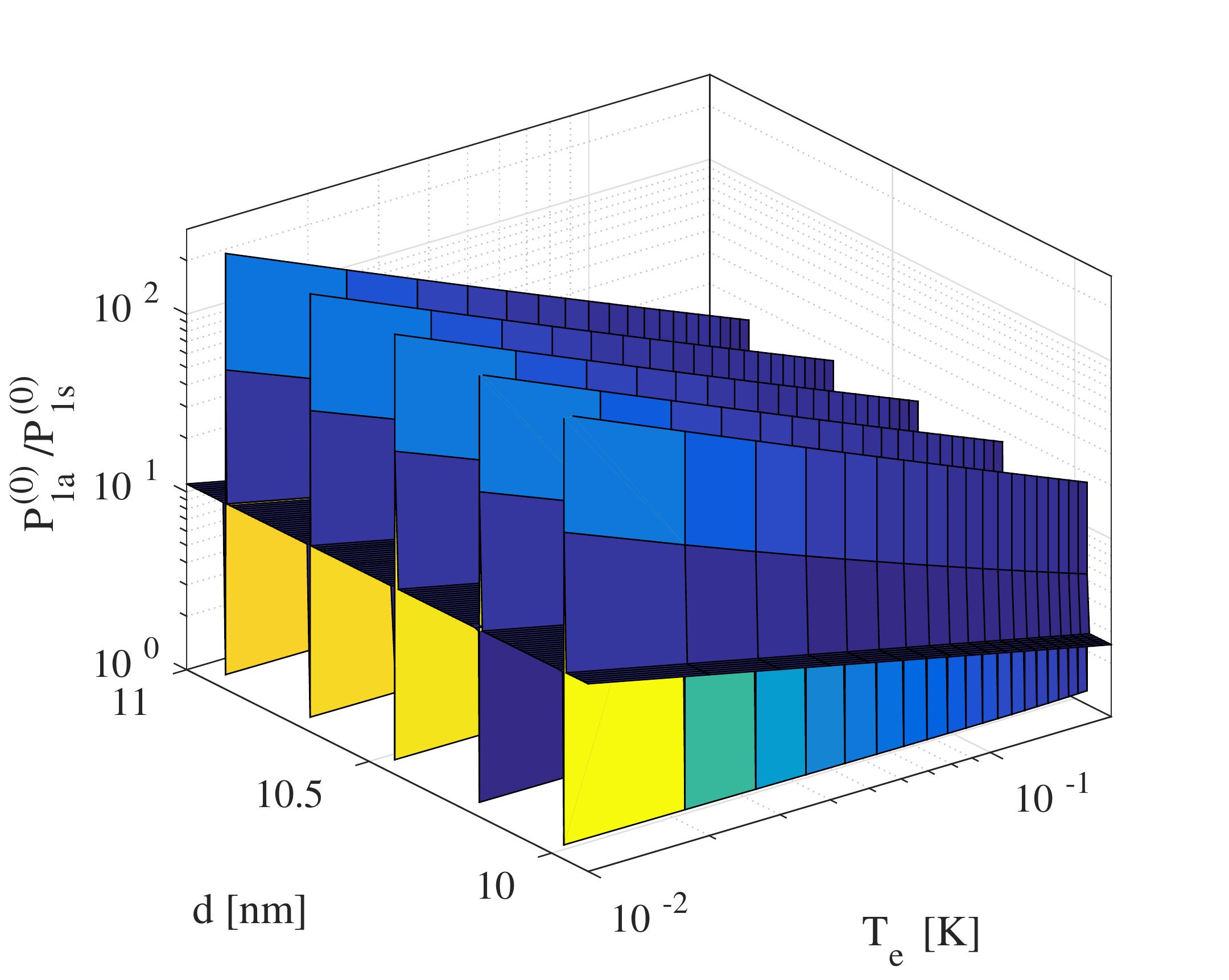}
  \caption{The ratio $P^{(0)}_a/P^{(0)}_s$ for $L=100$~nm. We see that the dominant contribution comes from the heat flux from electrons to the antisymmetric phonons.}
  \label{P1a_div_P1s}
\end{figure}

\begin{figure}[t]
\begin{center}
\includegraphics[width=8cm,keepaspectratio=true]{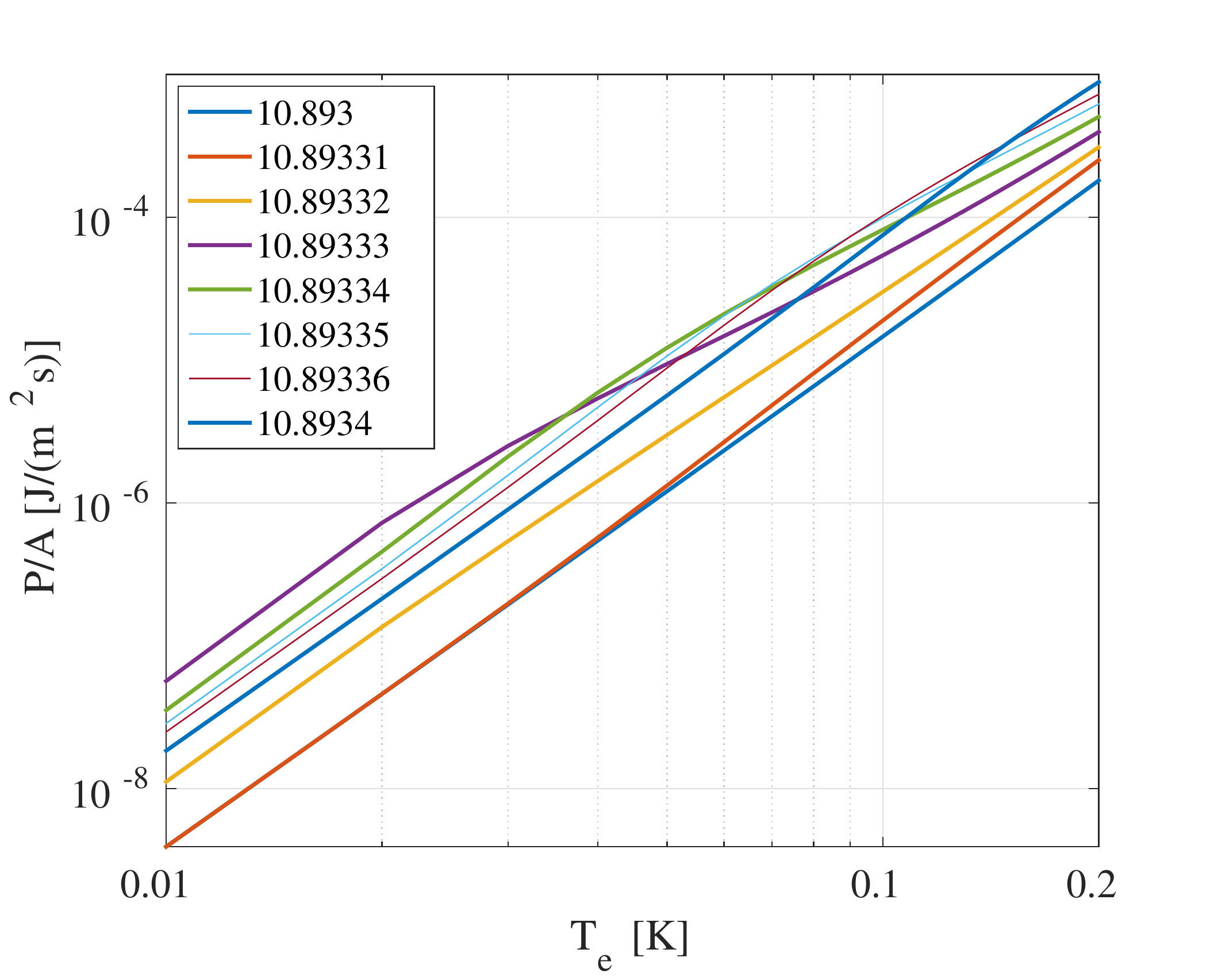}
\end{center}
\caption{(Color online) $P^{(0)}$ as a function of $T_e$ for different
values of $d$ (shown in the legend) and $L=100$~nm. The higher values of $P^{(0)}$ are from
the crests regions (see Fig.~\ref{log10_P0_vol_0o01_0o2K_9o9_11nm_L100nm}). We notice that the heat flux
cannot be written as $P^{(0)} \propto T_e^x$.}
\label{log10_P_tot_0o01_0o2K_9o9_11nm_L100nm}
\end{figure}

Adding all the contributions (\ref{Ps_em_S1S2S3}) and (\ref{Pa_em_sum}) we obtain the total heat power $P^{(0)}$ emitted from the electrons into the phonons system. $P^{(0)}$ is plotted in Fig. \ref{log10_P0_vol_0o01_0o2K_9o9_11nm_L100nm} and the ratio $P^{(0)}_a/P^{(0)}_s$ is plotted in Fig. \ref{P1a_div_P1s}. We see that the dominant contribution comes from the heat power to the antisymmetric modes. Therefore in the low temperature limit $P^{(0)}$ is described by Eq. (\ref{Pa_em_S1s_valey}) between the crests, so it is proportional to $T_e^{7/2}$. Along the crests, both $P^{(0)}_a$ and $P^{(0)}_s$ may contribute substantially, together with $P^{(0)}_a$, and there is no simple temperature dependence of $P^{(0)}$.

%

\begin{figure}[b]
\centering
\includegraphics[width=8cm]{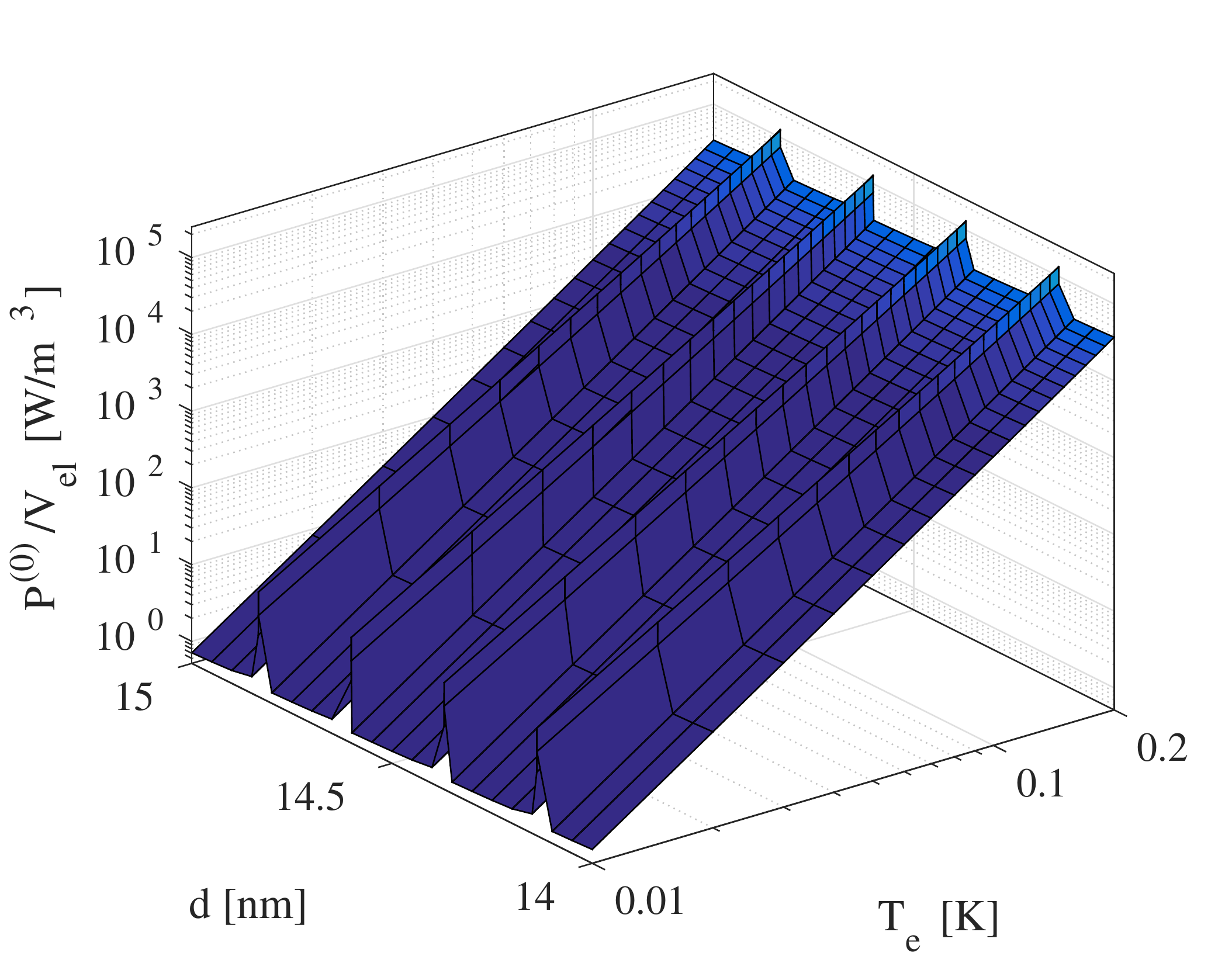}
\caption{$P^{(0)}(T_e,d)/V_{el}$ for $L = 45$~nm.}
\label{log10_P0_vol_0o01_0o2K_14_15nm_L45nm}
\end{figure}

For a rough comparison with experimental data published in Ref. [\onlinecite{PRL.99.145503.2007.Karvonen, JPhysConfSer.92.012043.2007.Karvonen}] we plot in Fig. \ref{log10_P0_vol_0o01_0o2K_14_15nm_L45nm} the heat power for a sample with $L = 45$~nm (like sample M1 in Ref. [\onlinecite{PRL.99.145503.2007.Karvonen}]). Although the experimental data is scattered, we observe a qualitative agreement.

\subsection{The heat flux from phonons to electrons} \label{sub_ph_to_el}

The heat power $P^{(1)}$ from phonon to electrons is calculated as $P^{(0)}$, with the exception that in the integrals over $\epsilon_{ph}$ one should use a Bose distribution corresponding to the temperature $T_{ph}$ instead of $T_e$.

\subsubsection{The contribution of the $s$ modes} \label{subsub_ph_to_el_s}

As in Section \ref{subsubsec_HeatFl_S}, we write
\begin{eqnarray}
  P_s^{(1)} &=& A \frac{m_e E_{F}^{2}}{36\pi^2 L \rho c_l^4 \hbar^5} \frac{1}{(1-J)^2} ( S^{(1)}_{1s} + S^{(1)}_{2s} + S^{(1)}_{3s} ) \nonumber\\
  &\equiv& P^{(1)}_{1s} + P^{(1)}_{2s} + P^{(1)}_{3s} , \label{Ps1_em_S1S2S3}
\end{eqnarray}
where
\begin{eqnarray}
  && S^{(1)}_{1s} = 2 c_l \sqrt{2 m_e J (1-J)} \sum_{n'} \int_0^\infty d\epsilon_{ph} \int_{\epsilon^{(s)}_{n'}}^\infty d\epsilon_e \nonumber \\
  && \times \frac{\epsilon_{ph}^2 \, n(\beta_{ph} \epsilon_{ph})}{\sqrt{\epsilon_e - \epsilon^{(s)}_{n'}(n',\epsilon_{ph})}} \big\{ f[\beta_e (\epsilon_e - \epsilon_{ph})] - f(\beta_e \epsilon_e) \big\} \nonumber \\
  &=& 2 c_l \sqrt{2 m_e J (1-J)} (k_BT_e)^{7/2} \sum_{n'} \int_0^\infty dx_{ph} \frac{x_{ph}^2}{e^{t x_{ph}} - 1} \nonumber \\
  &\times& \int\limits_0^\infty \frac{dx}{\sqrt{x}} \left\{ \frac{1}{e^{x - (y-x^{(s)}_{n'}(n',x_{ph},T_e)) - x_{ph}} + 1} \right. \nonumber \\
  && \left. - \frac{1}{e^{x - (y-x^{(s)}_{n'}(n',x_{ph},T_e))} + 1} \right\} \label{S1s1_approx}
\end{eqnarray}
where $t = T_e/T_{ph}$, and all the other quantities are the same as in Section \ref{subsubsec_HeatFl_S}.

\begin{figure}[t]
  \begin{center}
  \includegraphics[width=8cm]{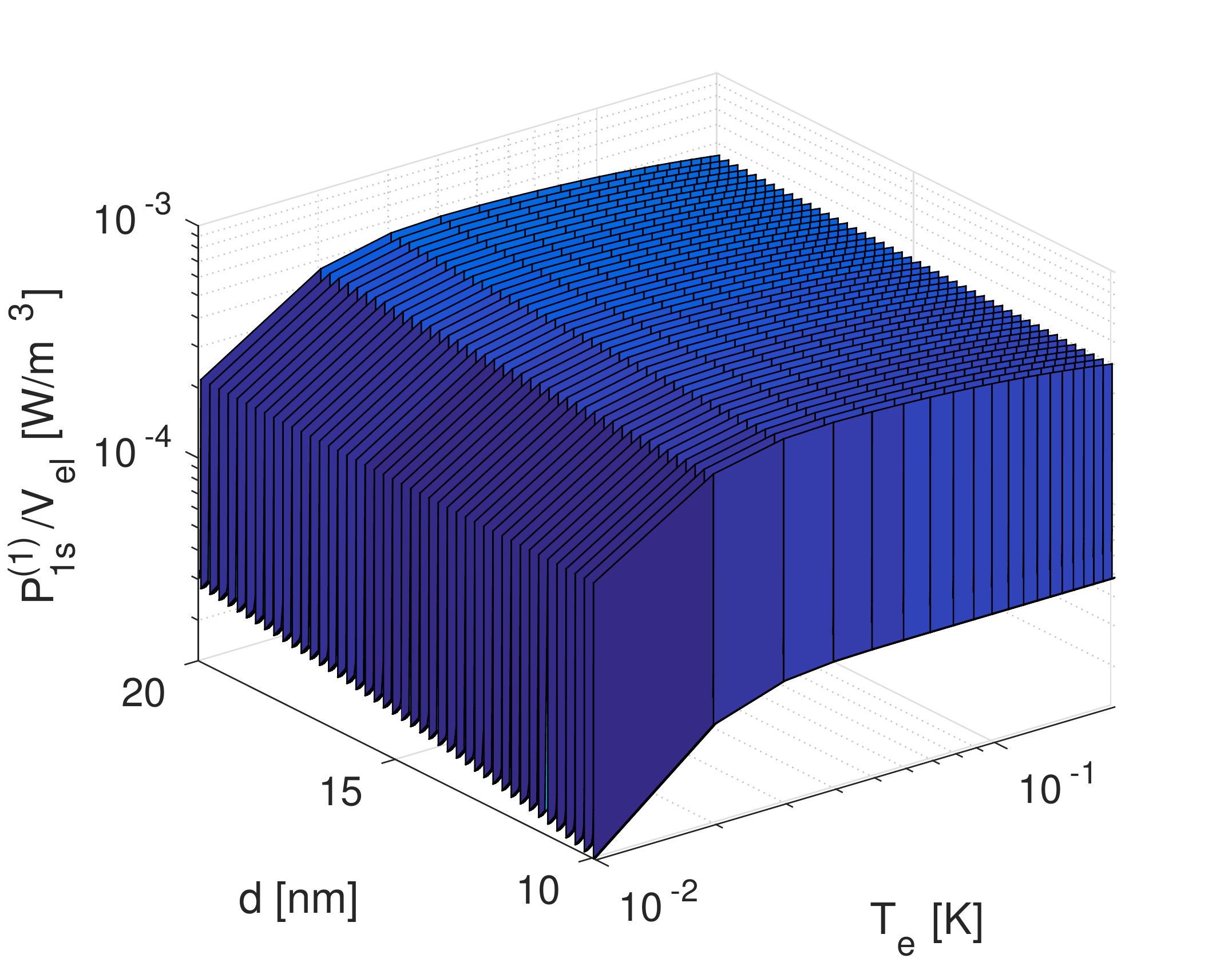} \\
  \includegraphics[width=8cm]{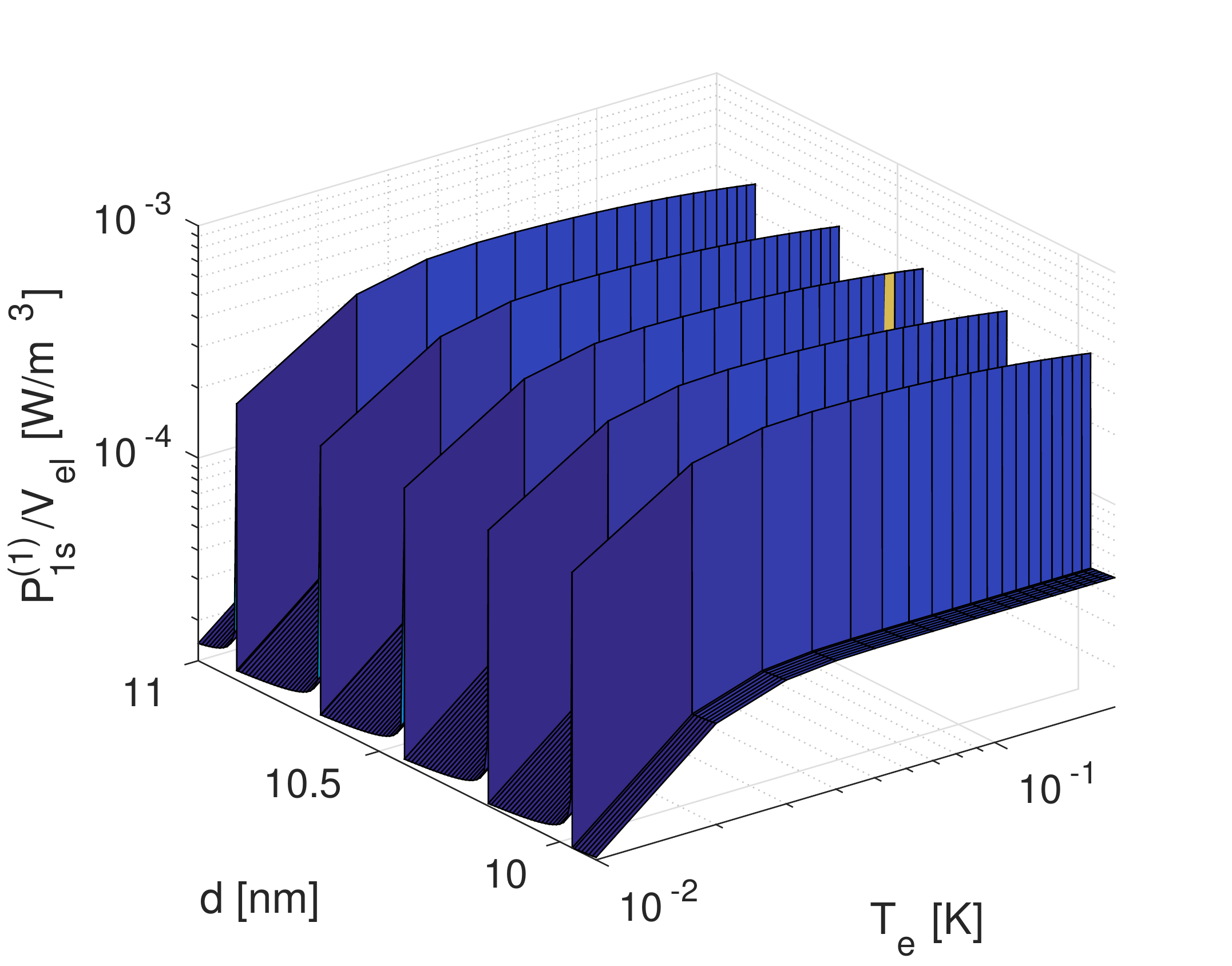}
  \caption{$P^{(1)}_{1s}/V_{el}$ (Eqs. \ref{Ps_em_S1S2S3} and \ref{S1s_approx}) vs $T_e$ and $d$. The lower plot is a zoom in the upper figure.}
  \label{P1_1s_0o01_2K_9o9_11o1nm}
  \end{center}
\end{figure}

From Eqs. (\ref{S1s1_approx}) we calculate $P^{(1)}_{1s}(T_e)$, which is plotted in Fig. \ref{P1_1s_0o01_2K_9o9_11o1nm} as a function of $T_e$ and $d$. We observe very sharp crests forming in narrow $d$ intervals, which correspond to $E_F \approx \epsilon_{n_F}$, as explained in Appendix \ref{app_S1s1}. In the regions between the crests
\begin{eqnarray}
  \frac{P^{(1)}_{1s}}{A} &=& (k_BT_{ph})^{4} \frac{\pi^2}{135 \sqrt{2}} \frac{m_e^{3/2} E_{F}^{2}}{\hbar^5 \rho c_l^3 L} \frac{J^{1/2}}{(1-J)^{3/2}} \nonumber \\
  && \times \sum_{n = 1}^{n_F} \frac{1}{\sqrt{E_F - \epsilon_n}} , \label{Ps1_pe_em_valley}
\end{eqnarray}
whereas on the crests the heat power may increase by more than one order of magnitude and is not described by a simple power law.

The contribution coming from the electrons which are scattered between sub-bands of indexes that differ by an even integer ($n'-n=2k$) is (see Eq. \ref{lim_Ss_n1n2_2k})
\begin{eqnarray}
  && S^{(1)}_{2s} = \sum_{n'} \sum_{n}^{n' - n = 2k} \int_0^\infty d\epsilon_{ph} \int_{\epsilon_1}^\infty d\epsilon_e \int_0^{2\pi} d\phi \, \epsilon_{ph}^3 \nonumber \\
  && \times S( n', n, s, q_{l} )  n(\beta_{ph} \epsilon_{ph}) \big\{ f[\beta_e (\epsilon_e - \epsilon_{ph})] - f(\beta_e \epsilon_e) \big\} \nonumber \\
  && \times \delta(\epsilon_{\bk_\parallel - \bq_\parallel,n} - \epsilon_{\bk_\parallel,n'} + \epsilon_{ph}) \nonumber \\
  && = \frac{\sqrt{2 m_e} d^4}{8 \pi^4 \hbar^4 c_l^3} \frac{(1 - 2J)^4}{[J(1 - J)]^{3/2}} (k_BT)^{15/2} \sum_{n'} \sum_{n}^{n' - n = 2k} \nonumber \\
  && \times \bigg[ \frac{ 1 }{(n' - n)^2} - \frac{1}{(n'+ n)^2} \bigg]^2 \int_0^\infty \frac{x_{ph}^6 \, dx_{ph}}{e^{t x_{ph}}-1} \nonumber \\
  && \times \int_{0}^\infty \frac{dx}{\sqrt{x}} \left\{ \frac{1}{e^{x - (y-x^{(s)}_{n'}) - x_{ph}} + 1} \right. \nonumber \\
  && \left. - \frac{1}{e^{x - (y-x^{(s)}_{n'})} + 1} \right\} \nonumber
\end{eqnarray}
whereas the rest of the terms in the summation (\ref{Ps1_em_S1S2S3}) are
\begin{eqnarray}
  && S^{(1)}_{3s} = \sum_{n',n}^{n'-n=2k+1} \int_0^\infty d\epsilon_{ph} \int_{\epsilon_1}^\infty d\epsilon_e \int_0^{2\pi} d\phi \, \epsilon_{ph}^3 S ( n^{\prime},n, s, q_{l} ) \nonumber \\
  && \times n(\beta_{ph} \epsilon_{ph}) \big\{ f[\beta_e (\epsilon_e - \epsilon_{ph})] - f(\beta_e \epsilon_e) \big\} \nonumber \\
  && \times \delta(\epsilon_{\bk_\parallel - \bq_\parallel,n} - \epsilon_{\bk_\parallel,n'} +\epsilon_{ph}) \nonumber \\
  && = \frac{ (L - d/2)^2 d^2 \sqrt{2 m_e}}{2 \pi^4 \hbar^4 c_l^3} \frac{(1-2J)^4}{[J(1-J)]^{3/2}} (k_BT_e)^{15/2} \nonumber \\
  && \times \sum_{n'} \sum_{n_2}^{n'-n_2 = 2k+1}  \bigg[ \frac{ 1 }{(n' - n_2)^2} -  \frac{ 1 }{(n' + n_2)^2} \bigg]^2 \nonumber \\
  && \times \int_0^\infty \frac{dx_{ph}\, x_{ph}^6}{e^{t x_{ph}} - 1} \int_{0}^\infty \frac{dx_e}{\sqrt{x}} \left[ \frac{1}{e^{x - (y - x^{(s)}_{n'}) - x_{ph}} + 1} \right. \nonumber \\
  && \left. - \frac{1}{e^{x - (y - x^{(s)}_{n'})} + 1} \right] . \nonumber 
\end{eqnarray}
Like in the case of $P^{(0)}$, the terms $P^{(1)}_{2s}$ and $P^{(1)}_{3s}$ do not contribute significantly to $P^{(1)}_s$.

\subsubsection{The contribution of the $a$ modes} \label{subsub_ph_to_el_a}

For the antisymmetric modes we have
\begin{eqnarray}
  P_a^{(1)} &=& \frac{2 A}{3\pi^2 L^3} \frac{m_e E_F^2}{\rho c_l^2 \hbar^3} \frac{J}{(1-J)} (S^{(1)}_{1a} + S^{(1)}_{2a} + S^{(1)}_{3a}) \nonumber \\
  &\equiv& P_{1a}^{(1)} + P_{2a}^{(1)} + P_{3a}^{(1)}, \label{P1a_em_sum}
\end{eqnarray}
where
\begin{eqnarray}
  && S^{(1)}_{1a} = \frac{1}{2} \sqrt{\frac{m_e}{2\hbar^3}} \left[\frac{3}{ J (1-J)}\right]^{1/4} \frac{(L-d)^2}{\sqrt{c_l L}} (k_BT_e)^{3} \sum_{n'} \nonumber \\
  && \times \int_0^\infty \frac{dx_{ph} \, x_{ph}^{3/2}}{e^{t x_{ph}} - 1} \int_{0}^\infty \frac{dx}{\sqrt{x}} \left\{ \frac{1}{e^{x - (y-x^{(a)}_{n'}(n',x_{ph})) - x_{ph}} + 1} \right. \nonumber \\
  && \left. - \frac{1}{e^{x - (y-x^{(a)}_{n'}(n',x_{ph}))} + 1} \right\} . \label{S1_1a_form1}
\end{eqnarray}
The variables are the same as before.

\begin{figure}[t]
  \centering
  \includegraphics[width=8cm,bb=0 0 607 438,keepaspectratio=true]{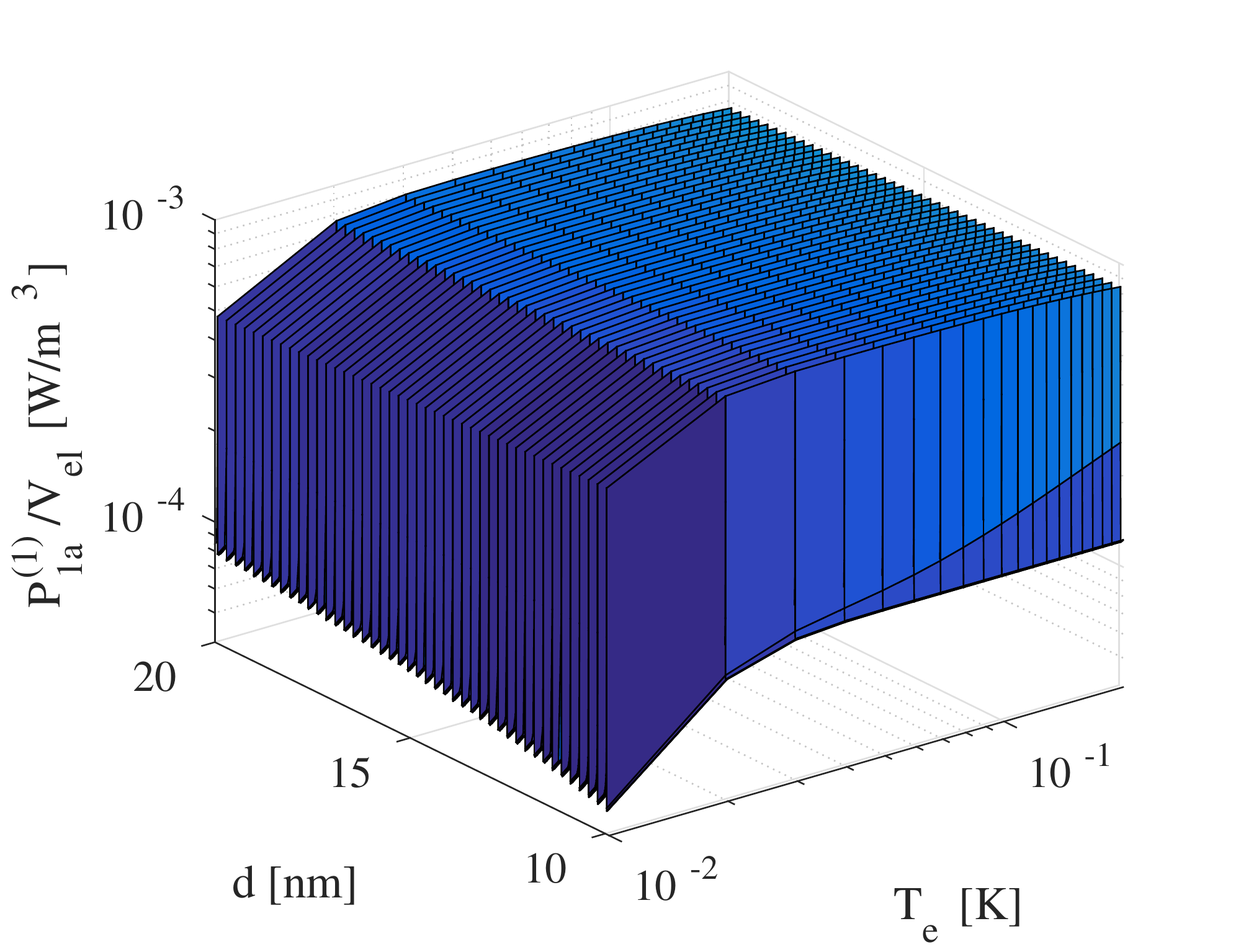} \\
  \includegraphics[width=8cm,bb=0 0 607 438,keepaspectratio=true]{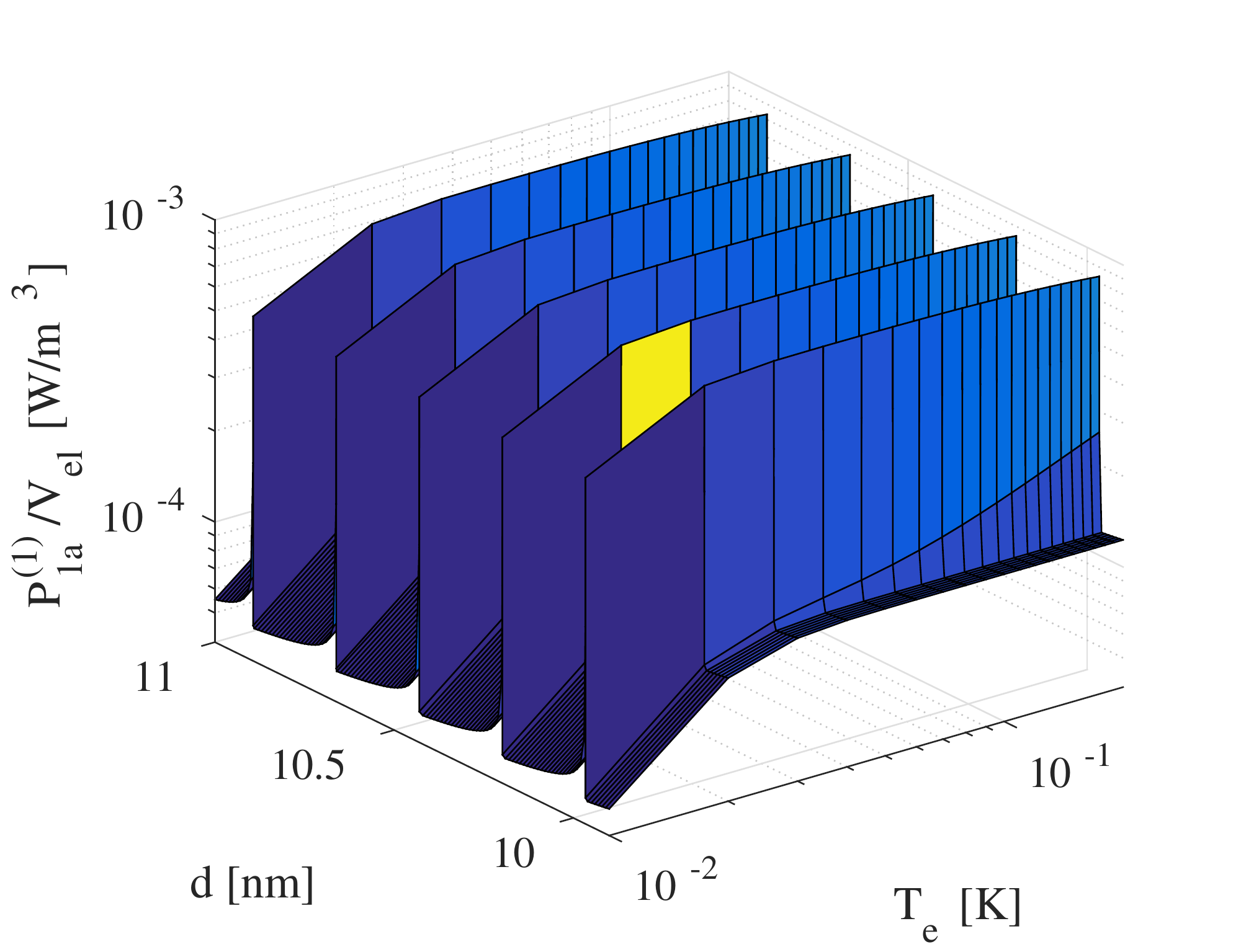}
  \caption{
  $P^{(1)}_{1a}/V_{el}$ vs $T_e$ and $d$. The lower plot is a zoom in the upper figure.}
  \label{P1_1a_0o01_2K_9o9_10o6nm_100nm}
\end{figure}

The function $P_{1a}/V_{el}$ is plotted in Fig. \ref{P1_1a_0o01_2K_9o9_10o6nm_100nm} and we notice similar crests as in Fig. \ref{P1_1s_0o01_2K_9o9_11o1nm}. The nature of these crests are studied in the Appendix \ref{app_S1a1}. In the regions between the crests
\begin{eqnarray}
  \frac{P^{(1)}_{1a}}{A} &\approx& (k_BT_{ph})^{7/2} \frac{3^{1/4} 5}{8 \sqrt{2} \pi^{3/2}} \zeta\left(\frac{7}{2}\right) \frac{m_e^{3/2} E_F^2 (L-d)^2}{\hbar^{9/2} \rho c_l^{5/2} L^{7/2}} \nonumber \\
  && \times \frac{J^{3/4}}{(1-J)^{5/4}} \sum_n^{n\le n_{n_F}} \frac{1}{\sqrt{E_F - \epsilon_n}} , \label{Pa1_pe_em_valley}
\end{eqnarray}
whereas on the crests the heat power may increase by more than one order of magnitude and is not described by a simple power law.

The other summations of Eq. (\ref{P1a_em_sum}) are
\begin{eqnarray}
  S^{(1)}_{2a} &=& \frac{9}{2 \pi^4} \sqrt{\frac{m_e}{2\hbar^7}} \frac{d^4 (L-d)^2}{(L c_l)^{5/2} J(1-J)} \left[ \frac{3}{J (1-J)} \right]^{1/4} \nonumber \\
  && \times \sum_{n',n}^{n'-n = 2k} \bigg\{ \frac{ 1 }{(n'-n)^2} - \frac{ 1 }{(n'+n)^2} \bigg\}^2 (k_BT_e)^5 \nonumber \\
  && \times \int_0^\infty \frac{dx_{ph} \, x_{ph}^{7/2}}{e^{t x_{ph}} -1} \int_{0}^\infty \frac{dx}{\sqrt{x}} \left\{ \frac{1}{e^{x - (y-x_1') - x_{ph}} + 1} \right. \nonumber \\
  && \left. - \frac{1}{e^{x - (y-x_1')} + 1} \right\} \label{S1_2a_form1}
\end{eqnarray}
and
\begin{eqnarray}
  S^{(1)}_{3a} &=&  \frac{4\sqrt{2}}{\pi^4} \sqrt{\frac{m_e}{\hbar^3}} \left[ \frac{3}{J (1-J)} \right]^{1/4} \frac{d^2}{\sqrt{L c_l}} (k_BT_e)^3 \nonumber \\
  && \times \sum_{n',n}^{n'-n = 2k+1}  \bigg\{ \frac{ 1 }{(n'-n)^2} - \frac{ 1 }{(n'+n)^2} \bigg\}^2 \nonumber \\
  && \times \int_0^\infty \frac{dx_{ph} \, x_{ph}^{3/2}}{e^{t x_{ph}} - 1} \int_{0}^\infty \frac{dx}{\sqrt{x}} \left[ \frac{1}{e^{x - (y-x_1') - x_{ph}} + 1} \right. \nonumber \\
  && \left. - \frac{1}{e^{x - (y-x_1')} + 1} \right] , \label{S1_3a_form1}
\end{eqnarray}
but they do not contribute significantly to the total heat power.

\begin{figure}[t]
  \centering
  \includegraphics[width=8cm,bb=0 0 607 438,keepaspectratio=true]{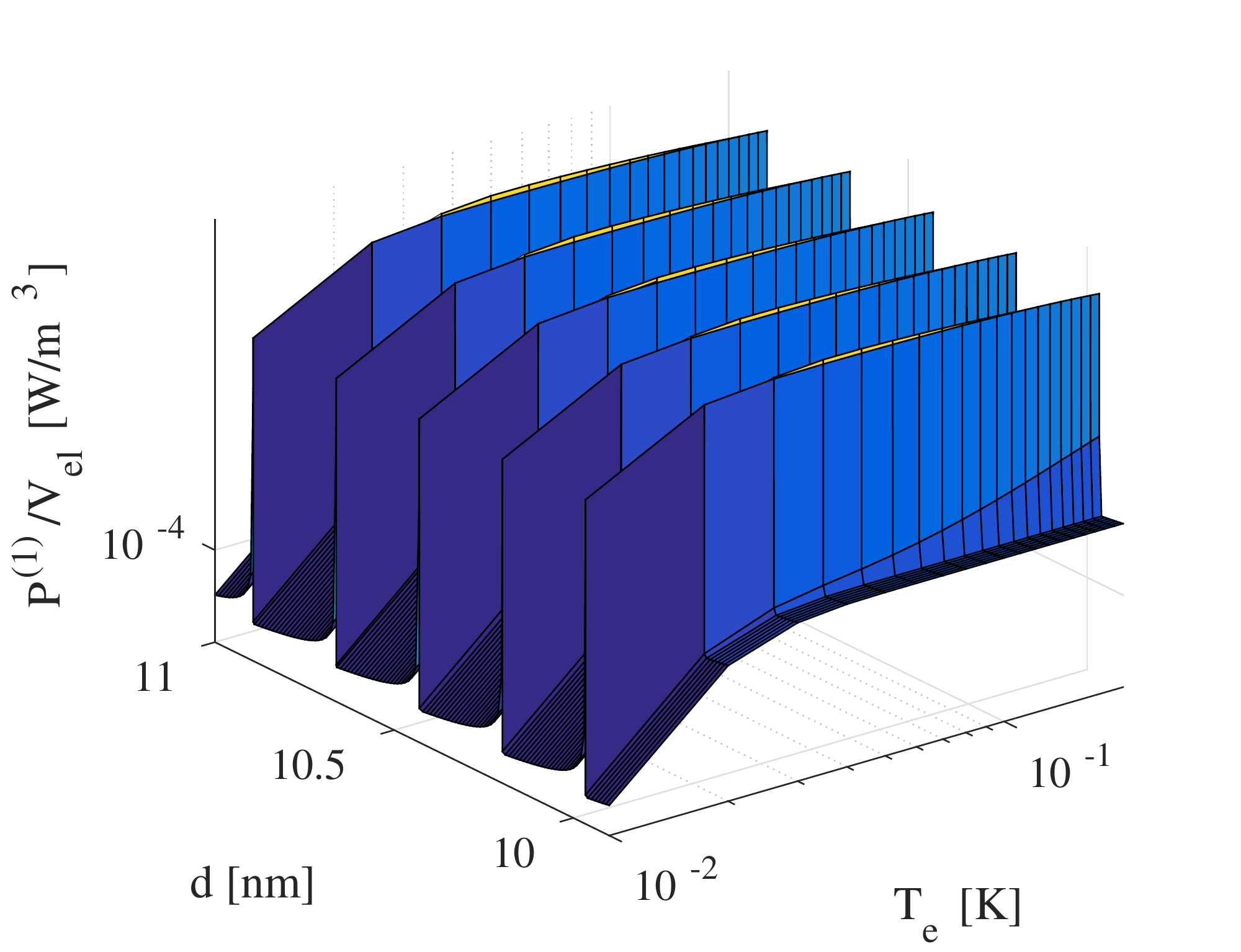}
  \caption{Total power flux per unit volume $P^{(1)}/V_{el}$ from the phonon system into the electron system for $L=100$~nm.}
  \label{log10_P1_tot_Tp0o2K_Te0o01_0o2K_9o9_11nm_L100nm}
\end{figure}

\begin{figure}[t]
  \centering
  \includegraphics[width=8cm,bb=0 0 607 438,keepaspectratio=true]{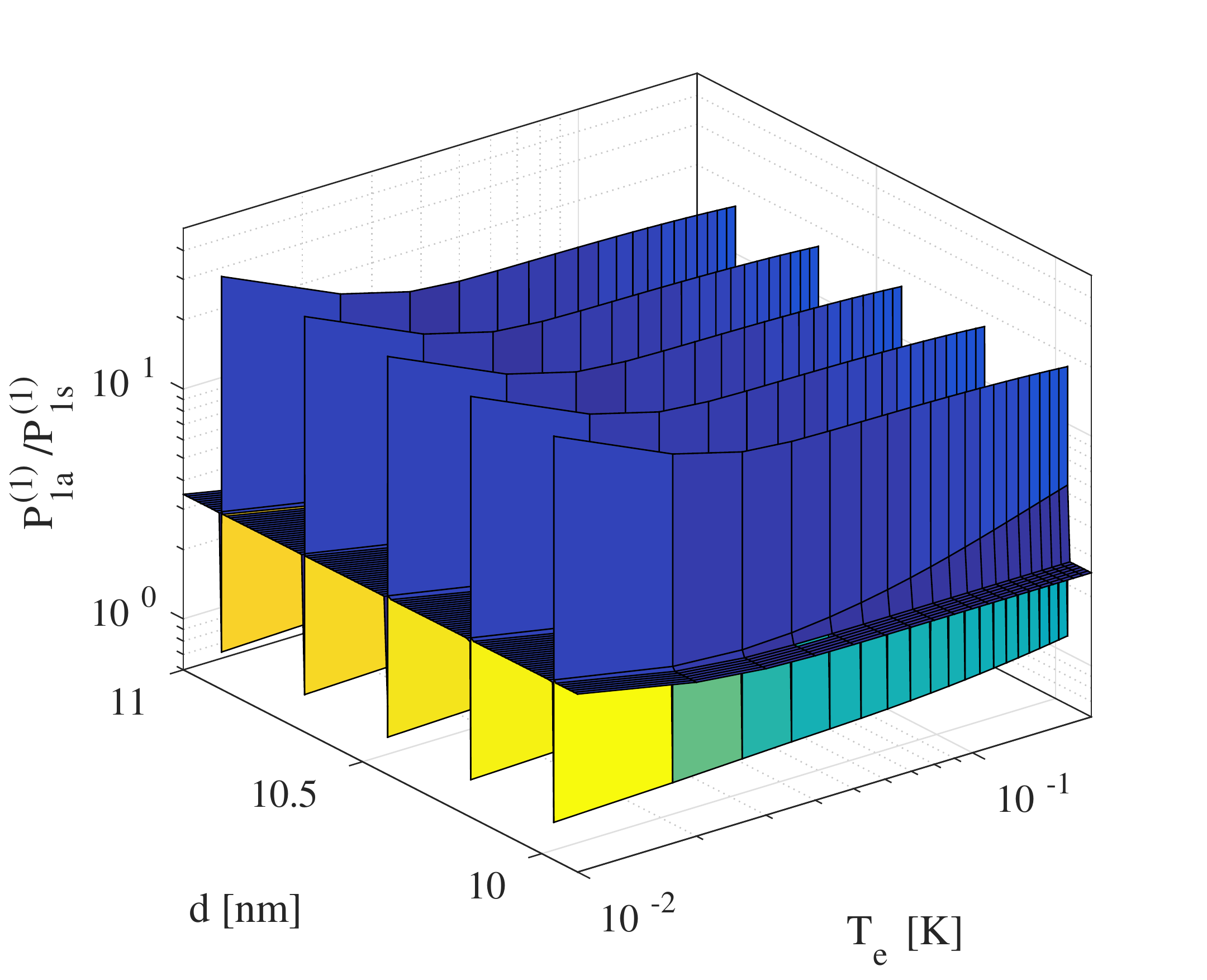}
  \caption{The ratio $P^{(1)}_a/P^{(1)}_s$ for $L=100$~nm. We see that the dominant contribution comes from the heat flux from electrons to the antisymmetric phonons.}
  \label{P1_1a_div_P1_1s}
\end{figure}

Adding all the contributions to $P^{(1)}$ we obtain the total heat power from the phonons to electrons system. $P^{(1)}$ is plotted in Fig. \ref{log10_P1_tot_Tp0o2K_Te0o01_0o2K_9o9_11nm_L100nm} and the ratio $P^{(1)}_a/P^{(1)}_s$ is plotted in Fig. \ref{P1_1a_div_P1_1s}. We observe in Fig. \ref{P1_1a_div_P1_1s} that $P^{(1)}_a$ has the dominant contribution and therefore between the crests $P^{(1)}$ is described by Eq. (\ref{Pa1_pe_em_valley}), with a temperature dependence $P^{(1)} \propto T_{ph}^{7/2}$. On the crests there is a combined contribution of both, $P^{(1)}_s$ and $P^{(1)}_a$.

\section{Conclusions}

We calculated the heat flux $P$ between electrons and phonons in a system which consists of a metallic film of thickness $d$ of the order of 10~nm deposited on an insulating SiN$_{x}$ membrane of thickness $L$ of the order of 100~nm (see Fig. \ref{fig1}). We described the electrons as a gas of free fermions confined in the metallic film.
The principal characteristic of the electron gas is that due to the quantization of the wavevector component perpendicular to the film surfaces the electron gas forms quasi 2D sub-bands, which are specific to a quantum well (QW) formalism.\cite{PhysRevB.51.9930.1995.Bannov,Stroscio_Dutta:book,PhysRevB.65.205315.2002.Glavin,JPhysCondMatt.20.035213.2008.Wu} In an accompanying paper we describe the electrons as a 3D Fermi gas.
We choose a range of parameters $L$, $T_e$ (electrons temperature), and $T_{ph}$ (lattice temperature) such that the phonon gas has a quasi two-dimensional distribution (for $L=100$~nm we took $Te, T_{ph} \le 0.2$~K).

The heat flux assumes the non-symmetric form $P = P^{(0)}(T_e) - P^{(1)}(T_e,T_{ph})$, where $P^{(0)}$ is the heat flux from electrons to phonons and $P^{(1)}$ is the heat flux from phonons to electrons.
Due to the sub-bands formation, $P$ exhibits very sharp crests approximately along the temperature axis in a $|P|$ vs $(d,T_e)$ or $|P|$ vs $(d,T_{ph})$ plot, separated by much more flat valleys.
In the valley regions, in the low temperature limit $P \propto T_e^{3.5} - T_{ph}^{3.5}$, whereas on the crests the heat power increases by more than one order of magnitude and $P$ does not obey a simple power law behavior.
%
%
Such a sharp oscillatory behavior is a characteristic of the QW description of the electronic states in the metallic films\cite{PhysRevB.51.9930.1995.Bannov,Stroscio_Dutta:book,PhysRevB.65.205315.2002.Glavin,JPhysCondMatt.20.035213.2008.Wu} and could be useful for applications, like thickness variation detection or surface contamination.
Moreover, by varying the thickness of the membrane one might vary dramatically the electron-phonon coupling and, through this, the thermalisation or the noise level in the system.

The power-law behavior $P \propto T_e^{3.5} - T_{ph}^{3.5}$ does not agree with the generally expected formula $P \propto T_e^{s+2} - T_{ph}^{s+2}$, where $s$ is the smaller of the dimensions of the electron and phonon gases.\cite{PhysRevB.81.245404.2010.Viljas}
On the other hand, in an experimental setup various factors, such surface contamination, thickness variations, or other imperfections might make it difficult to observe the strong oscillations of $P$ with the thickness of the film and eventually an averaging procedure might be more appropriate.

\begin{acknowledgments}

Discussions and comments from Profs. Yuri Galperin, Tero Heikkil\"a, Ilari Maasilta, and Dr. Thomas K\"uhn are gratefully acknowledged.
This work has been financially supported by CNCSIS-UEFISCDI (project IDEI
114/2011) and ANCS (project PN-09370102). Travel support from Romania-JINR Collaboration grants is gratefully acknowledged.

\end{acknowledgments}

\appendix

\section{The contribution $P^{(0)}_{1s}$} \label{app_S1s}

We analyze in more detail the term $P^{(0)}_{1s}$ (Eq. \ref{Ps_em_S1S2S3}). From Eq. (\ref{Ps_em_S1S2S3}) we have
\begin{eqnarray}
  P^{(0)}_{1s} &=& A \frac{m_e E_{F}^{2}}{36\pi^2 L \rho c_l^4 \hbar^5} \frac{1}{(1-J)^2} S^{(0)}_{1s} , \label{Ps_em_S1s}
\end{eqnarray}
where
\begin{eqnarray}
  S^{(0)}_{1s}(y, T_e) &=& 2 c_l \sqrt{2 m_e J (1-J)} (k_BT_e)^{7/2} \nonumber \\
  && \times \sum_{n} I_{S^{(0)}_{1s}}(y - x_n, T_e) , \label{app_S1s_def} \\
  I_{S^{(0)}_{1s}}(y,T_e) &=& \int_0^\infty \frac{dx_{ph} \, x_{ph}^2}{e^{x_{ph}} - 1} \int_0^\infty \frac{dx}{\sqrt{x}} \left\{ \frac{1}{e^{x - (y-\delta x^{(s)}_{n}) - x_{ph}} + 1} \right. \nonumber \\
  && \left. - \frac{1}{e^{x - (y-\delta x^{(s)}_{n})} + 1} \right\} , \label{def_I_1s_a}
\end{eqnarray}
and $\delta x^{(s)}_n \equiv \delta x^{(s)}_n(n,x_{ph},T_e)$.
We take separately the integral over $x$ and generalize it to
\begin{eqnarray}
  I_1(r,y) &=& \int_0^\infty dx\, x^{r-1} \left( \frac{1}{e^{x - (y-\delta x^{(s)}_{n}) - x_{ph}} + 1} \right. \nonumber \\
  && \left. - \frac{1}{e^{x - (y-\delta x^{(s)}_{n})} + 1} \right) \label{app_I1}
\end{eqnarray}
If $y-\delta x_{n'} \gg 1$ in Eq. (\ref{app_I1}) we can use the approximation
\begin{eqnarray}
  I_1 &\approx& y^{r-1} x_{ph} + \frac{r-1}{2} y^{r-2} x_{ph}^2 . \label{dif_Fdistr_approx}
\end{eqnarray}
Taking $r=1/2$ and plugging the highest order approximation back into (\ref{def_I_1s_a}), we get
\begin{equation}
  I_{S^{(0)}_{1s}} (y,T_e) \approx \frac{1}{\sqrt{y}} \int_0^\infty dx_{ph} \frac{x_{ph}^3}{(e^{x_{ph}} - 1)}
  = \frac{\pi^4}{15} \frac{1}{\sqrt{y}} . \label{approx_I1}
\end{equation}

\begin{figure}[t]
  \centering
  \includegraphics[width=54mm,bb=0 0 650 440]{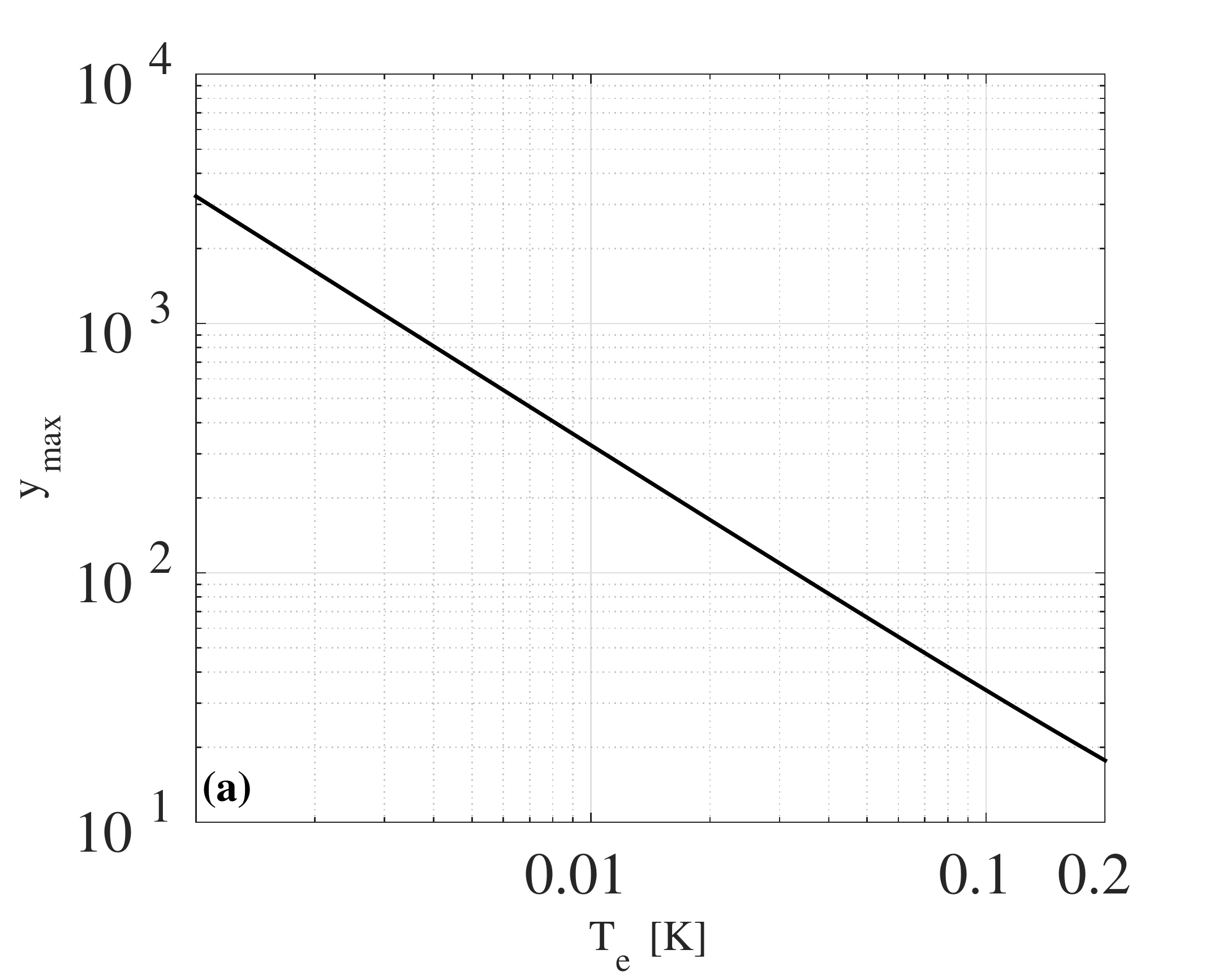} \\ \vspace{5mm}
  \includegraphics[width=54mm,bb=0 0 650 440]{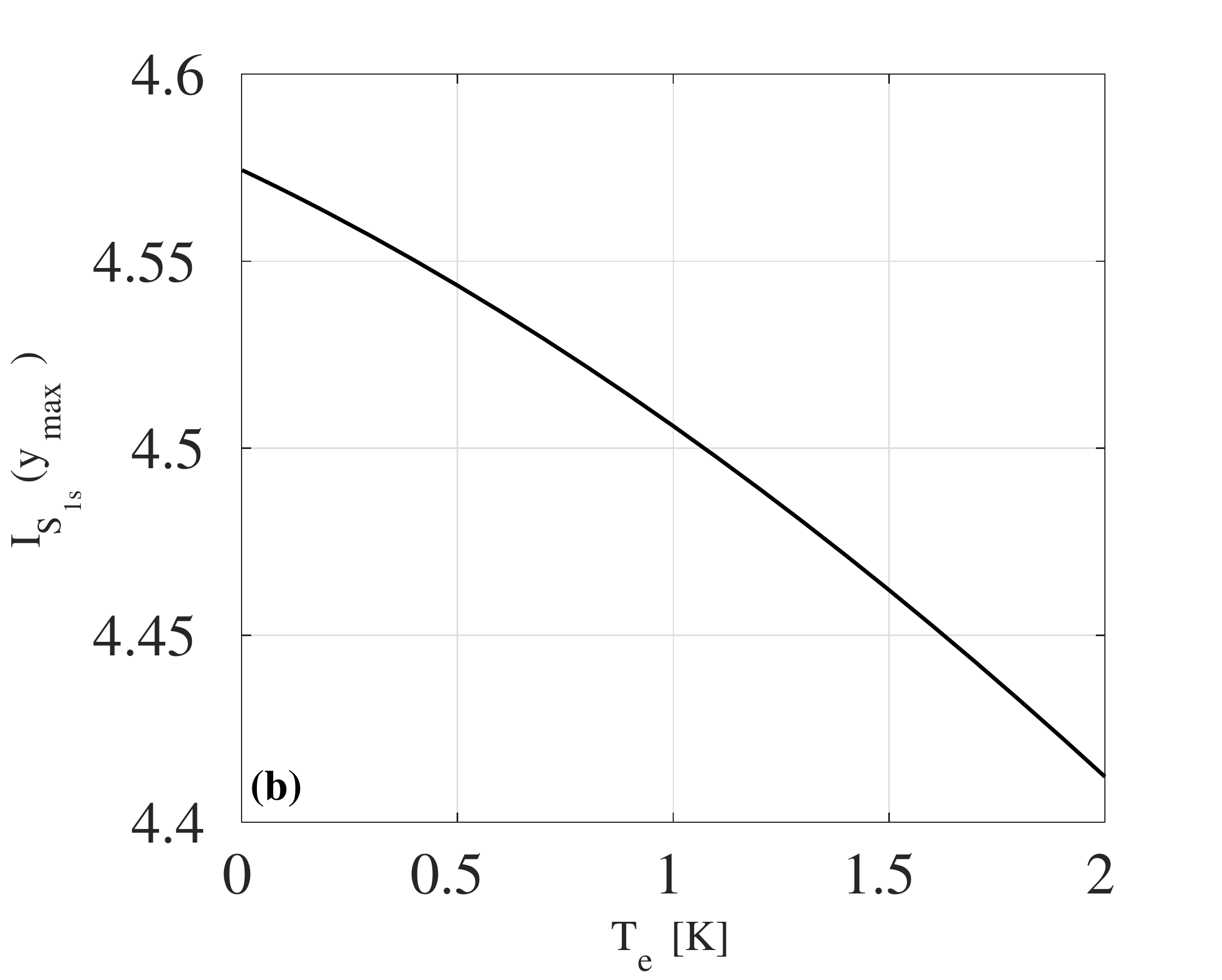} \\ \vspace{5mm}
  \includegraphics[width=54mm,bb=0 0 650 440]{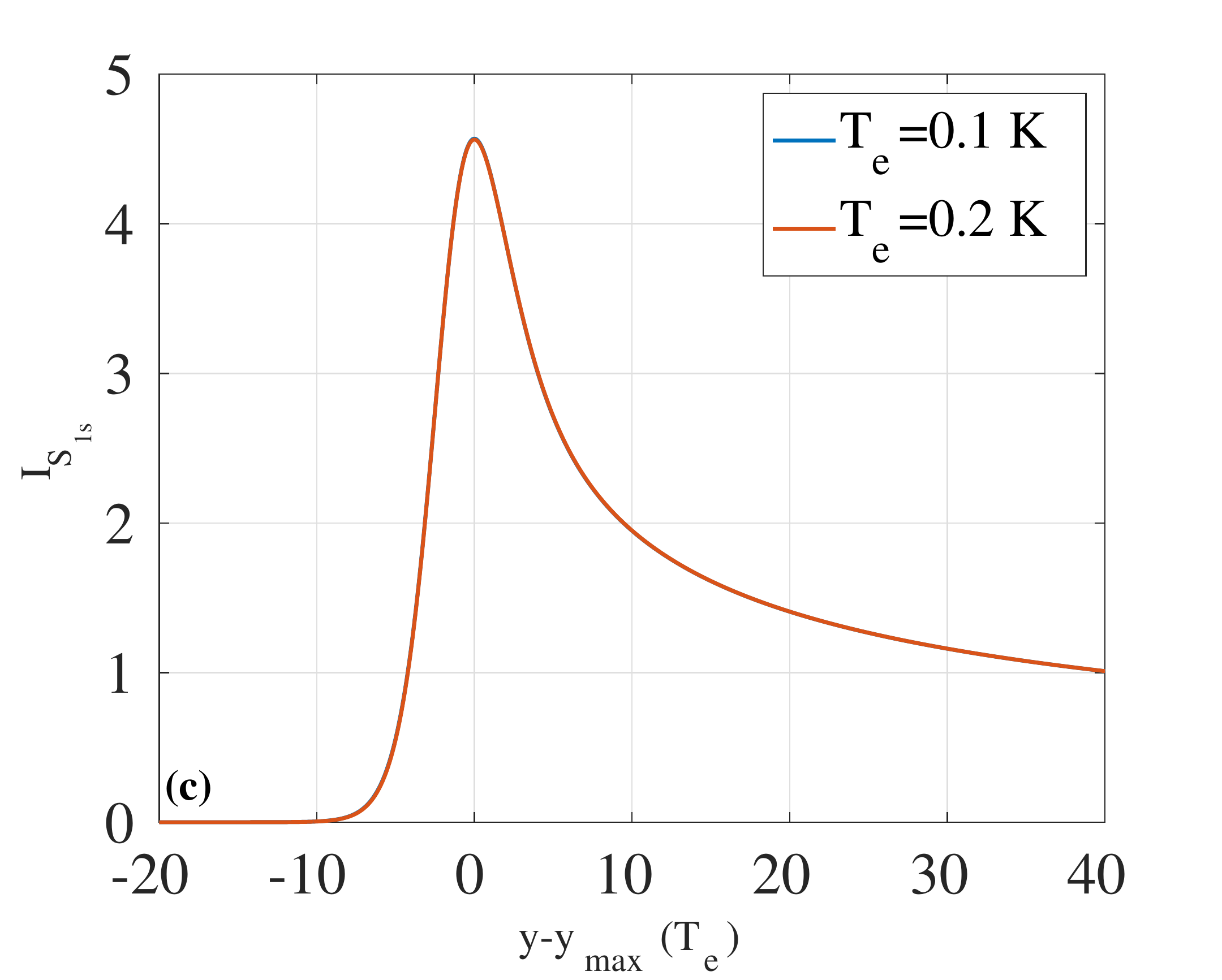}
  \caption{(Color online) (a) $y_{max}$ vs $T_e$, (b) $I_{1s}(y_{\rm max})$ vs. $T_e$ and (c) $I_{1s}[y-y_{\rm max}(T_e)]$ for $T_e=0.1$~K (blue curve) and $T_e=0.2$~K (red curve); the two curves in (c) are indistinguishable.}
  \label{y_max_vs_Te}
\end{figure}

The approximation (\ref{approx_I1}) cannot hold if $y-\delta x_{n'}$ is of the order 1. In such case we have to evaluate Eq. (\ref{def_I_1s_a}) explicitly and we notice that $I_{S^{(0)}_{1s}}(y,T_e)$ forms a maximum at $y = y_{\rm max}(T_e)$ (see Fig. \ref{y_max_vs_Te}). The shape of $I_{S^{(0)}_{1s}}$ as a function of $y-y_{\rm max}$ varies little with $T_e$ (Fig. \ref{y_max_vs_Te}~c) and the maximum value $I_{S^{(0)}_{1s}}(y_{\rm max})\approx 4.5$ slightly decreases with $T_e$ (Fig. \ref{y_max_vs_Te}~b).

In the limit $y \ll -1$, $I_{S^{(0)}_{1s}} (y,T_e)$ decays exponentially with $y$ and therefore we neglect all terms with $n>n_F+1$ from the summation (\ref{app_S1s_def}). The summation extends up to $n_F$ or $n_F+1$ and we may have three situations:
\begin{subequations} \label{S1s_approx_e123}
\begin{eqnarray}
  S^{(0)}_{1s} &\approx& 2 c_l \sqrt{2 m_e J (1-J)} \left\{ \frac{\pi^4}{15} (k_BT_e)^{4} \sum_{n = 1}^{n_F - 1} \frac{1}{\sqrt{E_F - \epsilon_n}} \right. \nonumber \\
  && \left. +  (k_BT_e)^{7/2} I_{S^{(0)}_{1s}} ( y-x_{n_F}) \right\} , \label{S1s_approx_e1}
\end{eqnarray}
when $y-x_{n_F} - \delta x^{(s)}_{n_F} \gtrsim 0$ and the approximation (\ref{approx_I1}) does not hold for $I_{S^{(0)}_{1s}} (y-x_{n_F},T_e)$,
\begin{equation}
  S^{(0)}_{1s} \approx \frac{2 \pi^4}{15} c_l \sqrt{2 m_e J (1-J)} (k_BT_e)^{4} \sum_{n = 1}^{n_F} \frac{1}{\sqrt{E_F - \epsilon_n}} , \label{S1s_approx_e2}
\end{equation}
when $y-x_{n_F} - \delta x^{(s)}_{n_F} \gg 1$ and $y-x_{n_F+1} \ll -1$, and finally,
\begin{eqnarray}
  S^{(0)}_{1s} &\approx& 2 c_l \sqrt{2 m_e J (1-J)} \left\{ \frac{\pi^4}{15} (k_BT_e)^{4} \sum_{n = 1}^{n_F} \frac{1}{\sqrt{E_F - \epsilon_n}} \right. \nonumber \\
  && \left. +  (k_BT_e)^{7/2} I_{S^{(0)}_{1s}} ( y-x_{n_F+1}) \right\} , \label{S1s_approx_e3}
\end{eqnarray}
\end{subequations}
when $y-x_{n_F+1} \lesssim 0$ and $I_{S^{(0)}_{1s}} (y-x_{n_F+1},T_e)$ is not negligible as compared to the whole summation in (\ref{app_S1s_def}).

We say that in cases (\ref{S1s_approx_e1}) and (\ref{S1s_approx_e3}) we are in the crests regions of $P_{1s}$, whereas in case (\ref{S1s_approx_e2}) $P_{1s} \propto T_e^{4}$ and the heat power may be more than one order of magnitude smaller than in cases (\ref{S1s_approx_e1}) and (\ref{S1s_approx_e3}).

\section{The contribution $S^{(0)}_{1a}$} \label{app_S1a}

From Eqs. (\ref{Pa_em_sum}) and (\ref{S1a_form1}) we write
\begin{eqnarray}
  P_{1a}^{(0)} &=& \frac{2 A}{3\pi^2 L^3} \frac{m_e E_F^2}{\rho c_l^2 \hbar^3} \frac{J}{(1-J)} S^{(0)}_{1a} , \label{P0_1a_em_sum} \\
  S^{(0)}_{1a} &=& \frac{1}{2} \sqrt{\frac{m_e}{2\hbar^3}} \left[\frac{3}{ J (1-J)}\right]^{1/4} \frac{(L-d)^2}{\sqrt{c_l L}} (k_BT_e)^{3} \nonumber \\
  && \times \sum_{n} I_{S^{(0)}_{1a}}(y - x^{(a)}_n)  . \label{S1a_form2}
\end{eqnarray}
where
\begin{eqnarray}
  I_{S^{(0)}_{1a}}(y) &=& \int_0^\infty \frac{dx_{ph} \, x_{ph}^{3/2}}{e^{x_{ph}} - 1} \int_{0}^\infty \frac{dx}{\sqrt{x}} \left\{ \frac{1}{e^{x - (y-\delta x^{(a)}_n) - x_{ph}} + 1} \right. \nonumber \\
  && \left. - \frac{1}{e^{x - (y-\delta x^{(a)}_n)} + 1} \right\} \label{def_I_1a}
\end{eqnarray}
and $x^{(a)}_n \equiv x^{(a)}_n(n,x_{ph})$ (Eq. \ref{epsp_np_a}).

\begin{figure}[t]
  \centering
  \includegraphics[width=10cm]{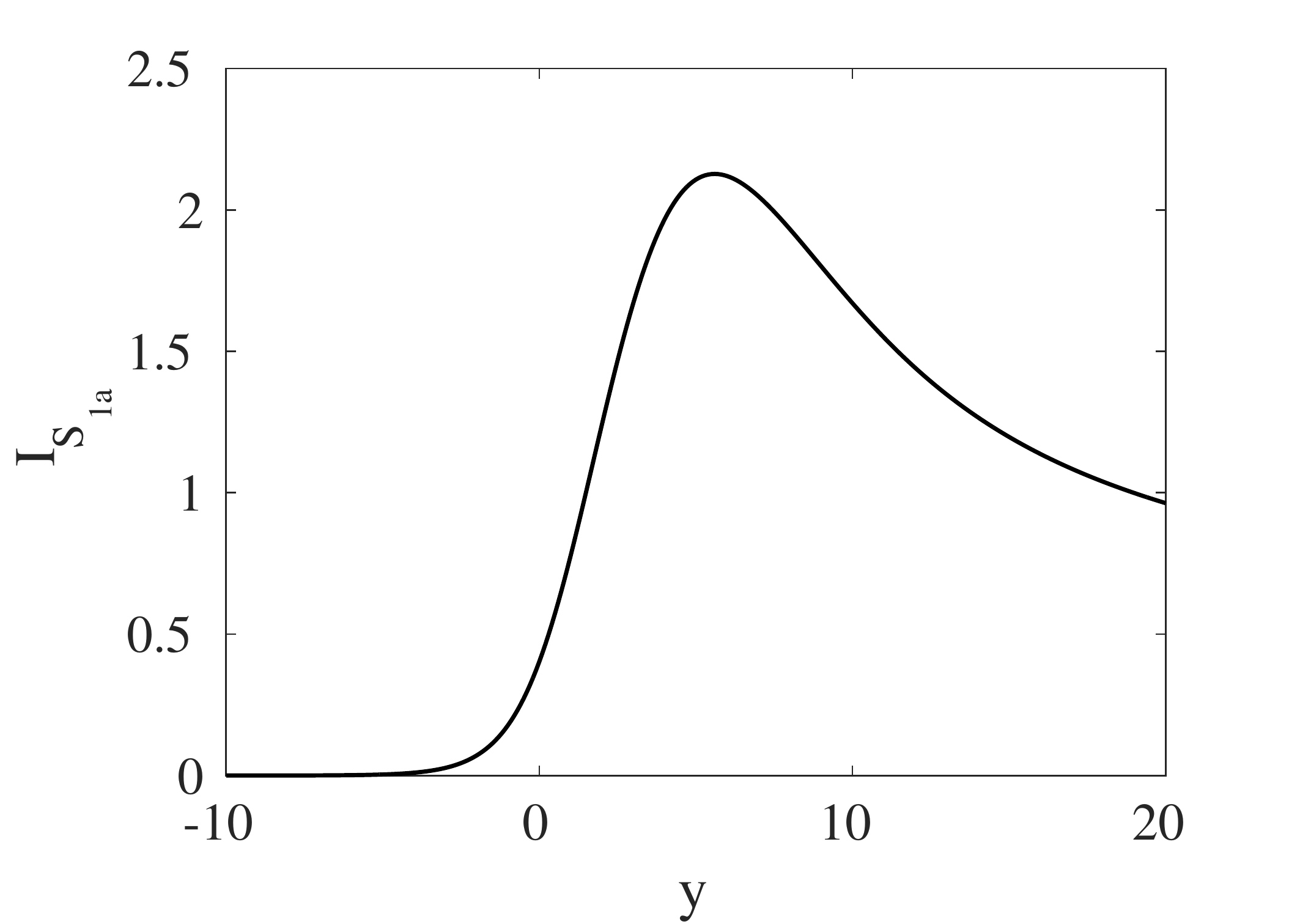}
  \caption{$I_{S^{(0)}_{1a}}$ vs $y$ has a maximum at $y \approx 5.6$. $I_{S^{(0)}_{1a}} (y')$ decays exponentially for $y \ll -1$ and is proportional to $T_e^{7/2}$ for .}
  \label{I_S_1a_vs_y}
\end{figure}

We observe that the integral over $x$ is similar to Eq. \ref{app_I1}. In this case $\delta x^{(a)}_n$ depends on the variables $n$ and $x_{ph}$, but not explicitly on $T_e$, like $\delta x^{(s)}_n(n,x_{ph},T_e)$. Therefore $I_{S^{(0)}_{1a}}(y)$ has a shape which is independent of $T_e$ (see Fig. \ref{I_S_1a_vs_y}) with a maximum at $y_{max} \approx 5.6$.

If $y \gg y_{\rm max}$ we can use the approximation (\ref{dif_Fdistr_approx}) to write
\begin{equation}
  I_{S^{(0)}_{1a}}(y) \approx \frac{1}{\sqrt{y - x_1'}} \int_0^\infty \frac{dx_{ph} \, x_{ph}^{5/2}}{e^{x_{ph}} - 1} = \frac{15 \sqrt{\pi}}{8} \zeta\left(\frac{7}{2}\right) , \label{def_I_1a_appr}
\end{equation}
where $\zeta(3.5) \approx 1.13$ is the Riemann's zeta function.

If $y \ll -1$, $I_{S^{(0)}_{1a}}(y)$ decays exponentially with $y$ and therefore the terms $I_{S^{(0)}_{1a}}(y-x^{(a)}_n)$ with $n>n_F+1$ will be omitted from the summation (\ref{def_I_1a}). The summation extends up to $n_F$ or $n_F+1$ and we have again three cases, like in Eqs. (\ref{S1s_approx_e123}):
\begin{subequations} \label{S1a_approx_e123}
\begin{eqnarray}
  S^{(0)}_{1a} &\approx& \frac{1}{2} \sqrt{\frac{m_e}{2\hbar^3}} \left[\frac{3}{ J (1-J)}\right]^{1/4} \frac{(L-d)^2}{\sqrt{c_l L}} \nonumber \\
  && \times \Bigg\{ (k_BT_e)^3 I_{S^{(0)}_{1a}}(y - x_{n_F}) \label{S1a_approx_e1} \\
  && + (k_BT_e)^{7/2} \frac{15\sqrt{\pi}}{8} \zeta\left( \frac{7}{2} \right) \sum_n^{n<n_{n_F}} \frac{1}{\sqrt{E_F - \epsilon_n}} \Bigg\} \nonumber
\end{eqnarray}
when $y-x_{n_F} - \delta x^{(a)}_{n_F} \gtrsim 0$ and the approximation (\ref{def_I_1a_appr}) does not hold for $I_{S^{(0)}_{1s}} (y-x_{n_F},T_e)$,
\begin{eqnarray}
  S^{(0)}_{1a} &\approx& \frac{15\sqrt{\pi}}{16} \zeta\left( \frac{7}{2} \right) \sqrt{\frac{m_e}{2\hbar^3}} \left[\frac{3}{ J (1-J)}\right]^{1/4} \frac{(L-d)^2}{\sqrt{c_l L}} \nonumber \\
  && \times (k_BT_e)^{7/2} \sum_n^{n\le n_{n_F}} \frac{1}{\sqrt{E_F - \epsilon_n}} \label{S1a_approx_e2}
\end{eqnarray}
when $y-x_{n_F} - \delta x^{(a)}_{n_F} \gg 1$ and $y-x_{n_F+1} \ll -1$, and finally,
\begin{eqnarray}
  S^{(0)}_{1a} &\approx& \frac{1}{2} \sqrt{\frac{m_e}{2\hbar^3}} \left[\frac{3}{ J (1-J)}\right]^{1/4} \frac{(L-d)^2}{\sqrt{c_l L}} \nonumber \\
  && \times \Bigg\{ (k_BT_e)^3 I_{S^{(0)}_{1a}}(y - x_{n_F}) \label{S1a_approx_e3} \\
  && + (k_BT_e)^{7/2} \frac{15\sqrt{\pi}}{8} \zeta\left( \frac{7}{2} \right) \sum_n^{n \le n_F} \frac{1}{\sqrt{E_F - \epsilon_n}} \Bigg\} , \nonumber
\end{eqnarray}
\end{subequations}
when $y-x_{n_F+1} \lesssim 0$ and $I_{S^{(0)}_{1s}} (y-x_{n_F+1},T_e)$ is not negligible as compared to the whole summation in (\ref{S1a_form2}).

In the cases (\ref{S1a_approx_e1}) and (\ref{S1a_approx_e3}), $d$ corresponds to crests regions of Fig. \ref{S_1a_0o01_2K_9o9_10o6nm_100nm}, whereas in case (\ref{S1a_approx_e2}), $d$ corresponds to the valley region and the heat power may be more than one order of magnitude smaller than in the crests.

\section{The contribution $S^{(1)}_{1s}$} \label{app_S1s1}

\begin{figure}[t]
  \centering
  \includegraphics[width=10cm,bb=0 0 650 440]{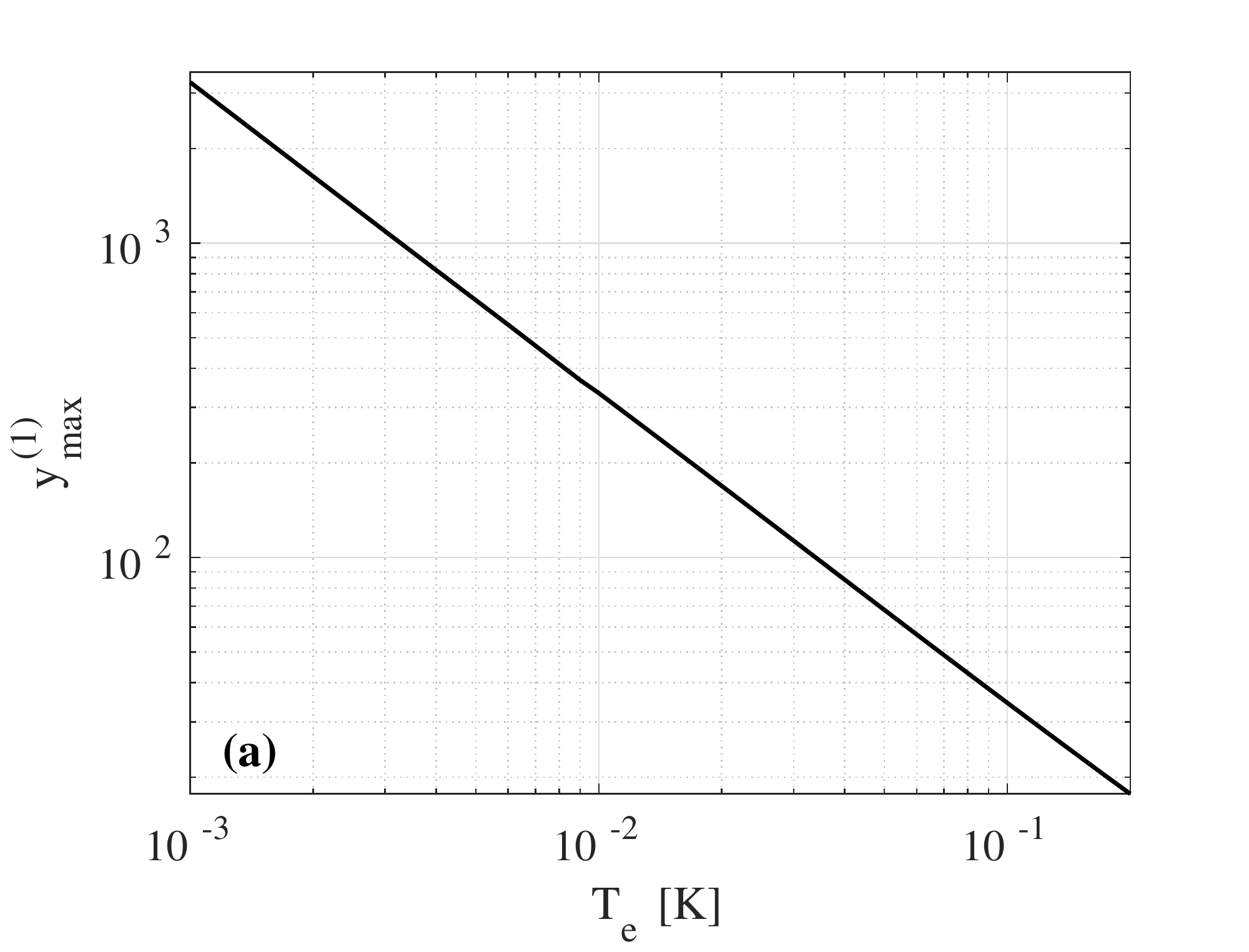} \\ \vspace{5mm}
  \includegraphics[width=10cm,bb=0 0 650 440]{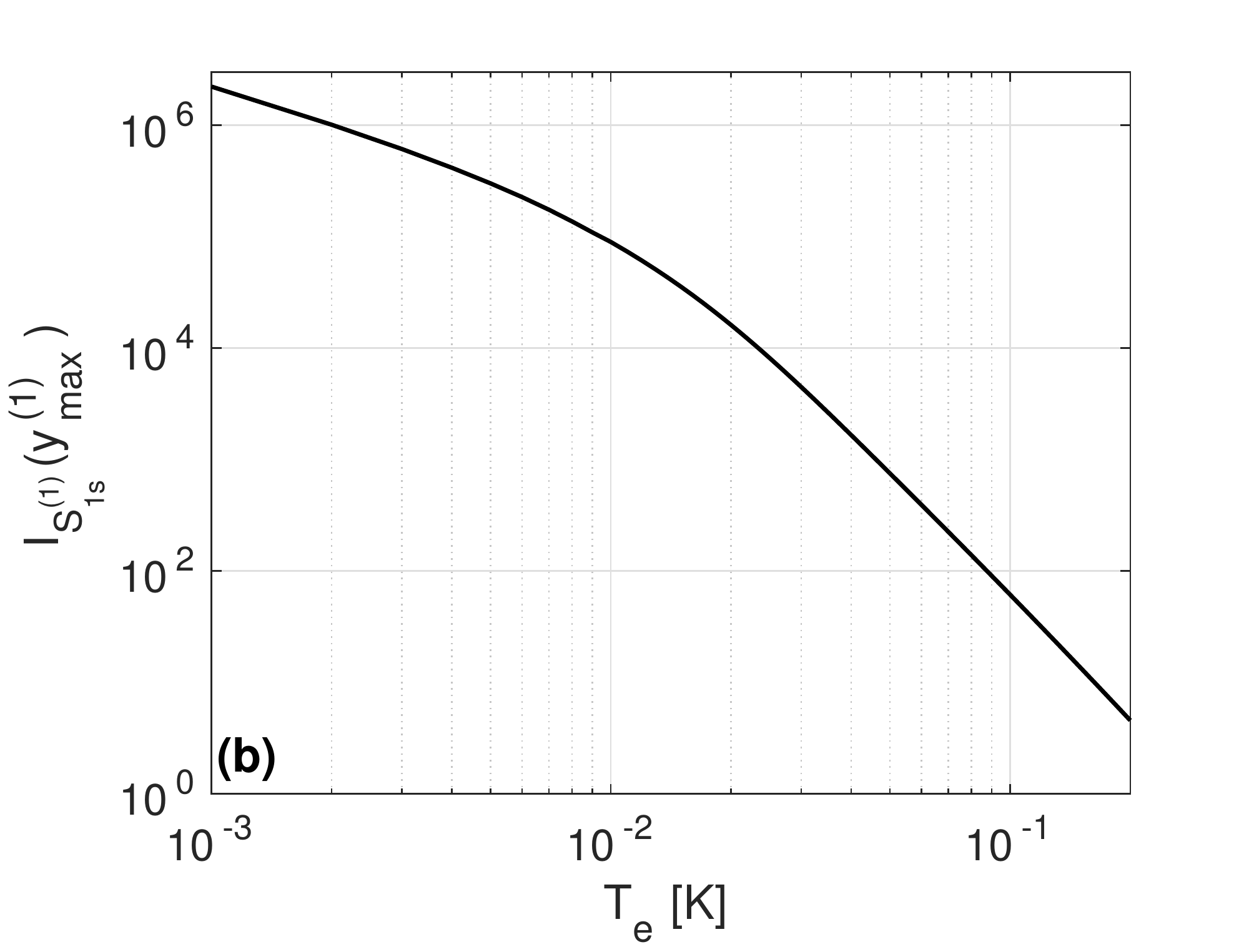}
  \caption{(Color online) (a) $y_{max}$ vs $T_e$ and (b) $I_{S^{(1)}_{1s}}(y_{\rm max})$ vs. $T_e$ for $T_{ph} = 0.2$~K.}
  \label{I1_S1s_y_max_vs_Te}
\end{figure}

From Eqs. (\ref{Ps1_em_S1S2S3}) and (\ref{S1s1_approx}) we have
\begin{eqnarray}
  P^{(1)}_{1s} &=& A \frac{m_e E_{F}^{2}}{36\pi^2 L \rho c_l^4 \hbar^5} \frac{1}{(1-J)^2} S^{(1)}_{1s} , \label{Ps1_pe_em_S1s}
\end{eqnarray}
where
\begin{eqnarray}
  S^{(1)}_{1s}(y, T_e, T_{ph}) &=& 2 c_l \sqrt{2 m_e J (1-J)} (k_BT_e)^{7/2} \nonumber \\
  && \times \sum_{n} I_{S^{(1)}_{1s}}(y - x_n, T_e, T_e/T_{ph}) , \label{app_S1s_pe_def}
\end{eqnarray}
\begin{eqnarray}
  && I_{S^{(1)}_{1s}}(y,T_e, t) = \int_0^\infty \frac{dx_{ph} \, x_{ph}^2}{e^{t x_{ph}} - 1} \int_0^\infty \frac{dx}{\sqrt{x}} \nonumber \\
  && \times \left\{ \frac{1}{e^{x - (y-\delta x^{(s)}_{n}) - x_{ph}} + 1} - \frac{1}{e^{x - (y-\delta x^{(s)}_{n})} + 1} \right\} , \label{def_I_1s1_a}
\end{eqnarray}
with similar notations as in Appendix \ref{app_S1s} and $t \equiv T_e/T_{ph}$.
From Eq. (\ref{def_I_1s1_a}) and using the results of Appendix \ref{app_S1s} we obtain in the limit $y-\delta x^{(s)}_{n} \gg 1$,
\begin{equation}
  I_{S^{(1)}_{1s}} (y,T_e,t) \approx \frac{1}{\sqrt{y}} \int_0^\infty dx_{ph} \frac{x_{ph}^3}{(e^{t x_{ph}} - 1)}
  = \frac{\pi^4}{15} \frac{1}{t^4 \sqrt{y}} . \label{approx_I1_ep}
\end{equation}
If $y$ in Eq. (\ref{def_I_1s1_a}) is of the order 1 (positive or negative), then we evaluate Eq. (\ref{def_I_1s1_a}) explicitly. Taking $T_{ph} = 0.2$, $I_{S^{(1)}_{1s}}(y,T_e,T_e/T_{ph})$ forms a maximum at $y = y^{(1)}_{\rm max}(T_e)$ (see Fig. \ref{I1_S1s_y_max_vs_Te}).


In the limit $y \ll -1$, $I_{S^{(1)}_{1s}} (y,T_e,T_e/T_{ph})$ decays exponentially with $y$ and therefore we neglect all terms with $n>n_F+1$ from the summation (\ref{app_S1s_def}).

The summation extends up to $n_F$ or $n_F+1$ and we may have three situations:
\begin{subequations} \label{Spe1s_approx_e123}
\begin{eqnarray}
  S^{(1)}_{1s} &\approx& 2 c_l \sqrt{2 m_e J (1-J)} \left\{ \frac{\pi^4}{15} (k_BT_{ph})^{4} \sum_{n = 1}^{n_F - 1} \frac{1}{\sqrt{E_F - \epsilon_n}} \right. \nonumber \\
  && \left. +  (k_BT_e)^{7/2} I_{S^{(0)}_{1s}} ( y-x_{n_F}) \right\} , \label{Spe1s_approx_e1}
\end{eqnarray}
when $y-x_{n_F} - \delta x^{(s)}_{n_F} \gtrsim 0$ and the approximation (\ref{approx_I1}) does not hold for $I_{S^{(1)}_{1s}} (y-x_{n_F},T_e)$,
\begin{equation}
  S^{(1)}_{1s} \approx \frac{2 \pi^4}{15} c_l \sqrt{2 m_e J (1-J)} (k_BT_{ph})^{4} \sum_{n = 1}^{n_F} \frac{1}{\sqrt{E_F - \epsilon_n}} , \label{Spe1s_approx_e2}
\end{equation}
when $y-x_{n_F} - \delta x^{(s)}_{n_F} \gg 1$ and $y-x_{n_F+1} \ll -1$, and finally,
\begin{eqnarray}
  S^{(1)}_{1s} &\approx& 2 c_l \sqrt{2 m_e J (1-J)} \left\{ \frac{\pi^4}{15} (k_BT_{ph})^{4} \sum_{n = 1}^{n_F} \frac{1}{\sqrt{E_F - \epsilon_n}} \right. \nonumber \\
  && \left. +  (k_BT_e)^{7/2} I_{S^{(1)}_{1s}} ( y-x_{n_F+1}) \right\} , \label{Spe1s_approx_e3}
\end{eqnarray}
\end{subequations}
when $y-x_{n_F+1} \lesssim 0$ and $I_{S^{(1)}_{1s}} (y-x_{n_F+1},T_e)$ is not negligible as compared to the whole summation in (\ref{app_S1s_def}).

Equations (\ref{Spe1s_approx_e1}) and (\ref{Spe1s_approx_e3}) describe the crests regions of $P^{(1)}_{1s}$, whereas in case (\ref{Spe1s_approx_e2}) $P_{1s} \propto T_{ph}^{4}$ and the heat power may be more than one order of magnitude smaller than in cases (\ref{Spe1s_approx_e1}) and (\ref{Spe1s_approx_e3}).

\section{The contribution $S^{(1)}_{1a}$} \label{app_S1a1}

From Eqs. (\ref{P1a_em_sum}) and (\ref{S1_1a_form1}) we write
\begin{eqnarray}
  P_{1a}^{(1)} &=& \frac{2 A}{3\pi^2 L^3} \frac{m_e E_F^2}{\rho c_l^2 \hbar^3} \frac{J}{(1-J)} S^{(1)}_{1a} , \label{P1a1_em_sum} \\
  S^{(1)}_{1a} &=& \frac{1}{2} \sqrt{\frac{m_e}{2\hbar^3}} \left[\frac{3}{ J (1-J)}\right]^{1/4} \frac{(L-d)^2}{\sqrt{c_l L}} (k_BT_e)^{3} \nonumber \\
  && \times \sum_{n} I_{S^{(1)}_{1a}}(y - x^{(a)}_n, T_e/T_{ph}) , \label{S1_1a_form2}
\end{eqnarray}
where
\begin{eqnarray}
  I_{S^{(1)}_{1a}}(y, t) &=& \int_0^\infty \frac{dx_{ph} \, x_{ph}^{3/2}}{e^{t x_{ph}} - 1} \int_{0}^\infty \frac{dx}{\sqrt{x}} \left\{ \frac{1}{e^{x - (y-\delta x^{(a)}_n) - x_{ph}} + 1} \right. \nonumber \\
  && \left. - \frac{1}{e^{x - (y-\delta x^{(a)}_n)} + 1} \right\} \label{def_I1_1a}
\end{eqnarray}
and keeping the rest of the notations like in Appendices \ref{app_S1a} and \ref{app_S1s1}.

\begin{figure}[t]
  \centering
  \includegraphics[width=10cm]{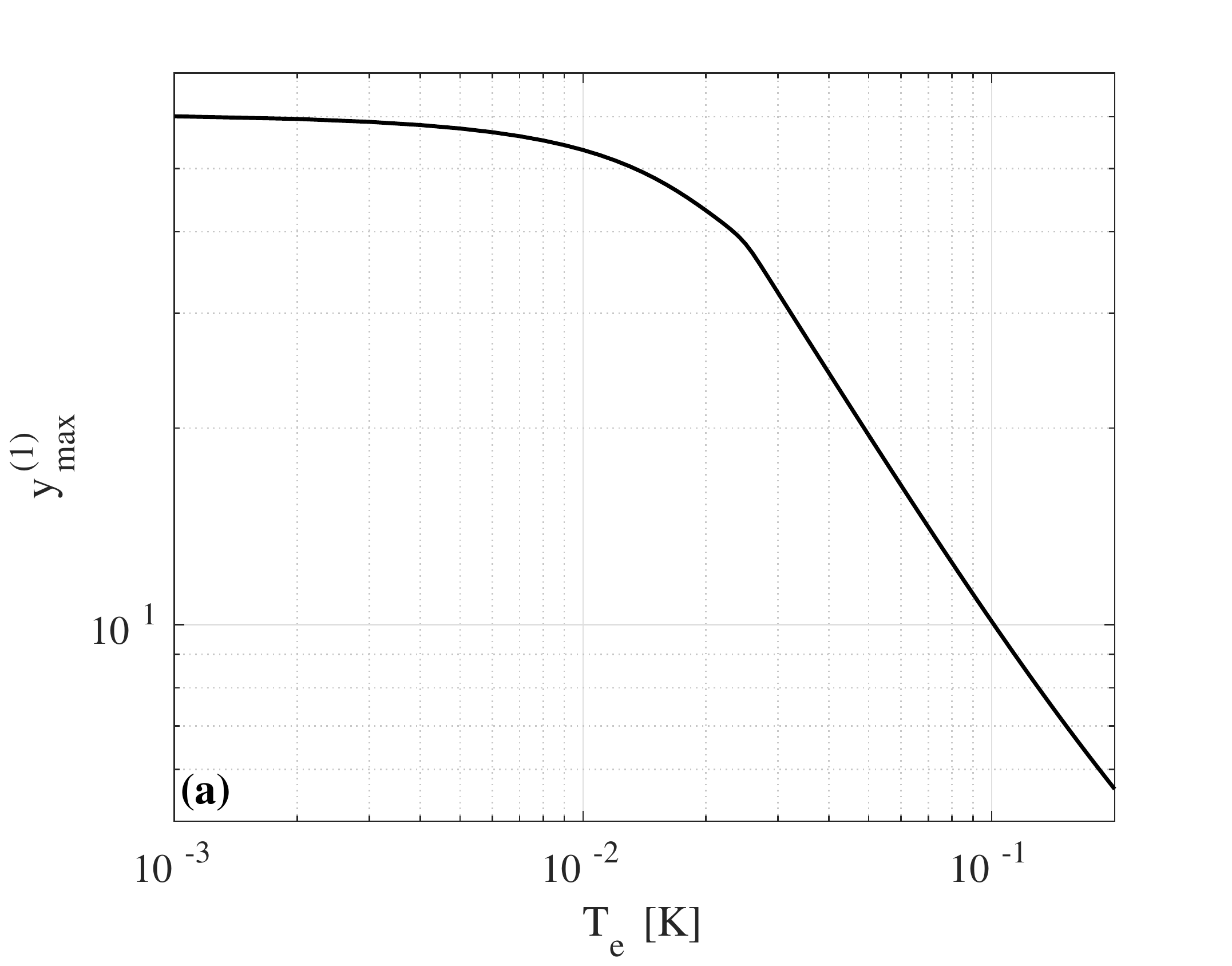} \\
  \includegraphics[width=10cm]{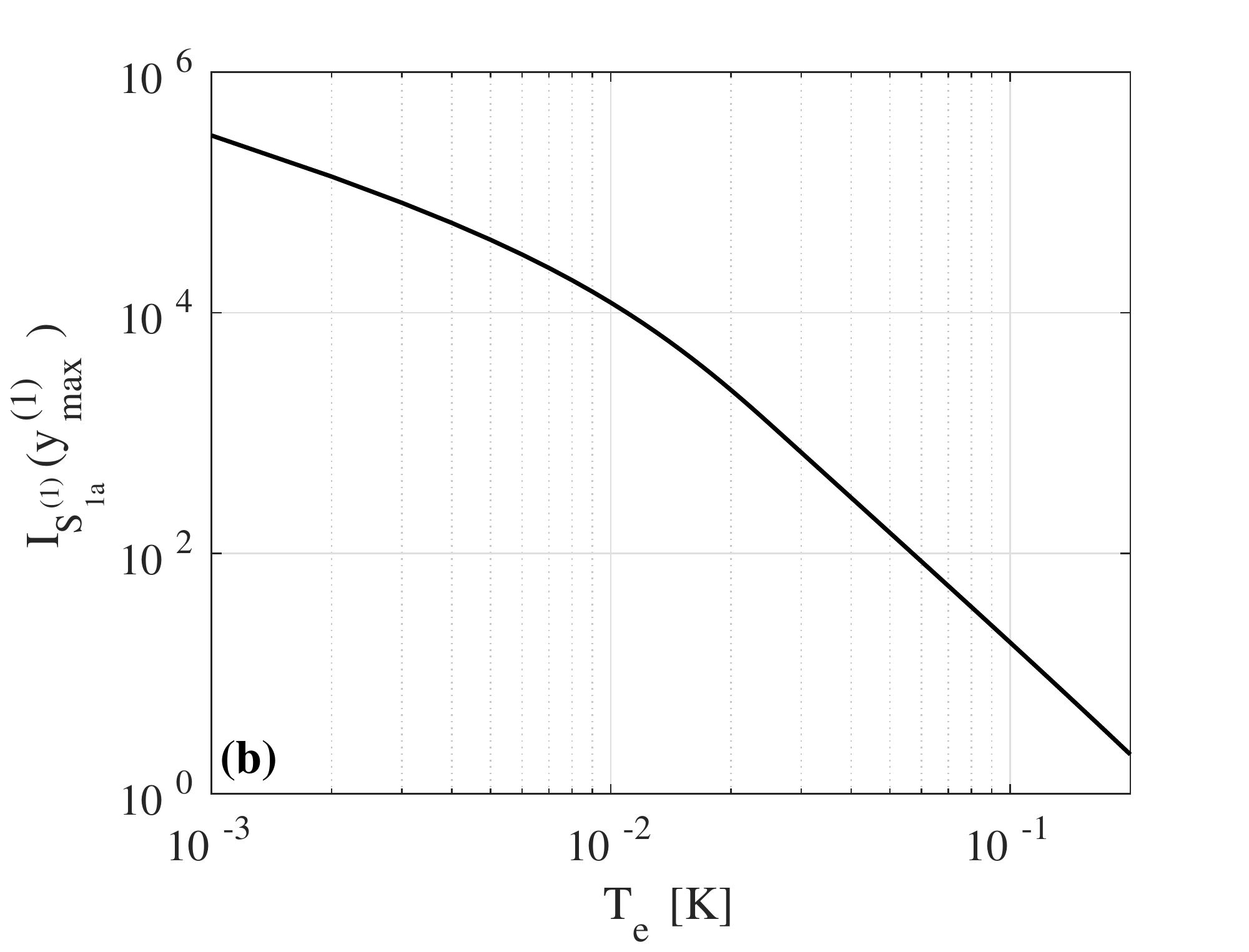}
  \caption{(Color online) (a) $y_{max}$ vs $T_e$ and (b) $I_{S^{(1)}_{1s}}(y_{\rm max})$ vs. $T_e$ for $T_{ph} = 0.2$~K.}
  \label{I_S1_1a_vs_y}
\end{figure}

$I_{S^{(1)}_{1a}}(y, t)$ has a maximum at $y = y^{(1)}_{max}$, which is plotted in Fig. \ref{I_S1_1a_vs_y}~(a) together with $I_{S^{(1)}_{1a}}(y^{(1)}_{max}, t)$ (Fig.~\ref{I_S1_1a_vs_y}~b). 
If $y \gg y_{\rm max}$ we can use the approximation (\ref{dif_Fdistr_approx}) to write
\begin{equation}
  I_{S^{(1)}_{1a}}(y,t) \approx \frac{1}{\sqrt{y - x_1'}} \int_0^\infty \frac{dx_{ph} \, x_{ph}^{5/2}}{e^{t x_{ph}} - 1} = \frac{15 \sqrt{\pi}}{8 t^{7/2}} \zeta\left(\frac{7}{2}\right) . \label{def_I1_1a_appr}
\end{equation}
If $y \ll -1$, $I_{S^{(1)}_{1a}}(y,t)$ decays exponentially and therefore the terms $I_{S^{(1)}_{1a}}(y-x^{(a)}_n)$ with $n>n_F+1$ will be omitted from the summation (\ref{def_I1_1a}).

The summation extends up to $n_F$ or $n_F+1$ and we have again three cases, like in Eqs. (\ref{S1s_approx_e123}):
\begin{subequations} \label{S1_1a_approx_e123}
\begin{eqnarray}
  S^{(1)}_{1a} &\approx& \frac{1}{2} \sqrt{\frac{m_e}{2\hbar^3}} \left[\frac{3}{ J (1-J)}\right]^{1/4} \frac{(L-d)^2}{\sqrt{c_l L}} \nonumber \\
  && \times \Bigg\{ (k_BT_e)^3 I_{S^{(1)}_{1a}}(y - x_{n_F}) \label{S1_1a_approx_e1} \\
  && + (k_BT_{ph})^{7/2} \frac{15\sqrt{\pi}}{8} \zeta\left( \frac{7}{2} \right) \sum_n^{n<n_{n_F}} \frac{1}{\sqrt{E_F - \epsilon_n}} \Bigg\} \nonumber
\end{eqnarray}
when $y-x_{n_F} - \delta x^{(a)}_{n_F} \gtrsim 0$ and the approximation (\ref{def_I1_1a_appr}) does not hold for $I_{S^{(1)}_{1s}} (y-x_{n_F},T_e)$,
\begin{eqnarray}
  S^{(1)}_{1a} &\approx& \frac{15\sqrt{\pi}}{16} \zeta\left( \frac{7}{2} \right) \sqrt{\frac{m_e}{2\hbar^3}} \left[\frac{3}{ J (1-J)}\right]^{1/4} \frac{(L-d)^2}{\sqrt{c_l L}} \nonumber \\
  && \times (k_BT_{ph})^{7/2} \sum_n^{n\le n_{n_F}} \frac{1}{\sqrt{E_F - \epsilon_n}} \label{S1_1a_approx_e2}
\end{eqnarray}
when $y-x_{n_F} - \delta x^{(a)}_{n_F} \gg 1$ and $y-x_{n_F+1} \ll -1$, and finally,
\begin{eqnarray}
  S^{(1)}_{1a} &\approx& \frac{1}{2} \sqrt{\frac{m_e}{2\hbar^3}} \left[\frac{3}{ J (1-J)}\right]^{1/4} \frac{(L-d)^2}{\sqrt{c_l L}} \nonumber \\
  && \times \Bigg\{ (k_BT_e)^3 I_{S^{(1)}_{1a}}(y - x_{n_F}) \label{S1_1a_approx_e3} \\
  && + (k_BT_{ph})^{7/2} \frac{15\sqrt{\pi}}{8} \zeta\left( \frac{7}{2} \right) \sum_n^{n \le n_F} \frac{1}{\sqrt{E_F - \epsilon_n}} \Bigg\} , \nonumber
\end{eqnarray}
\end{subequations}
when $y-x_{n_F+1} \lesssim 0$ and $I_{S^{(0)}_{1s}} (y-x_{n_F+1},T_e)$ is not negligible as compared to the whole summation in (\ref{S1a_form2}).

In the cases (\ref{S1_1a_approx_e1}) and (\ref{S1_1a_approx_e3}), $d$ corresponds to crests regions of Fig. \ref{S_1a_0o01_2K_9o9_10o6nm_100nm}, whereas in case (\ref{S1_1a_approx_e2}), $d$ corresponds to the valley region and the heat power may be more than one order of magnitude smaller than in the crests.


\begin{thebibliography}{29}
\expandafter\ifx\csname natexlab\endcsname\relax\def\natexlab#1{#1}\fi
\expandafter\ifx\csname bibnamefont\endcsname\relax
  \def\bibnamefont#1{#1}\fi
\expandafter\ifx\csname bibfnamefont\endcsname\relax
  \def\bibfnamefont#1{#1}\fi
\expandafter\ifx\csname citenamefont\endcsname\relax
  \def\citenamefont#1{#1}\fi
\expandafter\ifx\csname url\endcsname\relax
  \def\url#1{\texttt{#1}}\fi
\expandafter\ifx\csname urlprefix\endcsname\relax\def\urlprefix{URL }\fi
\providecommand{\bibinfo}[2]{#2}
\providecommand{\eprint}[2][]{\url{#2}}

\bibitem[{\citenamefont{Nguyen et~al.}(2014)\citenamefont{Nguyen, Meschke,
  Courtois, and Pekola}}]{PhysRevApplied.2.054001.2014.Nguyen}
\bibinfo{author}{\bibfnamefont{H.~Q.} \bibnamefont{Nguyen}},
  \bibinfo{author}{\bibfnamefont{M.}~\bibnamefont{Meschke}},
  \bibinfo{author}{\bibfnamefont{H.}~\bibnamefont{Courtois}}, \bibnamefont{and}
  \bibinfo{author}{\bibfnamefont{J.~P.} \bibnamefont{Pekola}},
  \bibinfo{journal}{Phys. Rev. Appl.} \textbf{\bibinfo{volume}{2}},
  \bibinfo{pages}{054001} (\bibinfo{year}{2014}).

\bibitem[{\citenamefont{Nahum et~al.}(1994)\citenamefont{Nahum, Eiles, and
  Martinis}}]{ApplPhysLett.65.3123.Nahum}
\bibinfo{author}{\bibfnamefont{M.}~\bibnamefont{Nahum}},
  \bibinfo{author}{\bibfnamefont{T.~M.} \bibnamefont{Eiles}}, \bibnamefont{and}
  \bibinfo{author}{\bibfnamefont{J.~M.} \bibnamefont{Martinis}},
  \bibinfo{journal}{Appl. Phys. Lett.} \textbf{\bibinfo{volume}{65}},
  \bibinfo{pages}{3123} (\bibinfo{year}{1994}).

\bibitem[{\citenamefont{Leivo et~al.}(1996)\citenamefont{Leivo, Pekola, and
  Averin}}]{ApplPhysLett.68.1996.1996.Leivo}
\bibinfo{author}{\bibfnamefont{M.~M.} \bibnamefont{Leivo}},
  \bibinfo{author}{\bibfnamefont{J.~P.} \bibnamefont{Pekola}},
  \bibnamefont{and} \bibinfo{author}{\bibfnamefont{D.~V.}
  \bibnamefont{Averin}}, \bibinfo{journal}{Appl. Phys. Lett.}
  \textbf{\bibinfo{volume}{68}}, \bibinfo{pages}{1996} (\bibinfo{year}{1996}).

\bibitem[{\citenamefont{Pekola et~al.}(2000{\natexlab{a}})\citenamefont{Pekola,
  Anghel, Suppula, Suoknuuti, Manninen, and Manninen}}]{APL76.2782.2000.Pekola}
\bibinfo{author}{\bibfnamefont{J.~P.} \bibnamefont{Pekola}},
  \bibinfo{author}{\bibfnamefont{D.~V.} \bibnamefont{Anghel}},
  \bibinfo{author}{\bibfnamefont{T.~I.} \bibnamefont{Suppula}},
  \bibinfo{author}{\bibfnamefont{J.~K.} \bibnamefont{Suoknuuti}},
  \bibinfo{author}{\bibfnamefont{A.~J.} \bibnamefont{Manninen}},
  \bibnamefont{and} \bibinfo{author}{\bibfnamefont{M.}~\bibnamefont{Manninen}},
  \bibinfo{journal}{Appl. Phys. Lett.} \textbf{\bibinfo{volume}{76}},
  \bibinfo{pages}{2782} (\bibinfo{year}{2000}{\natexlab{a}}).

\bibitem[{\citenamefont{Giazotto et~al.}(2006)\citenamefont{Giazotto,
  Heikkil{\"a}, Luukanen, Savin, and Pekola}}]{RevModPhys.78.217.2006.Giazotto}
\bibinfo{author}{\bibfnamefont{F.}~\bibnamefont{Giazotto}},
  \bibinfo{author}{\bibfnamefont{T.~T.} \bibnamefont{Heikkil{\"a}}},
  \bibinfo{author}{\bibfnamefont{A.}~\bibnamefont{Luukanen}},
  \bibinfo{author}{\bibfnamefont{A.~M.} \bibnamefont{Savin}}, \bibnamefont{and}
  \bibinfo{author}{\bibfnamefont{J.~P.} \bibnamefont{Pekola}},
  \bibinfo{journal}{Rev. Mod. Phys.} \textbf{\bibinfo{volume}{78}},
  \bibinfo{pages}{217} (\bibinfo{year}{2006}).

\bibitem[{\citenamefont{Muhonen et~al.}(2012)\citenamefont{Muhonen, Meschke,
  and Pekola}}]{RepProgrPhys.75.046501.2012.Muhonen}
\bibinfo{author}{\bibfnamefont{J.~T.} \bibnamefont{Muhonen}},
  \bibinfo{author}{\bibfnamefont{M.}~\bibnamefont{Meschke}}, \bibnamefont{and}
  \bibinfo{author}{\bibfnamefont{J.~P.} \bibnamefont{Pekola}},
  \bibinfo{journal}{Rep. Progr. Phys.} \textbf{\bibinfo{volume}{75}},
  \bibinfo{pages}{046501} (\bibinfo{year}{2012}).

\bibitem[{\citenamefont{Kauppila et~al.}(2013)\citenamefont{Kauppila, Nguyen,
  and Heikkil\"a}}]{PhysRevB.88.075428.2013.Kauppila}
\bibinfo{author}{\bibfnamefont{V.~J.} \bibnamefont{Kauppila}},
  \bibinfo{author}{\bibfnamefont{H.~Q.} \bibnamefont{Nguyen}},
  \bibnamefont{and} \bibinfo{author}{\bibfnamefont{T.~T.}
  \bibnamefont{Heikkil\"a}}, \bibinfo{journal}{Phys. Rev. B}
  \textbf{\bibinfo{volume}{88}}, \bibinfo{pages}{075428}
  (\bibinfo{year}{2013}).

\bibitem[{\citenamefont{Anghel and
  Kuzmin}(2003)}]{ApplPhysLett.82.293.2003.Anghel}
\bibinfo{author}{\bibfnamefont{D.~V.} \bibnamefont{Anghel}} \bibnamefont{and}
  \bibinfo{author}{\bibfnamefont{L.}~\bibnamefont{Kuzmin}},
  \bibinfo{journal}{Appl. Phys. Lett.} \textbf{\bibinfo{volume}{82}},
  \bibinfo{pages}{293} (\bibinfo{year}{2003}).

\bibitem[{\citenamefont{Pekola et~al.}(2000{\natexlab{b}})\citenamefont{Pekola,
  Manninen, Leivo, Arutyunov, Suoknuuti, Suppula, and
  Collaudin}}]{PhysicaBCondMatt.280.485.2000.Pekola}
\bibinfo{author}{\bibfnamefont{J.~P.} \bibnamefont{Pekola}},
  \bibinfo{author}{\bibfnamefont{A.~J.} \bibnamefont{Manninen}},
  \bibinfo{author}{\bibfnamefont{M.~M.} \bibnamefont{Leivo}},
  \bibinfo{author}{\bibfnamefont{K.}~\bibnamefont{Arutyunov}},
  \bibinfo{author}{\bibfnamefont{J.~K.} \bibnamefont{Suoknuuti}},
  \bibinfo{author}{\bibfnamefont{T.~I.} \bibnamefont{Suppula}},
  \bibnamefont{and}
  \bibinfo{author}{\bibfnamefont{B.}~\bibnamefont{Collaudin}},
  \bibinfo{journal}{Physica B: Cond. Matt.} \textbf{\bibinfo{volume}{280}}
  (\bibinfo{year}{2000}{\natexlab{b}}).

\bibitem[{\citenamefont{Anghel et~al.}(2001)\citenamefont{Anghel, Luukanen, and
  Pekola}}]{ApplPhysLett.78.556.2001.Anghel}
\bibinfo{author}{\bibfnamefont{D.~V.} \bibnamefont{Anghel}},
  \bibinfo{author}{\bibfnamefont{A.}~\bibnamefont{Luukanen}}, \bibnamefont{and}
  \bibinfo{author}{\bibfnamefont{J.~P.} \bibnamefont{Pekola}},
  \bibinfo{journal}{Appl. Phys. Lett.} \textbf{\bibinfo{volume}{78}},
  \bibinfo{pages}{556} (\bibinfo{year}{2001}).

\bibitem[{\citenamefont{Anghel and
  Pekola}(2001)}]{JLowTempPhys.123.197.2001.Anghel}
\bibinfo{author}{\bibfnamefont{D.~V.} \bibnamefont{Anghel}} \bibnamefont{and}
  \bibinfo{author}{\bibfnamefont{J.~P.} \bibnamefont{Pekola}},
  \bibinfo{journal}{J. Low Temp. Phys.} \textbf{\bibinfo{volume}{123}},
  \bibinfo{pages}{197} (\bibinfo{year}{2001}).

\bibitem[{\citenamefont{Leivo and
  Pekola}(1998)}]{ApplPhysLett.72.1305.1998.Leivo}
\bibinfo{author}{\bibfnamefont{M.~M.} \bibnamefont{Leivo}} \bibnamefont{and}
  \bibinfo{author}{\bibfnamefont{J.~P.} \bibnamefont{Pekola}},
  \bibinfo{journal}{Appl. Phys. Lett.} \textbf{\bibinfo{volume}{72}},
  \bibinfo{pages}{1305} (\bibinfo{year}{1998}).

\bibitem[{\citenamefont{Anghel et~al.}(1998)\citenamefont{Anghel, Pekola,
  Leivo, Suoknuuti, and Manninen}}]{PhysRevLett.81.2958.1998.Anghel}
\bibinfo{author}{\bibfnamefont{D.~V.} \bibnamefont{Anghel}},
  \bibinfo{author}{\bibfnamefont{J.~P.} \bibnamefont{Pekola}},
  \bibinfo{author}{\bibfnamefont{M.~M.} \bibnamefont{Leivo}},
  \bibinfo{author}{\bibfnamefont{J.~K.} \bibnamefont{Suoknuuti}},
  \bibnamefont{and} \bibinfo{author}{\bibfnamefont{M.}~\bibnamefont{Manninen}},
  \bibinfo{journal}{Phys. Rev. Lett.} \textbf{\bibinfo{volume}{81}},
  \bibinfo{pages}{2958} (\bibinfo{year}{1998}).

\bibitem[{\citenamefont{K{\"u}hn et~al.}(2004)\citenamefont{K{\"u}hn, Anghel,
  Pekola, Manninen, and Galperin}}]{PhysRevB.70.125425.2004.Kuhn}
\bibinfo{author}{\bibfnamefont{T.}~\bibnamefont{K{\"u}hn}},
  \bibinfo{author}{\bibfnamefont{D.~V.} \bibnamefont{Anghel}},
  \bibinfo{author}{\bibfnamefont{J.~P.} \bibnamefont{Pekola}},
  \bibinfo{author}{\bibfnamefont{M.}~\bibnamefont{Manninen}}, \bibnamefont{and}
  \bibinfo{author}{\bibfnamefont{Y.~M.} \bibnamefont{Galperin}},
  \bibinfo{journal}{Phys. Rev. B} \textbf{\bibinfo{volume}{70}},
  \bibinfo{pages}{125425} (\bibinfo{year}{2004}).

\bibitem[{\citenamefont{Wellstood et~al.}(1994)\citenamefont{Wellstood, Urbina,
  and Clarke}}]{PhysRevB.49.5942.1994.Wellstood}
\bibinfo{author}{\bibfnamefont{F.~C.} \bibnamefont{Wellstood}},
  \bibinfo{author}{\bibfnamefont{C.}~\bibnamefont{Urbina}}, \bibnamefont{and}
  \bibinfo{author}{\bibfnamefont{J.}~\bibnamefont{Clarke}},
  \bibinfo{journal}{Phys. Rev. B} \textbf{\bibinfo{volume}{49}},
  \bibinfo{pages}{5942} (\bibinfo{year}{1994}).

\bibitem[{\citenamefont{Stroscio and Dutta}(2004)}]{Stroscio_Dutta:book}
\bibinfo{author}{\bibfnamefont{M.~A.} \bibnamefont{Stroscio}} \bibnamefont{and}
  \bibinfo{author}{\bibfnamefont{M.}~\bibnamefont{Dutta}},
  \emph{\bibinfo{title}{Phonons in Nanostructures}} (\bibinfo{publisher}{CUP,
  United Kingdom}, \bibinfo{year}{2004}).

\bibitem[{\citenamefont{Karvonen and
  Maasilta}(2007{\natexlab{a}})}]{PRL.99.145503.2007.Karvonen}
\bibinfo{author}{\bibfnamefont{J.~T.} \bibnamefont{Karvonen}} \bibnamefont{and}
  \bibinfo{author}{\bibfnamefont{I.~J.} \bibnamefont{Maasilta}},
  \bibinfo{journal}{Phys. Rev. Lett.} \textbf{\bibinfo{volume}{99}},
  \bibinfo{pages}{145503} (\bibinfo{year}{2007}{\natexlab{a}}).

\bibitem[{\citenamefont{Karvonen and
  Maasilta}(2007{\natexlab{b}})}]{JPhysConfSer.92.012043.2007.Karvonen}
\bibinfo{author}{\bibfnamefont{J.~T.} \bibnamefont{Karvonen}} \bibnamefont{and}
  \bibinfo{author}{\bibfnamefont{I.~J.} \bibnamefont{Maasilta}},
  \bibinfo{journal}{J. Phys. Conf. Ser.} \textbf{\bibinfo{volume}{92}},
  \bibinfo{pages}{012043} (\bibinfo{year}{2007}{\natexlab{b}}).

\bibitem[{\citenamefont{DiTusa et~al.}(1992)\citenamefont{DiTusa, Lin, Park,
  Isaacson, and Parpia}}]{PhysRevLett.68.1156.1992.DiTusa}
\bibinfo{author}{\bibfnamefont{J.~F.} \bibnamefont{DiTusa}},
  \bibinfo{author}{\bibfnamefont{K.}~\bibnamefont{Lin}},
  \bibinfo{author}{\bibfnamefont{M.}~\bibnamefont{Park}},
  \bibinfo{author}{\bibfnamefont{M.~S.} \bibnamefont{Isaacson}},
  \bibnamefont{and} \bibinfo{author}{\bibfnamefont{J.~M.}
  \bibnamefont{Parpia}}, \bibinfo{journal}{Phys. Rev. Lett.}
  \textbf{\bibinfo{volume}{68}}, \bibinfo{pages}{1156} (\bibinfo{year}{1992}).

\bibitem[{\citenamefont{Qu et~al.}(2005)\citenamefont{Qu, Cleland, and
  Geller}}]{PhysRevB.72.224301.2005.Qu}
\bibinfo{author}{\bibfnamefont{S.-X.} \bibnamefont{Qu}},
  \bibinfo{author}{\bibfnamefont{A.~N.} \bibnamefont{Cleland}},
  \bibnamefont{and} \bibinfo{author}{\bibfnamefont{M.~R.}
  \bibnamefont{Geller}}, \bibinfo{journal}{Phys. Rev. B}
  \textbf{\bibinfo{volume}{72}}, \bibinfo{pages}{224301}
  (\bibinfo{year}{2005}).

\bibitem[{\citenamefont{Bannov et~al.}(1995)\citenamefont{Bannov, Aristov,
  Mitin, and Stroscio}}]{PhysRevB.51.9930.1995.Bannov}
\bibinfo{author}{\bibfnamefont{N.}~\bibnamefont{Bannov}},
  \bibinfo{author}{\bibfnamefont{V.}~\bibnamefont{Aristov}},
  \bibinfo{author}{\bibfnamefont{V.}~\bibnamefont{Mitin}}, \bibnamefont{and}
  \bibinfo{author}{\bibfnamefont{M.~A.} \bibnamefont{Stroscio}},
  \bibinfo{journal}{Phys. Rev. B} \textbf{\bibinfo{volume}{51}},
  \bibinfo{pages}{9930} (\bibinfo{year}{1995}).

\bibitem[{\citenamefont{Glavin et~al.}(2002)\citenamefont{Glavin, Pipa, Mitin,
  and Stroscio}}]{PhysRevB.65.205315.2002.Glavin}
\bibinfo{author}{\bibfnamefont{B.~A.} \bibnamefont{Glavin}},
  \bibinfo{author}{\bibfnamefont{V.~I.} \bibnamefont{Pipa}},
  \bibinfo{author}{\bibfnamefont{V.~V.} \bibnamefont{Mitin}}, \bibnamefont{and}
  \bibinfo{author}{\bibfnamefont{M.~A.} \bibnamefont{Stroscio}},
  \bibinfo{journal}{Phys. Rev. B} \textbf{\bibinfo{volume}{65}},
  \bibinfo{pages}{205315} (\bibinfo{year}{2002}).

\bibitem[{\citenamefont{Viljas and
  Heikkil\"a}(2010)}]{PhysRevB.81.245404.2010.Viljas}
\bibinfo{author}{\bibfnamefont{J.~K.} \bibnamefont{Viljas}} \bibnamefont{and}
  \bibinfo{author}{\bibfnamefont{T.~T.} \bibnamefont{Heikkil\"a}},
  \bibinfo{journal}{Phys. Rev. B} \textbf{\bibinfo{volume}{81}},
  \bibinfo{pages}{245404} (\bibinfo{year}{2010}).

\bibitem[{\citenamefont{Hekking et~al.}(2008)\citenamefont{Hekking, Niskanen,
  and Pekola}}]{PhysRevB.77.033401.2008.Hekking}
\bibinfo{author}{\bibfnamefont{F.~W.~J.} \bibnamefont{Hekking}},
  \bibinfo{author}{\bibfnamefont{A.~O.} \bibnamefont{Niskanen}},
  \bibnamefont{and} \bibinfo{author}{\bibfnamefont{J.~P.}
  \bibnamefont{Pekola}}, \bibinfo{journal}{Phys. Rev. B}
  \textbf{\bibinfo{volume}{77}}, \bibinfo{pages}{033401}
  (\bibinfo{year}{2008}).

\bibitem[{\citenamefont{Muhonen et~al.}(2009)\citenamefont{Muhonen, Niskanen,
  Meschke, Pashkin, Tsai, Sainiemi, Franssila, and
  Pekola}}]{ApplPhysLett.94.073101.2009.Muhonen}
\bibinfo{author}{\bibfnamefont{J.~T.} \bibnamefont{Muhonen}},
  \bibinfo{author}{\bibfnamefont{A.~O.} \bibnamefont{Niskanen}},
  \bibinfo{author}{\bibfnamefont{M.}~\bibnamefont{Meschke}},
  \bibinfo{author}{\bibfnamefont{Y.~A.} \bibnamefont{Pashkin}},
  \bibinfo{author}{\bibfnamefont{J.~S.} \bibnamefont{Tsai}},
  \bibinfo{author}{\bibfnamefont{L.}~\bibnamefont{Sainiemi}},
  \bibinfo{author}{\bibfnamefont{S.}~\bibnamefont{Franssila}},
  \bibnamefont{and} \bibinfo{author}{\bibfnamefont{J.~P.}
  \bibnamefont{Pekola}}, \bibinfo{journal}{Appl. Phys. Lett.}
  \textbf{\bibinfo{volume}{94}}, \bibinfo{pages}{073101}
  (\bibinfo{year}{2009}).

\bibitem[{\citenamefont{Wu et~al.}(2008)\citenamefont{Wu, Choi, Krupin,
  Rotenberg, Wu, and Qiu}}]{JPhysCondMatt.20.035213.2008.Wu}
\bibinfo{author}{\bibfnamefont{J.}~\bibnamefont{Wu}},
  \bibinfo{author}{\bibfnamefont{J.}~\bibnamefont{Choi}},
  \bibinfo{author}{\bibfnamefont{O.}~\bibnamefont{Krupin}},
  \bibinfo{author}{\bibfnamefont{E.}~\bibnamefont{Rotenberg}},
  \bibinfo{author}{\bibfnamefont{Y.~Z.} \bibnamefont{Wu}}, \bibnamefont{and}
  \bibinfo{author}{\bibfnamefont{Z.~Q.} \bibnamefont{Qiu}},
  \bibinfo{journal}{J. Phys. Cond. Matt.} \textbf{\bibinfo{volume}{20}},
  \bibinfo{pages}{035213} (\bibinfo{year}{2008}).

\bibitem[{\citenamefont{Anghel and
  K{\"u}hn}(2007)}]{JPhysA.40.10429.2007.Anghel}
\bibinfo{author}{\bibfnamefont{D.~V.} \bibnamefont{Anghel}} \bibnamefont{and}
  \bibinfo{author}{\bibfnamefont{T.}~\bibnamefont{K{\"u}hn}},
  \bibinfo{journal}{J. Phys. A: Math. Theor.} \textbf{\bibinfo{volume}{40}},
  \bibinfo{pages}{10429} (\bibinfo{year}{2007}),
  \bibinfo{note}{cond-mat/0611528}.

\bibitem[{\citenamefont{Auld}(1990)}]{Auld:book}
\bibinfo{author}{\bibfnamefont{B.~A.} \bibnamefont{Auld}},
  \emph{\bibinfo{title}{Acoustic Fields and Waves in Solids, 2nd Ed.}}
  (\bibinfo{publisher}{Robert E. Krieger Publishing Company},
  \bibinfo{year}{1990}).

\bibitem[{\citenamefont{Ziman}(1976)}]{Ziman:book}
\bibinfo{author}{\bibnamefont{Ziman}}, \emph{\bibinfo{title}{Electrons and
  Phonons}} (\bibinfo{publisher}{Harcourt College Publishers},
  \bibinfo{year}{1976}), ISBN \bibinfo{isbn}{0-03-083993-9}.

\end{thebibliography}

\end{document}